\documentclass[12pt]{article}
\title{Monotone Emulation of Computer Experiments}
\usepackage{amsmath,latexsym, graphics,graphicx,showexpl,amsthm,amsfonts,fullpage}
\usepackage{graphicx}
\usepackage{subcaption}
\usepackage{natbib}
\usepackage{ednote}
\usepackage[pdftex,bookmarks,colorlinks=true,linkcolor=black]{hyperref}
\usepackage{algpseudocode,algorithm}
\usepackage{authblk}
\usepackage{bbm}

\usepackage{graphicx,eso-pic,xcolor}
\makeatletter
\AddToShipoutPicture{%
\setlength{\@tempdimb}{.5\paperwidth}%
\setlength{\@tempdimc}{.5\paperheight}%
\setlength{\unitlength}{1pt}%
\put(\strip@pt\@tempdimb,\strip@pt\@tempdimc){%
    \makebox(-500,200){\rotatebox{90}{\textcolor[gray]{0.70}%
       {\Large \textsf{Preprint, June 2014}}}}
  }%
}

\author[1]{Shirin Golchi\thanks{sgolchi@sfu.ca}}
\author[1]{Derek R. Bingham}
\author[2]{Hugh Chipman}
\author[1]{David A. Campbell}
\affil[1]{Department of Statistics and Actuarial Science, Simon Fraser university, Burnaby, BC, Canada V5A1S6}
\affil[2]{Department of Mathematics and Statistics, Acadia University, Wolfville, NS  Canada  B4P 2R6}

\date{}
\begin{document}

\maketitle

\newcommand{\etab}   {\mbox{\boldmath${\eta}$}}
\newcommand{\tmnote}[1]{{\bluenote{#1}}}
\newcommand{\kth}{$k^{\mbox{\footnotesize th}}$ }
\newcommand{\jth}{$j^{\mbox{\footnotesize th}}$ }
\begin{abstract}
In statistical modeling of computer experiments, prior information is sometimes available about the underlying function. For example, the physical system simulated by the computer code may be known to be monotone with respect to some or all inputs. We develop a Bayesian approach to Gaussian process modeling capable of incorporating monotonicity information for computer model emulation. Markov chain Monte Carlo methods are used to sample from the posterior distribution of the process given the simulator output and monotonicity information. The performance of the proposed approach in terms of predictive accuracy and uncertainty quantification is demonstrated in a number of simulated examples as well as a real queuing system application.
\end{abstract}
\noindent%
\vfill

\newpage

\section{Introduction}

Deterministic computer models are commonly used to study complex physical phenomena in many areas of science and engineering.  Oftentimes, evaluating a computational model can be very time consuming, and thus the simulator is only exercised on a relatively small set of input values.  In such cases, a statistical surrogate model (i.e., an emulator) is used to make predictions of the computer model output at unsampled inputs.   To this end, a Gaussian process (GP) model is a popular choice for deterministic computer model emulation \citep{Sacks1989}. The reason for this rests largely on the ability of the GP to interpolate the output of the simulator, the flexibility of the GP as a non-parametric regression estimator and its ability to provide a basis for statistical inference for deterministic computer models. 

An important property of GP models that makes them attractive in some settings is that the partial derivative processes are also GPs with covariance functions that can be derived from the original process  (see for e.g. \cite{Rasmussen2006}, Section 9.4).  Indeed, when the simulator output includes derivatives, they can be used to improve the accuracy of the emulator \citep{Morris1993}.   In some applications, derivative information is available only in the form of qualitative information - for example, the computer model response is known to be monotone increasing or decreasing in some of the inputs. Incorporating the derivative information into the emulator in such cases  is more challenging because the derivative values are unknown. The problem of using the known monotonicity of the computer model response in one or more of the inputs to build a more accurate emulator is the main focus of this paper. Although our approach can be applied in situations with observation error, we focus primarily on GP emulation of computer experiments.

While a rich literature exists on monotone function estimation, interpolation of monotone functions with uncertainty quantification remains an understudied topic. Examples of related work are monotone smoothing splines, isotonic regression, etc. See for e.g. \cite{Ramsay98}, \cite{He98} and \cite{Dette06}. Also work has been done on incorporating constraints in general and monotonicity specially into Gaussian process regression as discussed below. In a related framework, constrained Kriging has been considered in the area of geostatistics \citep{Kleijnen13}. While some of the existing methods may be modified to be used in the noise-free set-up none of them directly address monotone interpolation. On the other hand, there exist tools for monotone interpolation that do not provide uncertainty estimates. See for e.g. \cite{Wolberg99}.

\cite{RV2010} considered monotonicity assumptions for GP models for noisy data. They incorporate the monotonicity information by placing {\em virtual} derivatives at pre-specified input locations, thereby encouraging the derivative process to be positive at these points. They use expectation-propagation techniques \citep{Minka2001} to approximate the joint marginal likelihood of function and derivative values. Point estimates for the hyper-parameters are obtained by maximizing this approximate likelihood.

In this paper we initially take an approach similar to \cite{RV2010} to build an emulator, given the computer model output and monotonicity information.  Our approach is different in two respects. First, we focus on deterministic computer experiments where interpolation of the simulator is a requirement. Constructing a monotone emulator is more challenging in the deterministic setting than the noisy setting.  The problem in our case lies in generating sample paths from the GP that obey monotonicity  and also interpolate the simulator output.  In the noisy setting, where the GP need not interpolate the observations, sampling from the GP is easier.  Second, we sample from the exact joint posterior distribution rather than relying on an approximation. In doing so, we provide fully Bayesian inference for all parameters of the emulator as well as the predicted function at unsampled inputs, instead of using plug-in estimates of the parameters, thereby underestimating the posterior predictive uncertainty. We also take advantage of the flexible parametrization of monotonicity information to facilitate efficient computation. We show that when the monotonicity constraints are more strict, virtually all the posterior mass of the distribution for the derivatives falls in $\Re^+$. The end result of the proposed approach is an emulator of the computer model that uses the monotonicity information and is more accurate than the standard GP.

An advantage of approaches which use derivative sign information at specific locations in the input space, is that they offer flexibility in incorporation of monotonicity information. For example, by specifying that derivatives are positive with respect to a particular variable, at particular locations, we have the flexibility to make predictions of the response that has monotone relationship with a predictor in just a subset of the range of the predictor.

A related Bayesian approach to monotone estimation of GP models has been independently developed in \cite{Wang2012}.  Similar to this paper, the sign of derivatives at user-specified locations is assumed known.  Two modeling approaches are taken: i) an indicator variable formulation, which can be seen as a limiting case of a probit link of \cite{RV2010} (also used in this paper), and ii) a conditional GP model, which allows zero or positive derivative values.  Unlike \cite{Wang2012}, who uses plug-in estimates of GP parameters, we conduct full inference for GP parameters as well as function and derivative values.  We also demonstrate applications with more than one input dimension.  The extension to higher-order derivatives which \cite{Wang2012} considers, is not developed here. While \cite{Wang2012} uses Gibbs sampling to sample from the marginal posterior with the covariance parameters held fixed at their maximum likelihood estimates, we employ a sequentially constrained Monte Carlo (SCMC) \citep{Golchi2014} algorithm that permits sampling from the full posterior in fairly high-dimensional scenarios.

\cite{Lin13} propose a GP based method for estimating monotone functions that relies on projecting sample paths obtained from a GP fit to the space of monotone functions. They use the pooled adjacent violators (PAV) algorithm to obtain approximate projections. While the projections of interpolating GP sample paths into the space of monotone functions are not generally guaranteed to interpolate the function evaluations, the PAV algorithm can be modified to generate monotone interpolants. However, there are two drawbacks to this method; firstly, inference cannot be made about the GP model parameters by projecting GP sample paths since the posterior distribution of the covariance parameters is affected by the projection. Secondly, sample paths generated by the PAV algorithm are often non-smooth since monotonicity is gained by flattening the ridges and valleys resulting in flat segments followed by occasional rises. ``Box-like" credible intervals are obtained from the projected interpolants with truncation from below and above to exclude violating sample paths. These credible intervals can remain unchanged for a range of different coverage probabilities, making their interpretation difficult. 

The article is organized as follows.  Section 2 presents two motivating examples.  The first illustration demonstrates a situation where the unconstrained GP performs poorly. We also describe a motivating application involving a queuing system. New methodology for GP emulation of monotone computer models is proposed in Section 3. The monotone interpolation variant of the SCMC algorithm that samples from the full posterior is explained in Section 4. In Section 5, some insight and guidelines on the construction of the derivative design is provided. Two simulated examples illustrate the performance of the method in Section 6, and a simulation study is described in Section 7.  The proposed method is applied to the queuing system application in Section 8, and we finish the paper with some concluding remarks in the subsequent section.

\section{Motivating examples} 
\subsection{Illustrative example}\label{Ex}

\begin{figure}[h!]
        \centering
        \begin{subfigure}[b]{0.4\textwidth}
                \centering
                \includegraphics[width=\textwidth]{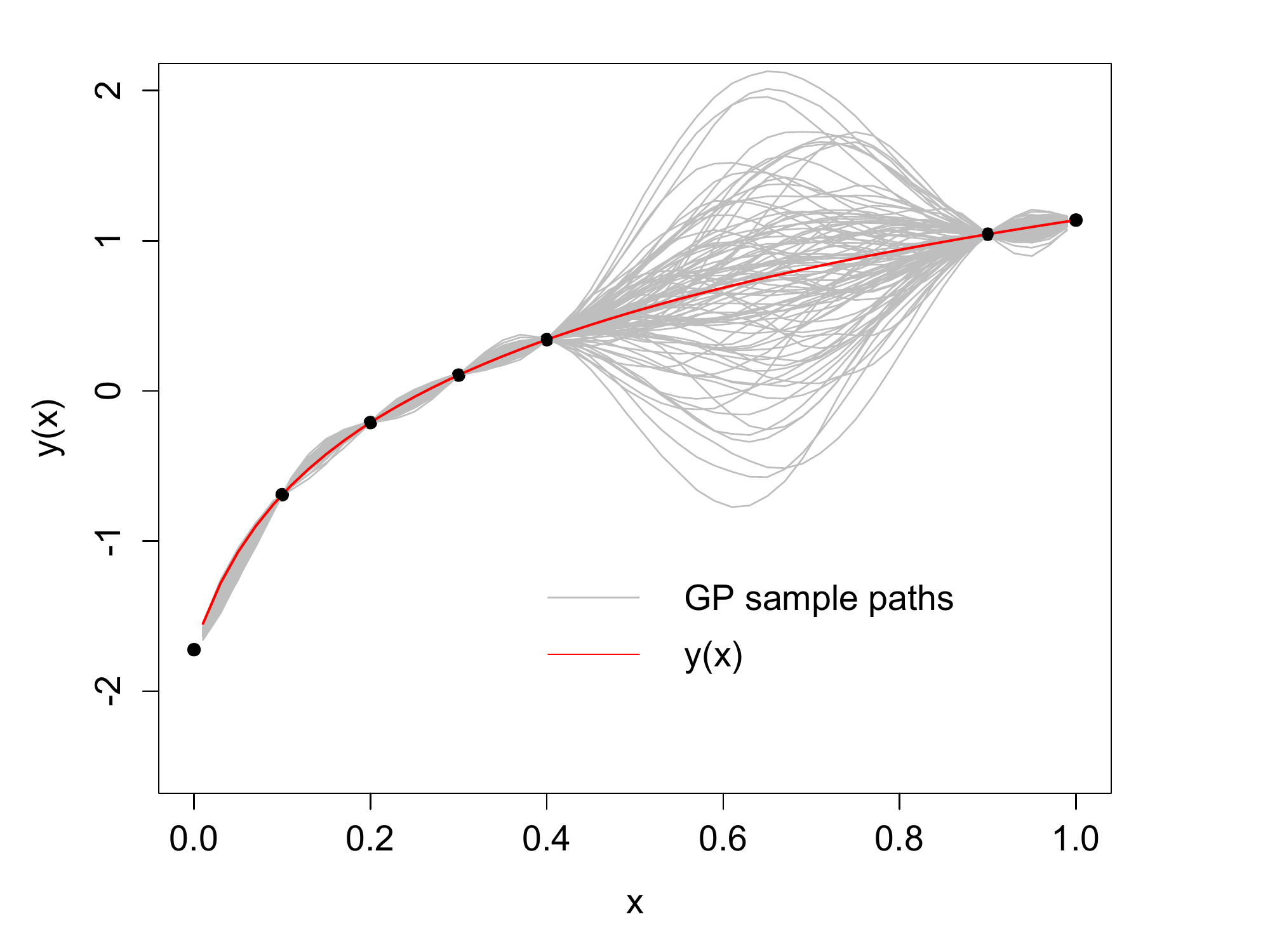}
                \caption{}
                \label{polya}
        \end{subfigure}\\

        \begin{subfigure}[b]{0.4\textwidth}
                \centering
                \includegraphics[width=\textwidth]{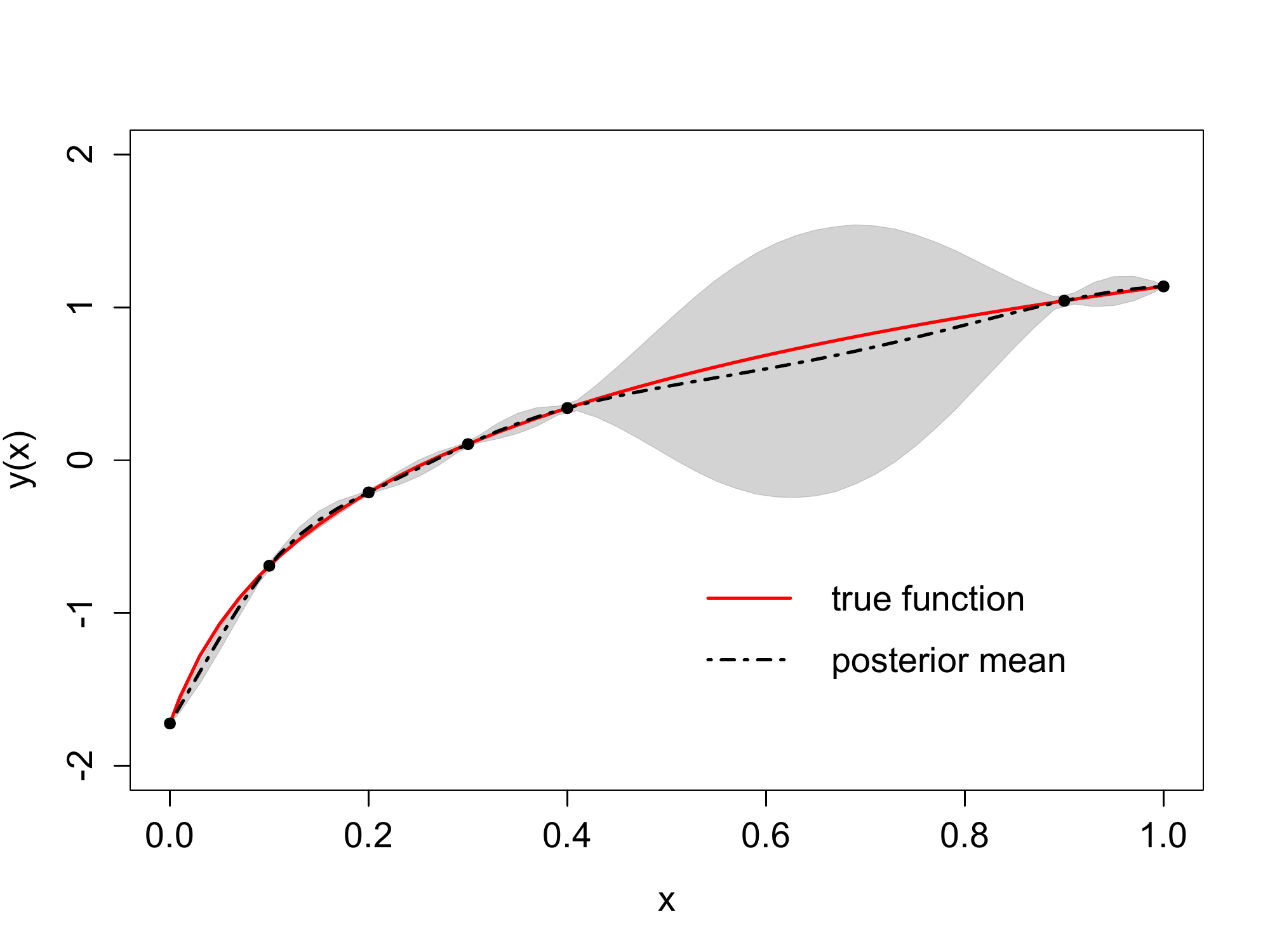}
                \caption{}
                \label{polyb}
        \end{subfigure}~
        \begin{subfigure}[b]{0.4\textwidth}
                \centering
                \includegraphics[width=\textwidth]{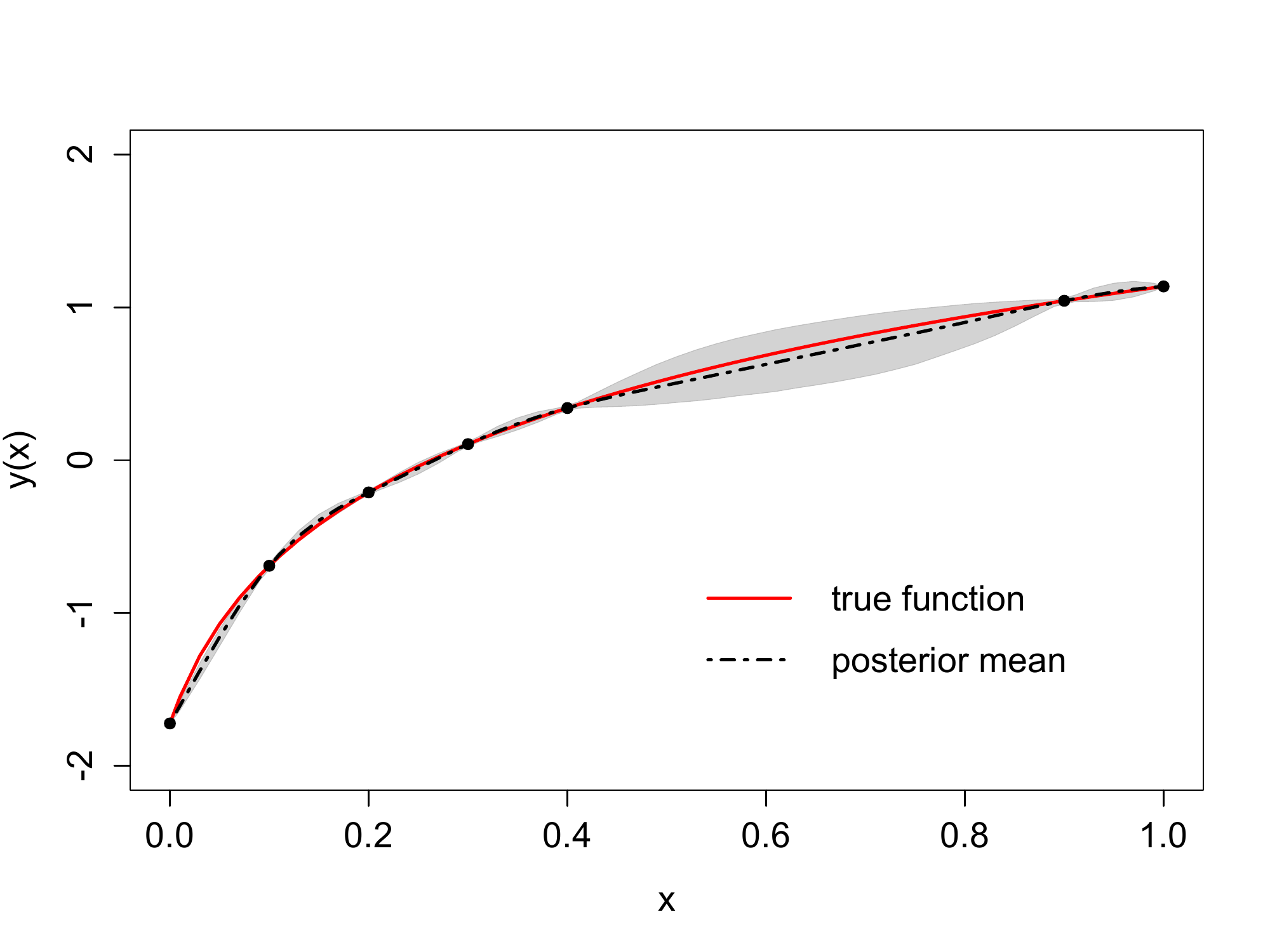}
                \caption{}
                \label{polyc}
        \end{subfigure}

        \caption{Illustrative Example: (a) the true function plotted against $x$, the 7 evaluated points of the function, and 100 sample paths taken from the GP posterior (b) posterior mean and 95\% credible intervals from the GP model together with the true function (c) posterior mean and 95\% credible intervals from the constrained model together with the true function }\label{poly}
\end{figure}

Consider the simple monotone function in Figure~\ref{polya}, sampled at $n=7$ design points. A GP model provides estimates of the function evaluations and uncertainties at unsampled points (Figure~\ref{polyb}) (we leave the details of GP models until Section~\ref{gpr}).

This example illustrates several issues. Suppose that the function is known to be monotone increasing.   Since the GP model does not incorporate monotonicity information, the posterior predictive distribution in Figure~\ref{polya}  includes sample paths (the gray curves) that decrease.  Furthermore, 95\% credible intervals in the range $x \in (0.4, 0.9)$ are relatively wide (Fiqure~\ref{polyb}), reflecting the wide variety of sample paths suggested by the GP model. While it is unlikely that an experimenter would run a design with such a large gap between the fifth and sixth design points (i.e., $0.4 < x < 0.9$), sizeable gaps are likely to occur in practice in higher input dimensions - especially when the usual run-size recommendations for computer model emulation (e.g., \cite{Leoppky09}) are adopted.

If the experimenter knows beforehand that the computational model is monotone (increasing in this example), this information should be used to rule out some of the non-monotone proposals for the emulator. We will see (Section~\ref{monmodel}) that using our proposed methodology, most of the posterior mass can be concentrated on increasing functions and the posterior predictive uncertainty can be reduced (Figure~\ref{polyc}). 

\subsection{Queuing system application}\label{queue}

The example that motivated this work is a computer networking queuing system in \cite{Bambos04} and \cite{Ranjan08}. Consider a single server and two first-in-first-out queues of jobs arriving stochastically at rates $x_1$ and $x_2$ (denoted as  $\lambda_1$ and $\lambda_2$ in \cite{Ranjan08}). The queues have infinite capacity buffers for the jobs waiting to be served.  To make decisions on which queue to serve, a server allocation/scheduling policy is used to maximize the average amount of work processed by the system per unit of time \citep{Bambos04}.   As mentioned by \cite{Ranjan08}, the queuing system also captures essential dynamics of many practical systems, such as data/voice transmissions in wireless networks or multi-product manufacturing systems.

A performance measure of the system is the average delay for the jobs to be served, as a function of the arrival rates $x_1$ and $x_2$. This average delay is not available in closed form and is evaluated by a computational model that simulates the system. Details of the simulator are discussed in \cite{Ranjan08}.  

An important characteristic of this queuing process is that the average delay is a monotone increasing function of the job arrival rates. The increase is negligible for small $x_1$ and $x_2$, yielding a nearly flat function, but eventually the average delay increases exponentially as the arrival rate(s) increase. With such behavior in the response surface, predicting the average delay at unsampled input rates is a challenge.   The monotone GP model introduced in Section~\ref{monmodel} will serve as a guide for improving inference in this context.

\section{Gaussian process models with monotonicity information}

We begin this section with a brief review of GP models and use of derivatives in GP regression. Next, new methodology for emulation of monotone computer models and a sequential Monte Carlo (SMC) algorithm for this setting are introduced.  

\subsection{Gaussian process regression}\label{gpr}

Let $y(.)$ be the function encoded in the computer model that is evaluated at $n$ design points (or input locations) given by the rows of the $n \times d$ design matrix $\mathbf{X}=(\mathbf{x}_1,\ldots,\mathbf{x}_n)^T$, where $\mathbf{x}_i  \in \Re^d$.  It is standard to assume that the response surface for the computer model is a realization of a GP \citep{Sacks1989}.  In practice, this amounts to placing a prior distribution over the class of functions produced by the simulator.  

Denote the vector of computer model outputs as $\mathbf{y}=(y_1,\ldots,y_n)^T$, where $y_i=y(\mathbf{x}_i)$ ($i=1, \ldots , n$). We specify $y(\mathbf{x})$ be a zero mean (without loss of generality) GP with an anisotropic, stationary covariance function. Therefore, $\mathbf{y}$ is a realization of a mean zero multivariate normal distribution with a covariance matrix $R_{n\times n}$ whose $(i,j)$th element is given by
\begin{equation*}
\label{covfnct}
\mbox{Cov}(y_i,y_j) = r(\mathbf{x}_i,\mathbf{x}_j)
= \sigma^2\prod_{k=1}^{d}g(|x_{ik}-x_{jk}|,l_k),   
\end{equation*}
where $ \boldsymbol{\it l}=(l_1,\ldots,l_d)$ is the vector of length scale parameters in each dimension and $\sigma^2$ is the constant variance. We choose $g(.)$ from the Matern class of covariance functions that are of the form
$$g_{\lambda}(|x_{ik}-x_{jk}|,l_k)=\frac{2^{1-\lambda}}{\Gamma(\lambda)}(\frac{\sqrt{2\lambda}}{l_k}|x_{ik}-x_{jk}|)^\lambda K_\lambda(\frac{\sqrt{2\lambda}}{l_k}|x_{ik}-x_{jk}|),$$
where $\Gamma$ is the gamma function, $K$ is the modified Bessel function of the second kind, and $\lambda$ is a non-negative parameter. The Matern correlation function is $t$-times mean square differentiable if and only if $\lambda>t$. We choose $\lambda=\frac{5}{2}$.

This choice of the Matern covariance function over the commonly used squared exponential family \citep{Sacks1989} avoids numerical instability, often observed when inverting the covariance matrix, by removing the restriction of infinite differentiability. Note that twice mean square differentiability is a requirement to be able to obtain the covariance function for the derivative process and is likely to be the level of smoothness one can safely assume in practice.

Following the above specification for the GP, the conditional distribution of $y(.)$ at any unsampled input $\mathbf{x}^*$, given the evaluations and the covariance parameters, is given by
\begin{eqnarray}
\label{postpred}
[y(\mathbf{x}^*)|\mathbf{y},\mathbf{\boldsymbol{\it l}},\sigma^2]={\cal N}(\mu(\mathbf{x}^*),\tau^2(\mathbf{x}^*)),
\end{eqnarray}
where
\begin{eqnarray*}
\mu(\mathbf{x}^*)=r(\mathbf{x}^*,\mathbf{X})R^{-1}\mathbf{y},
\end{eqnarray*}

\begin{eqnarray*}
\tau^2(\mathbf{x}^*)=r(\mathbf{x}^*,\mathbf{X})R^{-1}r(\mathbf{X},\mathbf{x}^*),
\end{eqnarray*}
and $r(\mathbf{x}^*,\mathbf{X})$ is the $n$-vector of correlations between the observation at $\mathbf{x}^*$ and the observations at the design points, $\mathbf{X}$.
The mean $\mu(\mathbf{x}^*)$ is often used as the prediction of the computer model response at $\mathbf{x}^*$  (i.e., $\hat{y}(\mathbf{x}^*)=\mu(\mathbf{x}^*)$) \citep{Jones98}.

We take a Bayesian approach to make inference about the GP parameters, $\mathbf{\boldsymbol{\it l}}$ and $\sigma^2$, instead of replacing them by their maximum likelihood point estimates \citep{Jones98}. Assuming a prior distribution, $[\mathbf{\boldsymbol{\it l}},\sigma^2]$, the joint posterior distribution of the GP parameters and $y(\mathbf{x}^*)$ is given by 
\begin{eqnarray}
\label{jointpostpred}
[\mathbf{\boldsymbol{\it l}},\sigma^2,y(\mathbf{x}^*)|\mathbf{y}]\propto [y(\mathbf{x}^*)|\mathbf{\boldsymbol{\it l}},\sigma^2,\mathbf{y}][\mathbf{y}|\mathbf{\boldsymbol{\it l}},\sigma^2][\mathbf{\boldsymbol{\it l}},\sigma^2],
\end{eqnarray}
where the first term on the right hand side of (\ref{jointpostpred}) is given by (\ref{postpred}) and the second term is the likelihood. We delay discussion of the specific choices of prior distributions until Section~\ref{examples}.

\subsection{Derivatives of a Gaussian process}

In the context of computer model emulation,  a more efficient emulator can be obtained by combining the model response and derivatives when the derivatives are observed  \citep{Morris1993}.   Results relating the GP and observed partial derivatives are presented in this section.  Of course, in our setting the partial derivatives are not observed, but we treat the derivatives as unobserved latent variables in the next section - doing so will allow us to incorporate monotonicity information into the emulator.

Consider the partial derivative of the GP, $y(\mathbf{x})$, with respect to $x_k$ (i.e., the \kth input dimension). Denote this derivative function by $y^{\prime}_k(\mathbf{x})=\frac{\partial}{\partial x_k}y(\mathbf{x})$. Since differentiation is a linear operator, the partial derivatives of a GP are also GPs \citep{Morris1993}. 

The means and covariances of the derivative function can be obtained by differentiating the mean and covariance of $y(\mathbf{x})$ (\cite{Parzen62},page 83), i.e.,
\begin{eqnarray*}
\mbox{E}[y^{\prime}_k(\mathbf{x})]=\frac{\partial}{\partial x_{k}}\mbox{E}[y(\mathbf{x})],
\end{eqnarray*}

\begin{eqnarray}
\label{corder}
\mbox{Cov}[y^{\prime}_k(\mathbf{x}_i),y(\mathbf{x}_j)]=r_{1k}(\mathbf{x}_i,\mathbf{x}_j)=\frac{\partial}{\partial x_{ik}}r(\mathbf{x}_i,\mathbf{x}_j),
\end{eqnarray}
and 
\begin{eqnarray*}
\mbox{Cov}[y^{\prime}_k(\mathbf{x}_i),y^{\prime}_k(\mathbf{x}_j)]=r_{2k}(\mathbf{x}_i,\mathbf{x}_j)=\frac{\partial^2}{\partial x_{ik} \partial x_{jk}}r(\mathbf{x}_i,\mathbf{x}_j).
\end{eqnarray*}
The analytic formulas for the derivatives of the Matern covariance function that we use are given in Appendix~\ref{derivatives}. Let  $\mathbf{X}'=(\mathbf{x}'_1,\ldots,\mathbf{x}'_p)^T$ be a set of input points where derivatives have been observed (note that $\mathbf{X}^{\prime}$ does not necessarily have any points in common with $\mathbf{X}$).  Furthermore, denote the vector of partial derivatives in the \kth input dimension, observed at $\mathbf{X}^{\prime}$  as $\mathbf{y}'_k=(y^{\prime}_{k,1},\ldots,y^{\prime}_{k,p})^T$, where $y^{\prime}_{k,i}=y^{\prime}_k(\mathbf{x}'_i)$ ($i=1, \ldots , p$).  The joint distribution of simulator evaluations, $\mathbf{y}$, at the design matrix, $\mathbf{X}$, and partial derivatives, $\mathbf{y}'_k$, at a set of points given by the matrix $\mathbf{X}'$, given the GP parameters follows a multivariate Gaussian distribution,
\begin{eqnarray}
\label{GP}
[\mathbf{y},\mathbf{y}^{\prime}_k(\mathbf{X}^{\prime})|\mathbf{\boldsymbol{\it l}},\sigma^2]={\cal N}(\mu,\Sigma),
\end{eqnarray}
where
\begin{eqnarray*}
\mu=\begin{pmatrix}
     \mbox{E}[\mathbf{y}]\\
     \mbox{E}[\mathbf{y}'_k]
     \end{pmatrix},
\end{eqnarray*}
and
\begin{eqnarray}
\label{covmat}
\Sigma=\begin{pmatrix} 
r(\mathbf{X},\mathbf{X}) & r_{1k}(\mathbf{X},\mathbf{X}^{\prime}) \\
r_{1k}(\mathbf{X}^{\prime},\mathbf{X}) & r_{2k}(\mathbf{X}^{\prime},\mathbf{X}^{\prime})
\end{pmatrix}.
\end{eqnarray}

Note that the four blocks of $\Sigma$ are covariance matrices whose components are obtained by applying the covariance functions to the points in $\mathbf{X}$ and $\mathbf{X}'$. Partial derivatives with respect to more than one dimension could be included analogously. 

\subsection{Incorporating monotonicity information}\label{monmodel}

In some computer model applications, the simulator output includes derivatives with respect to some of the input dimensions  \citep{Morris1993}.  In our setting, the experimenter knows beforehand that the simulator response is monotone in some, or all, of the inputs, but only knows the sign of the derivatives. The magnitude of the derivatives are unknown.  

Consider, for example, the one-dimensional illustration given in Section~\ref{Ex}. Suppose that in addition to the seven function evaluations, it is also known that $\frac{\partial}{\partial{x_k}}y(x)>0$, but the values of the partial derivative are unknown. Figure 1(b) shows the prediction intervals using the usual GP, while Figure 1(c) illustrates the prediction intervals using the methodology we will propose shortly.  Clearly, the prediction intervals are smaller when the derivative information is used in the predictions.   

In this section, we propose methodology to construct a more efficient emulator than the usual GP when the response is monotone in one or more of the inputs, but where the derivative process is unobserved.
 We consider the case where the function is strictly increasing with respect to the \kth input. The strictly decreasing case is handled similarly by replacing $y_k^{\prime}$ with $-y_k^{\prime}$.

With no constraints imposed, the derivatives of the GP, outlined in the previous section, take values in $\Re$. \cite{RV2010} proposed a method to impose the positivity constraint on the derivatives at a set of $p$ specified inputs $\mathbf{X}'=(\mathbf{x}'_1, \ldots , \mathbf{x}'_p)^T$.   Although the constraints at $p$ input values do not guarantee monotonicity everywhere, with enough well-placed points, monotone functions are made highly probable. They used a probit function that links the monotonicity information with the derivative values that are treated as unobserved latent variables. Building on their notation, let $m_k(\mathbf{x}')$ be a binary random variable that is equal to 1 when the derivative in the \kth input dimension is positive at $\mathbf{x}'$ and 0 otherwise. The $m_k$'s are linked to the corresponding latent partial derivatives through the following probit function,
\begin{eqnarray}
\label{monlik}
P(m_k(\mathbf{x^{\prime}})=1|y^{\prime}_k(\mathbf{x^{\prime}}),\tau)=\Phi(\tau y^{\prime}_k(\mathbf{x^{\prime}})),
\end{eqnarray} 
where $\Phi$ is the standard normal cumulative distribution function and $\tau>0$ controls the strictness of monotonicity constraints. A small value of $\tau$ relaxes the monotonicity constraints by allowing positive conditional probability for the event $m_k(\mathbf{x}')=1$ given negative derivative values. When $\tau=0$ the events $m_k(\mathbf{x}')=0$ and $m_k(\mathbf{x^{\prime}})=1$ will have equal conditional probabilities regardless of the value of $y'_k(\mathbf{x}')$, which corresponds to the usual  unconstrained GP,
\begin{eqnarray*}
P(m_k(\mathbf{x^{\prime}})=1|y^{\prime}_k(\mathbf{x^{\prime}}),\tau=0)=1-P(m_k(\mathbf{x^{\prime}})=0|y^{\prime}_k(\mathbf{x^{\prime}}),\tau=0)=\frac{1}{2}.
\end{eqnarray*}
At the other extreme, as $\tau\rightarrow \infty$ the conditional probability of the event $m=1$ given that $y^{\prime}$ is positive is $1$ and it is 0, otherwise, i.e., (\ref{monlik}) will approach a deterministic step function of $y'$, taking a steep step at $y'=0$,
\begin{eqnarray}
\label{mon}
\lim_{\tau\rightarrow \infty}P(m_k(\mathbf{x^{\prime}})=1|y^{\prime}_k(\mathbf{x^{\prime}}),\tau)=1-\lim_{\tau\rightarrow \infty}P(m_k(\mathbf{x^{\prime}})=0|y^{\prime}_k(\mathbf{x^{\prime}}),\tau)=\mathbbm{1}_{(0,\infty)}(y^{\prime}_k),
\end{eqnarray}
where $\mathbbm{1}_{A}(\cdot)$ is the indicator function of set $A$.  Consequently, under the above set-up, the constrained and unconstrained GPs can be viewed as opposite extremes of the same model.

Ideally, the parameter $\tau$ should be chosen to be as large as possible so that the probability of a negative derivative is close to 0. In practice, there is a trade-off between the strictness of the monotonicity constraint and the ease of sampling from the posterior predictive distribution. We use the monotonicity parameter, $\tau$, to facilitate sampling from the target posterior distribution using a variant of the sequential Monte Carlo samplers (see Section~\ref{SMC}).  

The idea behind the proposed approach is to augment the computer experiment with monotonicity information at a set of $p$ locations ($\mathbf{X}'$) in the input space to encourage the emulator to be monotone. That is, the aim is to estimate the simulator output while monotonicity information in the form of $m_k(\mathbf{x})$ is used.   Next, (\ref{monlik}) is used to link the monotonicity information to the derivative process, that in turn is connected to the emulator using the results in the previous section.

For the time being, we condition on the GP parameters $\mathbf{\boldsymbol{\it l}}$ and $\sigma^2$. Let $\mathbf{y}^*=(y(\mathbf{x}^*_1), \ldots , y(\mathbf{x}^*_s))^T$ be the vector of computer model responses at the prediction locations, $\mathbf{X}^*=(\mathbf{x}_1^*, \ldots , \mathbf{x}_s^*)^T$, and let $\mathbf{m}_k=(m_k(\mathbf{x}'_1),\dots,m_k(\mathbf{x}'_p))$.
 Using the above link function, the joint posterior distribution of $(\mathbf{y}^*,\mathbf{y}'_k)$ given the simulator output and monotonicity information can be written as
\begin{eqnarray*}
[\mathbf{y}^*,\mathbf{y}'_k\mid \mathbf{m}_k,\mathbf{y},\mathbf{\boldsymbol{\it l}},\sigma^2,\tau]\propto [\mathbf{y}^*,\mathbf{y}'_k\mid \mathbf{\boldsymbol{\it l}},\sigma^2][\mathbf{m}_k,\mathbf{y}\mid \mathbf{y}^*,\mathbf{y}'_k,\mathbf{\boldsymbol{\it l}},\sigma^2,\tau].
\end{eqnarray*}
Given $\mathbf{y}^*$, $\mathbf{y}'_k$ and the GP parameters, $\mathbf{y}$ and $\mathbf{m}_k$ are assumed to be independent of each other, i.e.,
\begin{eqnarray*}
[\mathbf{m}_k,\mathbf{y}\mid \mathbf{y}^*,\mathbf{y}'_k,\mathbf{\boldsymbol{\it l}},\sigma^2,\tau]=[\mathbf{m}_k\mid \mathbf{y}^*,\mathbf{y}'_k,\mathbf{\boldsymbol{\it l}},\sigma^2,\tau][\mathbf{y}|\mathbf{y}^*,\mathbf{y}'_k,\mathbf{\boldsymbol{\it l}},\sigma^2].
\end{eqnarray*}
Also, given the derivatives, $\mathbf{y}'_k$, $\mathbf{m}_k$ is assumed to be independent of $\mathbf{y^*}$ and the GP parameters, 
\begin{eqnarray*}
[\mathbf{m}_k\mid \mathbf{y}^*,\mathbf{y}'_k,\mathbf{\boldsymbol{\it l}},\sigma^2,\tau]=[\mathbf{m}_k\mid \mathbf{y}'_k,\tau].
\end{eqnarray*}
Therefore, the posterior predictive distribution can be simplified to the following form:
\begin{eqnarray}
\label{postpredyyd}
[\mathbf{y}^*,\mathbf{y}'_k\mid \mathbf{m}_k,\mathbf{y},\mathbf{\boldsymbol{\it l}},\sigma^2,\tau]\propto [\mathbf{y}^*,\mathbf{y}'_k\mid \mathbf{\boldsymbol{\it l}},\sigma^2][\mathbf{m}_k\mid \mathbf{y}'_k,\tau][\mathbf{y}\mid \mathbf{y}^*,\mathbf{y}'_k,\mathbf{\boldsymbol{\it l}},\sigma^2],
\end{eqnarray}
where $[\mathbf{y}^*,\mathbf{y}'_k|\mathbf{\boldsymbol{\it l}},\sigma^2]$ is the GP and $[\mathbf{y}|\mathbf{y}^*,\mathbf{y}'_k,\mathbf{\boldsymbol{\it l}},\sigma^2]$ is the GP likelihood. 

It follows from (\ref{mon}) that as $\tau\rightarrow \infty$ the support of $(\mathbf{y}^*,\mathbf{y}'_k)$ will be $\Re\times \Re^+$, i.e., 
\begin{eqnarray*}
\lim_{\tau \rightarrow \infty} [\mathbf{y}^*,\mathbf{y}'_k|\mathbf{m}_k,\mathbf{y},\mathbf{\boldsymbol{\it l}},\sigma^2,\tau]\propto [\mathbf{y}^*,\mathbf{y}'_k|\mathbf{y},\mathbf{\boldsymbol{\it l}},\sigma^2]\prod_{i=1}^{p}\mathbbm{1}_{(0,\infty)}(y'_{ik}).
\end{eqnarray*}

\cite{RV2010} used an expectation propagation technique that amounts to choosing a parametric approximating distribution from the exponential family and sequentially minimizing the Kullback-Leibler divergence between this approximate distribution and the target distribution. The approximating distribution is chosen to suit the nature and domain of the parameter of interest \citep{Minka2001}. The approximating family of distributions in \cite{RV2010} is chosen to be Gaussian. When the underlying function is obviously monotone, i.e., when derivatives are reasonably large positive numbers, a Gaussian distribution, although not consistent with the model assumptions in principle, may serve as a reasonable approximation. However, when derivatives are rather small and distributed near zero, the performance of the Gaussian family as the approximating family is questionable. We will overcome this drawback by directly sampling (as an alternative to approximation) from the exact posterior predictive distribution using a SMC algorithm. 

Another respect in which our approach is different from \cite{RV2010} is that we focus on deterministic computer experiments where there are two important objectives: interpolating the simulator output and providing valid credible intervals that reflect the deterministic nature of the simulator. The interpolation requirement could be considered as an extra constraint on the model since it restricts the function space and increases the difficulty of sampling from the target distribution. The noisy version of the problem is much easier to approach from a sampling point of view since sample paths are more likely to satisfy monotonicity when they do not need to necessarily interpolate the evaluations.

Finally, as mentioned earlier, instead of replacing the GP parameters, $\mathbf{\boldsymbol{\it l}}$ and $\sigma^2$, with their maximum likelihood estimates in the model we make Bayesian inference about these parameters, i.e., we sample from  the joint distribution of the GP parameters and $(\mathbf{y}^*,\mathbf{y}'_k)$ given the simulator output and monotonicity information. This joint posterior distribution is given by 
\begin{eqnarray}
\label{jpostpredyyd}
[\mathbf{\boldsymbol{\it l}},\sigma^2,\mathbf{y}^*,\mathbf{y}'_k\mid \mathbf{m}_k,\mathbf{y},\tau]\propto [\mathbf{\boldsymbol{\it l}},\sigma^2][\mathbf{y}^*,\mathbf{y}'_k\mid \mathbf{\boldsymbol{\it l}},\sigma^2][\mathbf{m}_k\mid \mathbf{y}'_k,\tau][\mathbf{y}\mid \mathbf{y}^*,\mathbf{y}'_k,\mathbf{\boldsymbol{\it l}},\sigma^2].
\end{eqnarray}
In the following section we explain the generalization to the case where derivatives with respect to more than one dimension are included.

\subsection{Including partial derivatives with respect to more than one input}\label{two-dim-der}

Before putting all of the pieces together for estimating the statistical model parameters and making predictions, we  extend the methodology to incorporate monotonicity in more than one dimension.  Statistical inference using MCMC for this model is presented in the next section.

Let $d_m\leq d$ be the number of inputs where the computer model is known to be monotone. Without loss of generality, let the computer model response be monotone in the first $d_m$ input dimensions.  Denote the locations where the monotonicity information is placed for each of the $d_m$ dimensions as $\mathbf{X'}_i$ ($i=1, \ldots , d_m$).   Note that the locations of the partial derivatives in each monotone input dimension does not have to be the same.  Furthermore, the number of derivative locations, $p_i$,  also does not have to be equal for each of the $d_m$ inputs (i.e., $\mathbf{X'}_i$ is a $p_i \times d$ matrix). Consequently, the  $d_m$ vectors of latent partial derivatives may be of different lengths. Let $\mathbf{y'}_i$ ($i=1, \ldots , d_m$) be the vector of unobserved partial derivatives at locations $\mathbf{X'}_i$. The joint distribution of $\mathbf{y}$ and the $d_m$ vectors of partial derivatives is given by

\begin{eqnarray}
\label{GP1}
[\mathbf{y},\mathbf{y}^{\prime}_1,\ldots,\mathbf{y}^{\prime}_{d_m}|\mathbf{\boldsymbol{\it l}},\sigma^2]={\cal N}(v,\Lambda),
\end{eqnarray}
where
\begin{eqnarray*}
v=\begin{pmatrix}
     \mbox{E}(\mathbf{y})\\
    \mbox{E}(\mathbf{y}'_1)\\
     \vdots \\
     \mbox{E}(\mathbf{y}'_{d_m})
     \end{pmatrix},
\end{eqnarray*}
and
\begin{eqnarray}
\label{covmat}
\Lambda=\begin{pmatrix} 
r(\mathbf{X},\mathbf{X}) & r_{11}(\mathbf{X},\mathbf{X}'_1) & \cdots & r_{1d_m}(\mathbf{X},\mathbf{X}'_{d_m}) \\
r_{11}(\mathbf{X}'_1,\mathbf{X}) & r_{21}(\mathbf{X}'_1,\mathbf{X}'_1) & \cdots & r_{2d_m}(\mathbf{X}'_1,\mathbf{X}'_{d_m})\\
\vdots & \vdots & \ddots & \vdots \\
r_{1d_m}(\mathbf{X}'_{d_m},\mathbf{X}) & r_{21}(\mathbf{X}'_{d_m},\mathbf{X}'_1) & \cdots & r_{2d_m}(\mathbf{X}'_{d_m},\mathbf{X}'_{d_m})
\end{pmatrix}.
\end{eqnarray}

The model in (\ref{GP1}) is a slightly more elaborate version of (\ref{GP}) and (\ref{jpostpredyyd}) where partial derivatives in multiple dimensions are included.  Furthermore, the joint covariance matrix includes correlations between not only the model responses and the partial derivatives, but also among the derivatives. 

The adaptation of a model for observed derivatives, (\ref{GP1}), to latent derivatives with observable signs is an immediate extension of the development in Section~\ref{monmodel}. We have not found it necessary to include different numbers of points where we observed the derivatives in each dimension, but we note that it can be done.  In the examples in Sections 4-6, we incorporate monotonicity information at equal numbers of points in each monotone dimension.

\section{Sequentially Constrained Monte Carlo for Posterior Sampling}\label{SMC}

In this section we outline the sampling algorithm adapted from \cite{Golchi2014} for the unobserved function values, $\mathbf{y}^*=y(\mathbf{X^*})$, derivative values, $\mathbf{y}'=y(\mathbf{X'})$, and covariance parameter values, $\mathbf{\boldsymbol{\it l}},\sigma^2$, from the posterior distribution, (\ref{jpostpredyyd}). While Bayesian inference for unconstrained GP regression can be made using basic MCMC algorithms based on Metropolis-Hastings steps, the addition of the constraints to the model creates sampling challenges that disqualify the existing MCMC algorithms as efficient sampling tools. The main reason is that under the constrained model, given the covariance parameters, $y$ and $y'$ are no longer Gaussian processes and their distribution cannot be obtained in closed form. Therefore, inference about the function and its derivatives is made based on point-wise sampling. Under point-wise sampling, the dimension of the state space gets large with the size of the prediction set and the number of locations at which derivatives are constrained. MCMC sampling from such a space will be inefficient.

Another factor that contributes to the difficulty of sampling from (\ref{jpostpredyyd}) is the restriction imposed on the support of the posterior distribution by the monotonicity constraints; an MCMC approach based on accept/reject steps is likely to be very inefficient when the support of the proposal distribution is very different from that of the target distribution. While the constraints define an explicit soft constraint on the derivative function space, the effect of this constraint on the covariance parameters' distribution is not obvious and therefore it is not trivial to define a proposal distribution that is likely to generate values for these parameters with high posterior probability.

To overcome these difficulties we use a variant of Sequential Monte Carlo (SMC) samplers \citep{Moral06}. SMC samplers take advantage of a sequence of distributions $\{\pi_{t}\}_{t=0}^{T}$ that bridge between a distribution that is straightforward to generate from (e.g. the prior) and the target distribution. \cite{Golchi2014} introduced a variant of SMC, referred to as the sequentially constrained Monte Carlo (SCMC), for the case that sampling from the target distribution is challenging due to imposition of constraints. The filtering sequence is defined based on the strictness of constraints. Starting from an initial distribution under which the constraints are relaxed fully or to an extent that sampling is feasible, the restriction is imposed gradually on the distributions to eventually obtain a sample from the fully constrained target distribution. In the following we describe the SCMC algorithm tailored for sampling from (\ref{jpostpredyyd}). 

As mentioned in Section~\ref{monmodel}, the rigidity of the monotonicity constraint is controlled by the parameter $\tau$. Larger values of $\tau$ more strictly constrain the partial derivatives to be positive at selected points. We use this property to define the filtering sequence of distributions. By specifying an increasing schedule over the monotonicity parameter, we move particles sampled from an unconstrained GP posterior towards the target model that has a large enough monotonicity parameter, say $\tau_T$. Let the vector of parameters defining each particle be denoted by $\etab=(\boldsymbol{\it l},\sigma^2,\mathbf{y}^*,\mathbf{y}'_k)$. The $t^{\mbox{th}}$ posterior in the sequence is given by,
$$\pi_{t}(\etab)\propto [\boldsymbol{\it l},\sigma^2][\mathbf{y}^*,\mathbf{y}'_k|\mathbf{y},\boldsymbol{\it l},\sigma^2][\mathbf{y}|\boldsymbol{\it l},\sigma^2]\prod_i{\Phi(\tau_{t}y'_k(\mathbf{x}'_i))};$$
where
$$0=\tau_0 <\tau_1<\ldots<\tau_T\rightarrow \infty.$$

\begin{algorithm}[h!]
	\caption{Sequential Monte Carlo for monotone emulation}\label{alg1}
	\begin{algorithmic}[1]
	\renewcommand{\algorithmicrequire}{\textbf{Input:}}
	\renewcommand{\algorithmicensure}{\textbf{Return:}}
	\Require a sequence of constraint parameters $\tau_t$, $t=1,\ldots,T$,

Forward transition kernels $K_t$,

proposal distributions $Q_{1t}$ and $Q_{2t}$ for $\boldsymbol{\it l}$ and $\sigma^2$,

proposal step adjustment parameter $q_{t}$.\\
	
	Generate an initial sample $\left(\boldsymbol{\it l},\sigma^{2},\mathbf{y}^*,\mathbf{y}'_k\right)^{1:N}_0\sim \pi_0$\\
	$W_1^{1:N}\gets \frac{1}{N}$
	\For{$t:=1, \ldots, T-1$}

	\begin{itemize}
\item	  $W_t^i\gets \frac{\tilde{w}_t^i}{\sum \tilde{w}_t^i }$ where $\tilde{w}_t^i=\frac{\prod_i{\Phi\left(\tau_{t}y'_k\left(\mathbf{x}'_i\right)\right)}}{\prod_i{\Phi\left(\tau_{t-1}y'_k\left(\mathbf{x}'_i\right)\right)}}$, $i=1,\ldots,N$
	
 \item           Resample the particles $\left(\boldsymbol{\it l},\sigma^{2},\mathbf{y}^*,\mathbf{y}'_k\right)_{t}^{1:N}$ with weights $W_t^{1:N}$ and $W_{t}^{1:N}\gets \frac{1}{N}$
	
\item	 Sample $\left(\boldsymbol{\it l},\sigma^{2},\mathbf{y}^*,\mathbf{y}'_k\right)_{t+1}^{1:N}\sim K_{t}\left(\left(\boldsymbol{\it l},\sigma^{2},\mathbf{y}^*,\mathbf{y}'_k\right)_{t}^{1:N},.\right)$ through the following steps
\begin{itemize}
\renewcommand{\labelitemi}{ }
\For{$i:=1\ldots,N$}
\begin{itemize}

\item $\left(\boldsymbol{\it l}^i_{t},\sigma^{2i}_{t},\mathbf{y}_{t}^{*i},\mathbf{y}_{t}^{'i}\right) \gets \left(\boldsymbol{\it l}^i_{t-1},\sigma^{2i}_{t-1},\mathbf{y}_{t-1}^{*i},\mathbf{y}_{t-1}^{'i}\right)$;
	
\item propose $\boldsymbol{\it l}^{\text{new}}\sim Q_{1t}\left(.|\boldsymbol{\it l}_{t}^i\right)$ and

	$\boldsymbol{\it l}_{t}^i\gets \boldsymbol{\it l}^{\text{new}}$ with probability $p=\min\{1,\frac{\pi_t\left(\boldsymbol{\it l}^{\text{new}},\sigma^{2i}_{t},\mathbf{y}_{t}^{*i},\mathbf{y}_{t}^{'i}\right)}{\pi_t\left(\boldsymbol{\it l}^i_{t},\sigma^{2i}_{t},\mathbf{y}_{t}^{*i},\mathbf{y}_{t}^{'i}\right)}\}$;

\item propose $\sigma^{2\text{new}}\sim Q_{2t}\left(.|\sigma_{t}^{2i}\right)$ and

	$\sigma_{t}^{2i}\gets \sigma^{2\text{new}}$ with probability $p=\min\{1,\frac{\pi_t\left(\boldsymbol{\it l}^i_{t},\sigma^{2\text{new}},\mathbf{y}_{t}^{*i},\mathbf{y}_{t}^{'i}\right)}{\pi_t\left(\boldsymbol{\it l}^i_{t},\sigma^{2i}_{t},\mathbf{y}_{t}^{*i},\mathbf{y}_{t}^{'i}\right)}\}$;

\item	propose $\left(\mathbf{y}^*,\mathbf{y}'\right)^{\text{new}}\sim {\cal N}\left(\left(\mathbf{y}^*,\mathbf{y}'\right)_{t}^i,q_{t}\Lambda_{\boldsymbol{\boldsymbol{\it l}}^i_{t}}\right)$ where $\Lambda_{{\boldsymbol{\it l}}^i_{t}}$ is the correlation matrix with correlation parameter vector $\boldsymbol{\it l}^i_{t}$ and

	$\left(\mathbf{y}^*,\mathbf{y}'\right)_{t}^i\gets \left(\mathbf{y}^*,\mathbf{y}'\right)^{\text{new}}$ with probability $p=\min\{1,\frac{\pi_t\left(\boldsymbol{\it l}^i_{t},\sigma^{2i}_{t},\mathbf{y}^{*\text{new}},\mathbf{y}^{'{\text{new}}}\right)}{\pi_t\left(\boldsymbol{\it l}^i_{t},\sigma^{2i}_{t},\mathbf{y}_{t}^{*i},\mathbf{y}_{t}^{'i}\right)}\}$
\end{itemize}
	\EndFor
\end{itemize}
	\end{itemize}
	\EndFor
	
	\Ensure Particles $\left(\boldsymbol{\it l},\sigma^{2},\mathbf{y}^*,\mathbf{y}'_k\right)_T^{1:N}$.
	\end{algorithmic}
	\end{algorithm}

The sequential Monte Carlo algorithm tailored for monotone interpolation is given in Algorithm~\ref{alg1}. In step 1 of Algorithm~\ref{alg1}, $\pi_0$ is chosen to be an unconstrained GP model corresponding to $\tau=0$, that fully relaxes the positivity constraint on the derivatives. However, MCMC is needed to generate a sample from the unconstrained model. Typical Metropolis within Gibbs algorithms used to sample from a GP posterior can be found in the literature. See for example \cite{Berger01}.

The simplified form of the incremental weights, $\tilde{w}^i_{t}$ is the result of the choice of the forward transition kernels $K_{t}$ as an MCMC kernel of the invariant distribution $\pi_{t}$ in the sampling step \citep{Moral06}. The proposal distributions, $Q_{1t}$ and $Q_{2t}$ used in the sampling step are chosen to generate adequate values under $\pi_{t}$ and depend on $t$ in the generic SCMC algorithm. In the current application the same proposal distribution can be used at all time points, $t$. However, the proposal step size (the variance of the proposal distribution) is adjusted when progressing through the filtering sequence to prevent particle degeneracy by keeping the acceptance rate in the sampling moves above a lower threshold. Existing guidelines for adjusting proposal moves in the MCMC literature can be used. The proposal step size parameters $q_{t}$ are chosen in the same manner. We discuss the specific choices made for our examples in Section~\ref{examples}.

The increasing schedule over the monotonicity parameter, $\tau$, should be determined such that the transition from $t$ to $t+1$ is made effectively. To this end, the distributions $\pi_{t}$ and $\pi_{t+1}$ should be close enough so that there is an overlap between samples taken from the two distributions. The effective sample size (ESS) can be used to measure the closeness between two consecutive distributions based on a sample of weighted particles,
$$\text{ESS}_t=\frac{1}{\sum_{i=1}^N W_{t}^i}.$$
Based on pilot runs of the algorithm with a smaller number of particles, the sequence of monotonicity parameters can be specified such that the ESS is above a certain threshold (e.g. half of the sample size). Of course, as mentioned by \cite{Moral06}, the effectiveness of the SMC depends on the length of the sequence, $T$, as well as the number of particles, $N$. The fact that parallelized computation is fairly straightforward for SMC samplers allows for using large values for $T$ and $N$ when needed.

\section{The Derivative Input Set } \label{der-design}

The GP model for derivatives in Section~\ref{gpr} and the mechanism for including monotonicity information via derivatives in Section~\ref{monmodel} make the assumption that derivative information is available at a ``derivative input set", $\mathbf{X}^{\prime}$.  At each $\mathbf{x}' \in \mathbf{X}^{\prime}$, we assume the soft constraint $m_k(\mathbf{x}')=1$, which via (\ref{monlik}) induces a probability that $y'_k(\mathbf{x}')>0$.  That is, instead of assuming that the derivative of the GP is positive everywhere, the model assumes that there is a (large) probability that the derivative is positive at a specified set of points.

To estimate the monotone GP, the derivative input set must be specified. The SCMC algorithm described above permits working with relatively large derivative sets since SCMC is proved to be stable as the dimensionality of the state space increases \citep{Beskos12}. Therefore, using a space filling design to construct a derivative set that assures imposition of the monotonicity constraints uniformly over the input space is a reasonable strategy. However, one may restrict the derivative set to regions that probability of occurrence of negative derivatives is high, as \cite{RV2010} suggest, to save computation time. 

\begin{figure}[htbp]

  \centering
   \begin{subfigure}[b]{0.4\textwidth}
                \centering
                \includegraphics[width=\textwidth]{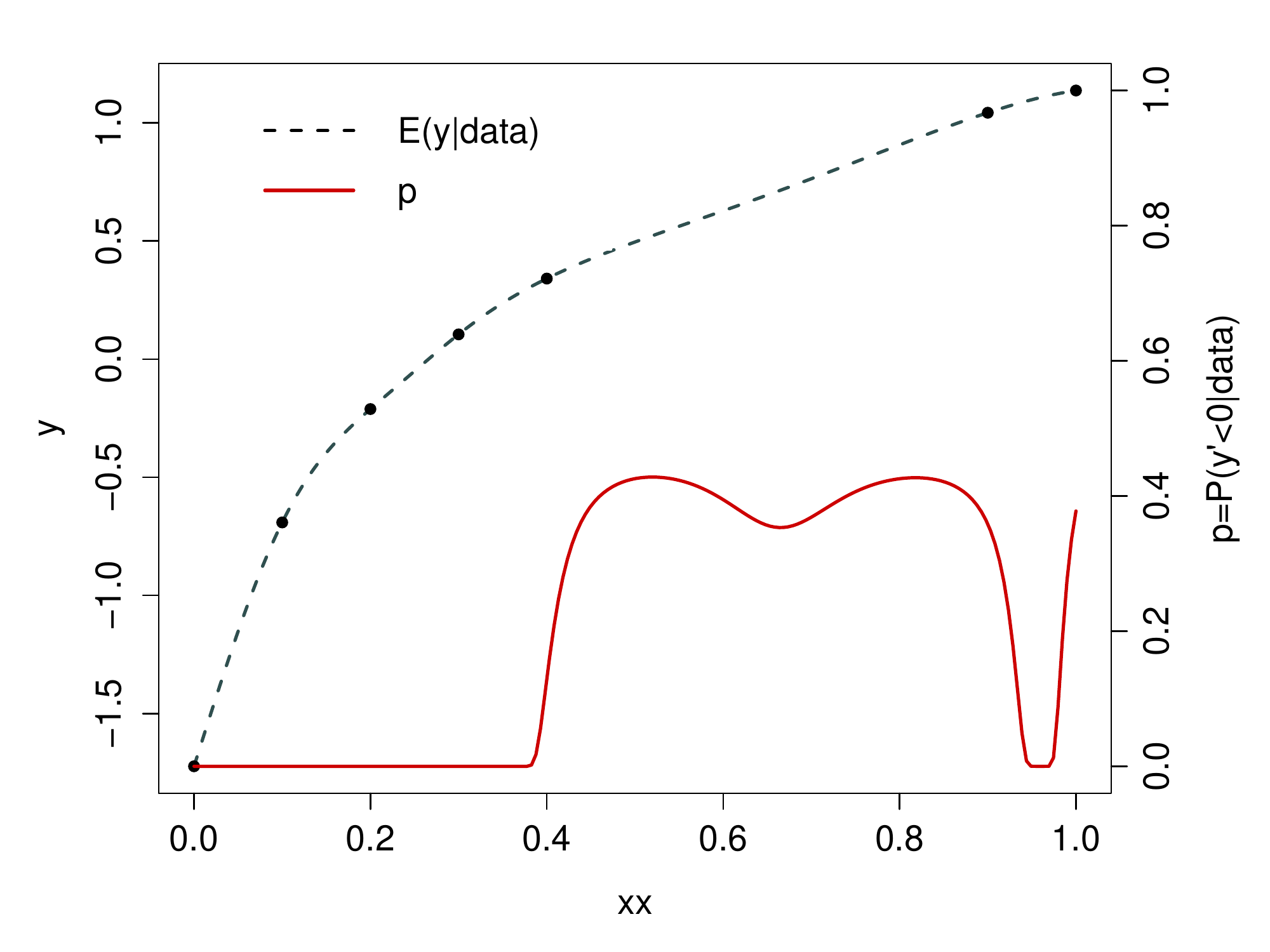}
                \caption{}
                \label{logdera}
        \end{subfigure}
    \begin{subfigure}[b]{0.4\textwidth}
                \centering
                \includegraphics[width=\textwidth]{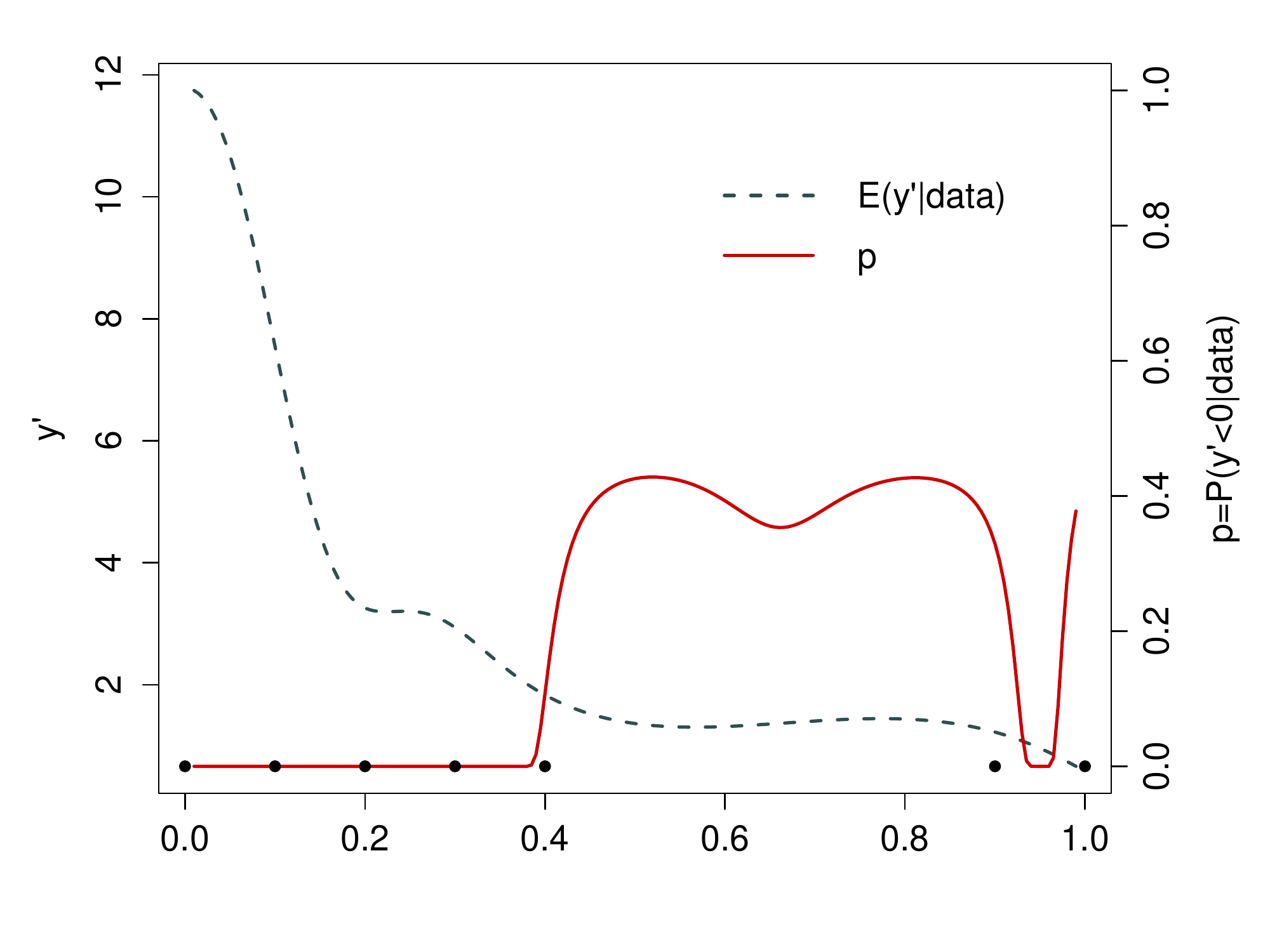}
                \caption{}
                \label{logderb}
        \end{subfigure}

  \caption{Example 1. (a) GP mean and the probability of negative derivatives, (b) mean of the GP derivative and the probability of negative derivatives.}
\label{logder}
\end{figure}

Under the unconstrained GP model the derivatives are Gaussian processes, and the point-wise probability of negative derivatives can be obtained analytically for fixed covariance parameters.  For Example 1, Figures~\ref{logdera} and \ref{logderb} show this probability together with the mean and expected derivative function of the Gaussian process evaluated at the posterior mean of the covariance parameters.

\cite{RV2010} recommend sequential addition of points where the probability of occurrence of negative derivatives is the largest. \cite{Wang2012} developed a sequential algorithm for selection of the derivative locations based on this idea with an upper bound on the size of the derivative set predetermined with consideration of computational limitation. \cite{Wang2012} use maximum likelihood estimates for the covariance parameters and are able to iteratively calculate the probability of negative derivatives analytically under the constrained model to determine where to place the next derivative point. Implementation of this algorithm in the fully Bayesian framework is not trivial. Online construction of derivative design that can be assembled into the SCMC algorithm, based on empirical estimates of the probability of negative derivatives at each step is the topic of current research.

\begin{figure}[h!]
\label{seq-der}
        \centering
        \begin{subfigure}[b]{0.23\textwidth}
                \centering
                \includegraphics[width=\textwidth]{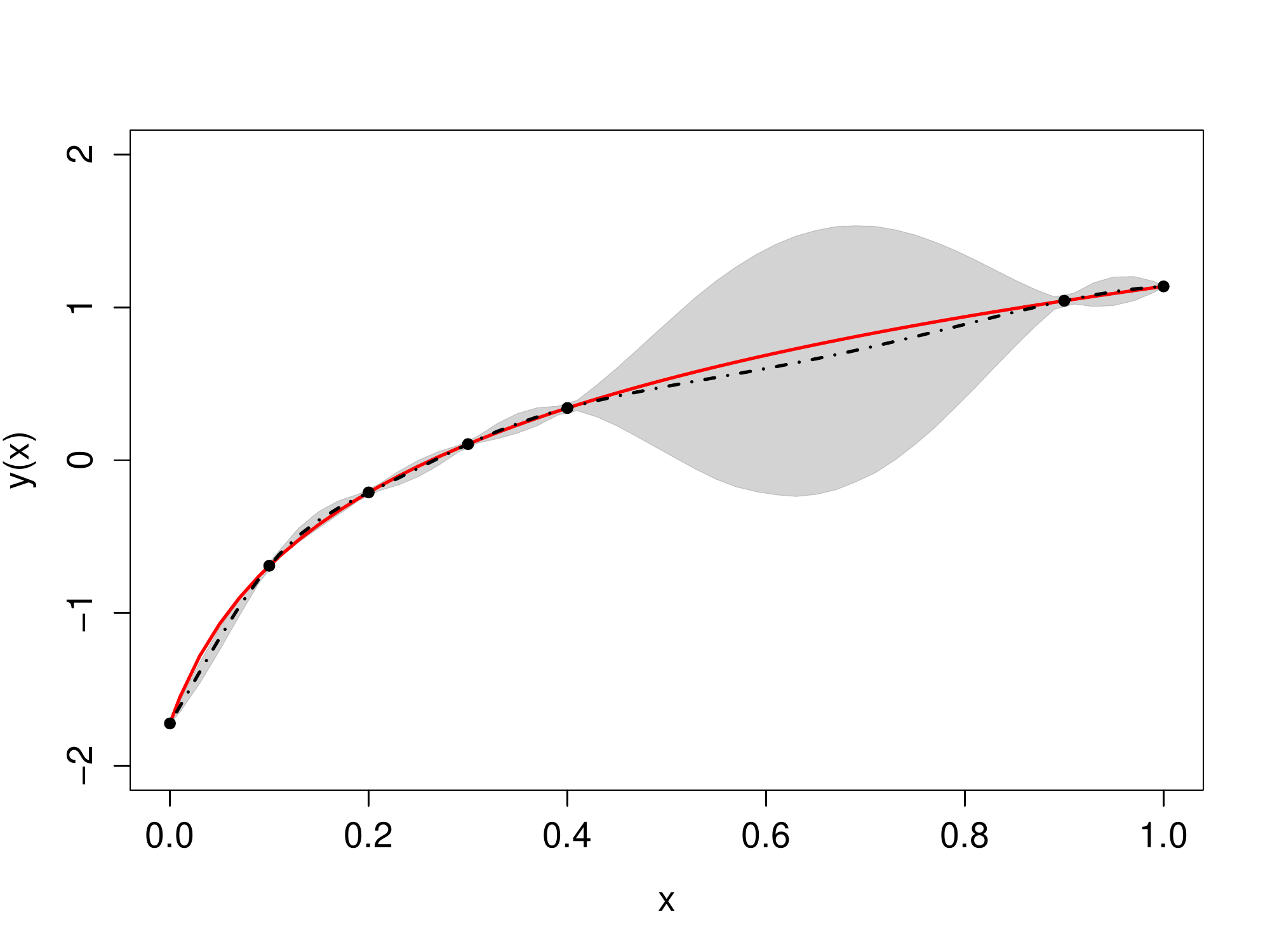}
                \caption{}
                \label{t0}
        \end{subfigure}
        \begin{subfigure}[b]{0.23\textwidth}
                \centering
                \includegraphics[width=\textwidth]{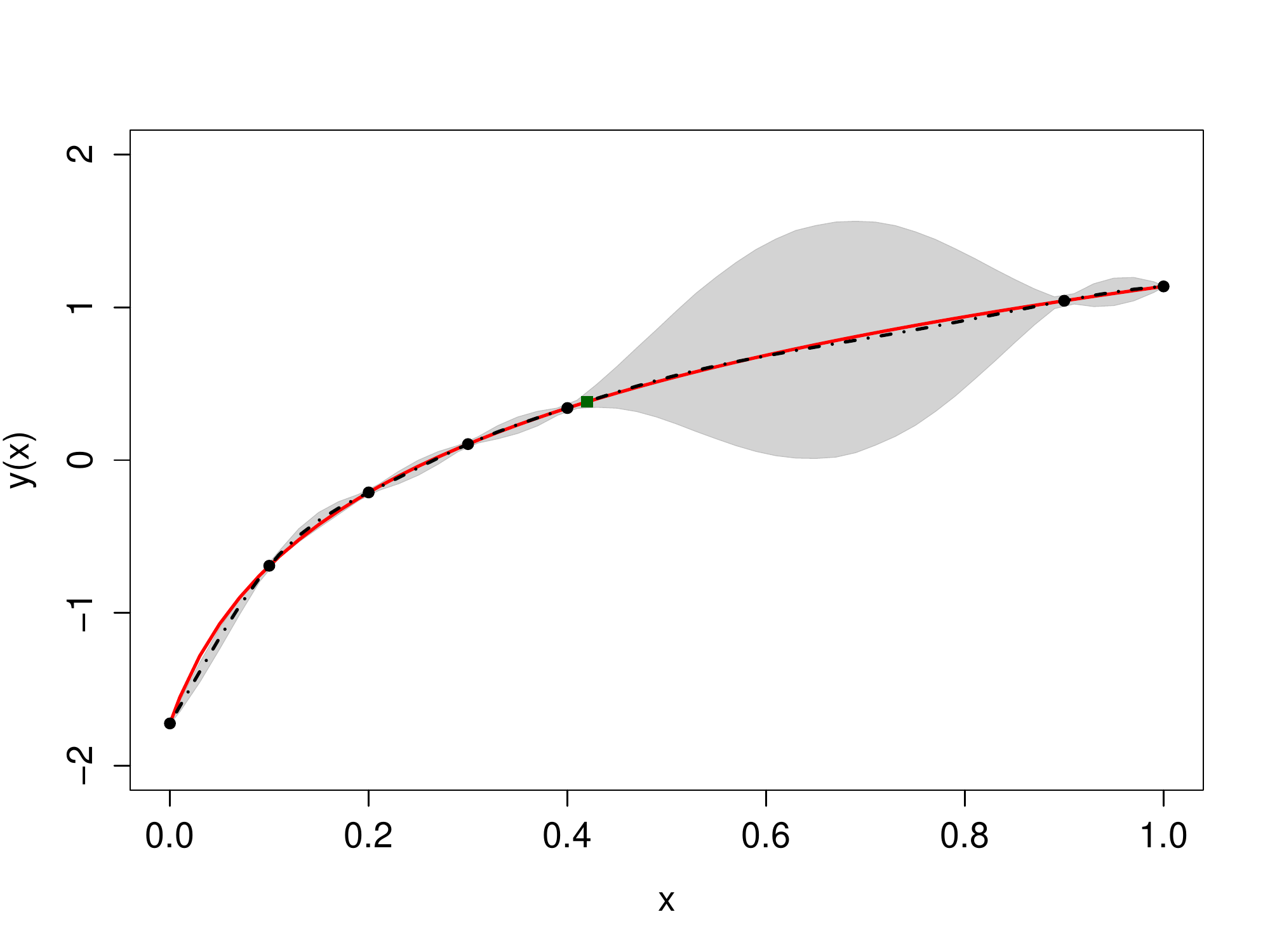}
                \caption{}
                \label{t0}
        \end{subfigure}
        \begin{subfigure}[b]{0.23\textwidth}
                \centering
                \includegraphics[width=\textwidth]{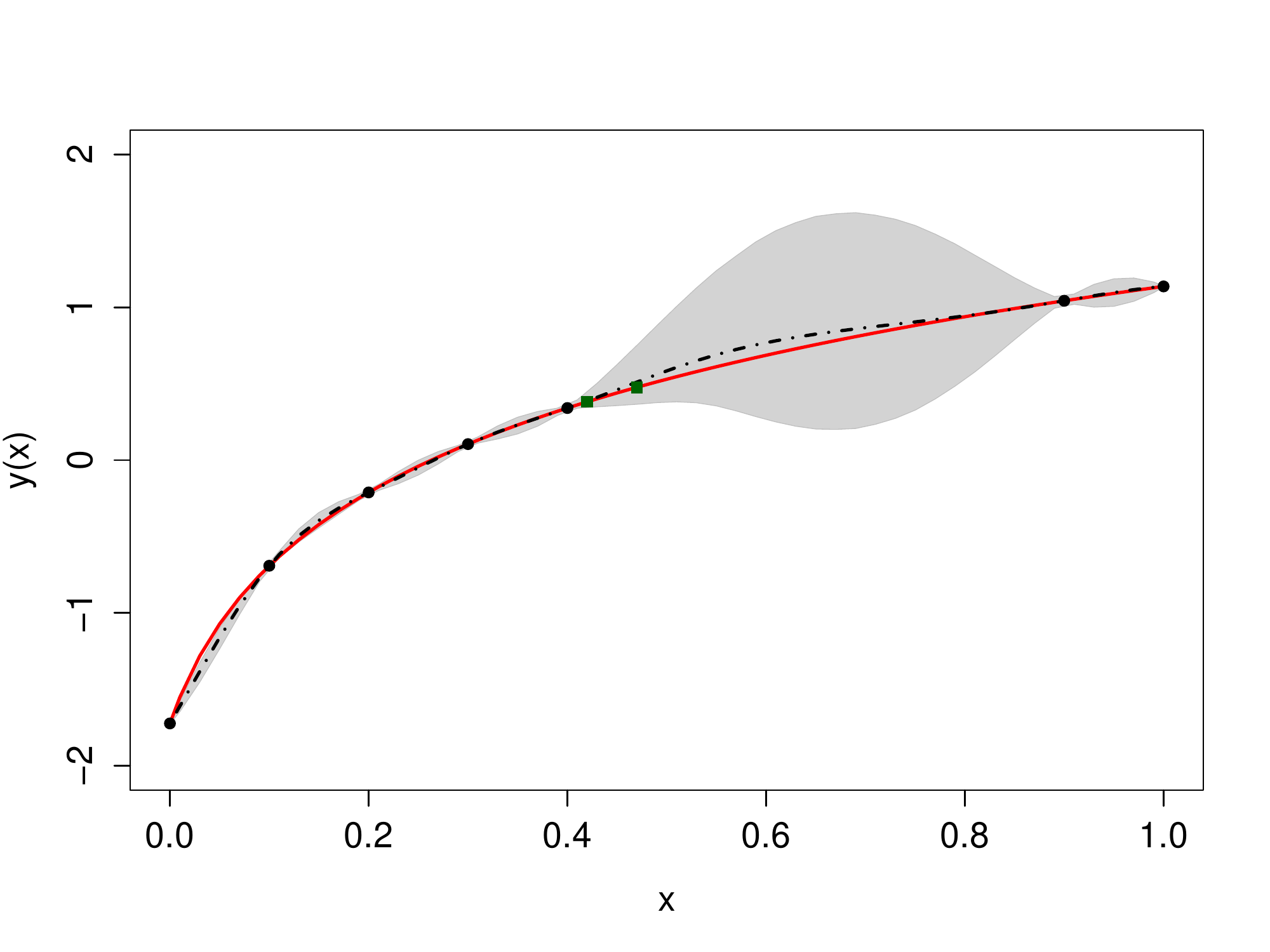}
                \caption{}
                \label{tau1}
        \end{subfigure}
 \begin{subfigure}[b]{0.23\textwidth}
                \centering
                \includegraphics[width=\textwidth]{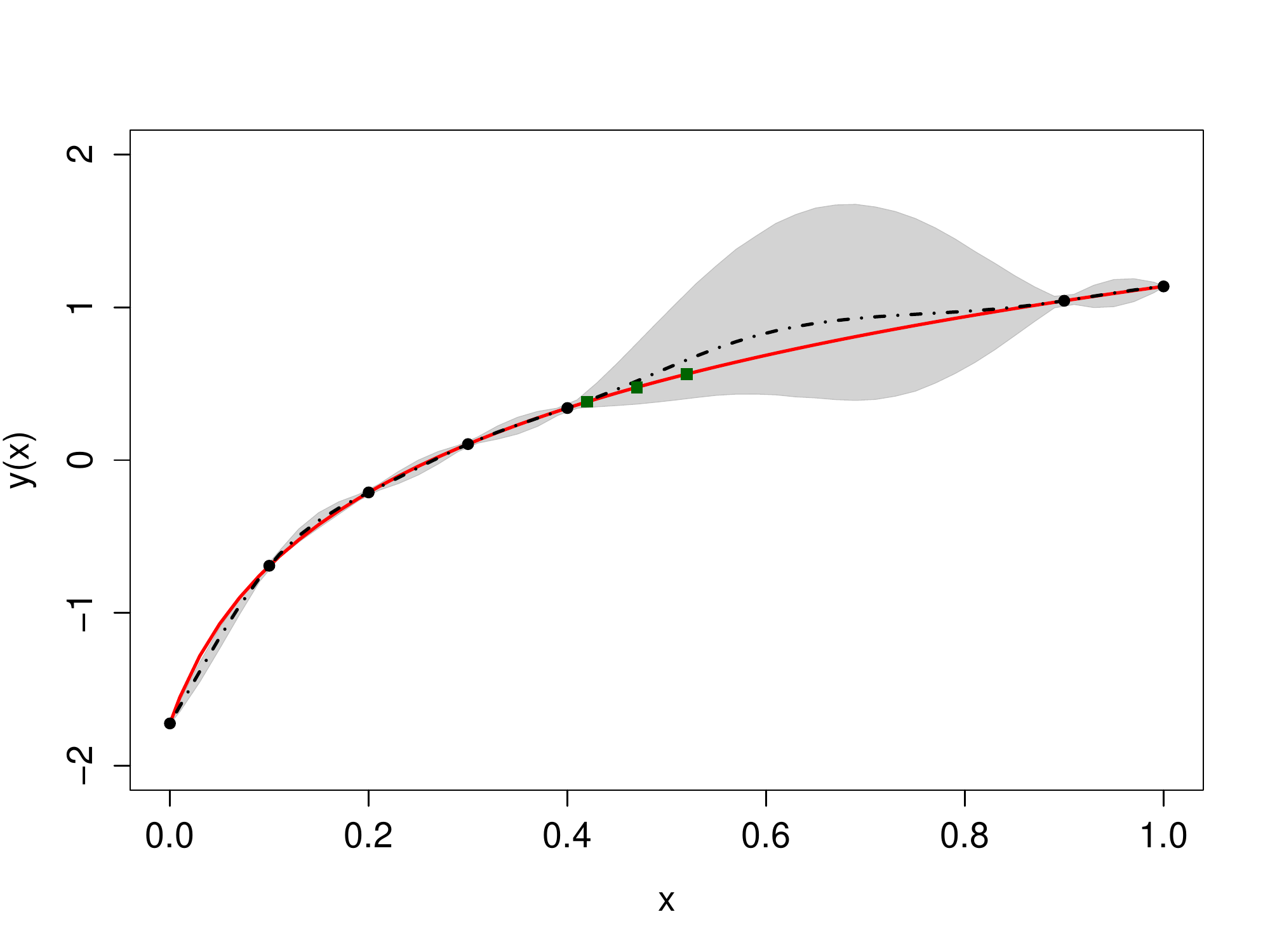}
                \caption{}
                \label{tau2}
        \end{subfigure}\\
        \begin{subfigure}[b]{0.23\textwidth}
                \centering
                \includegraphics[width=\textwidth]{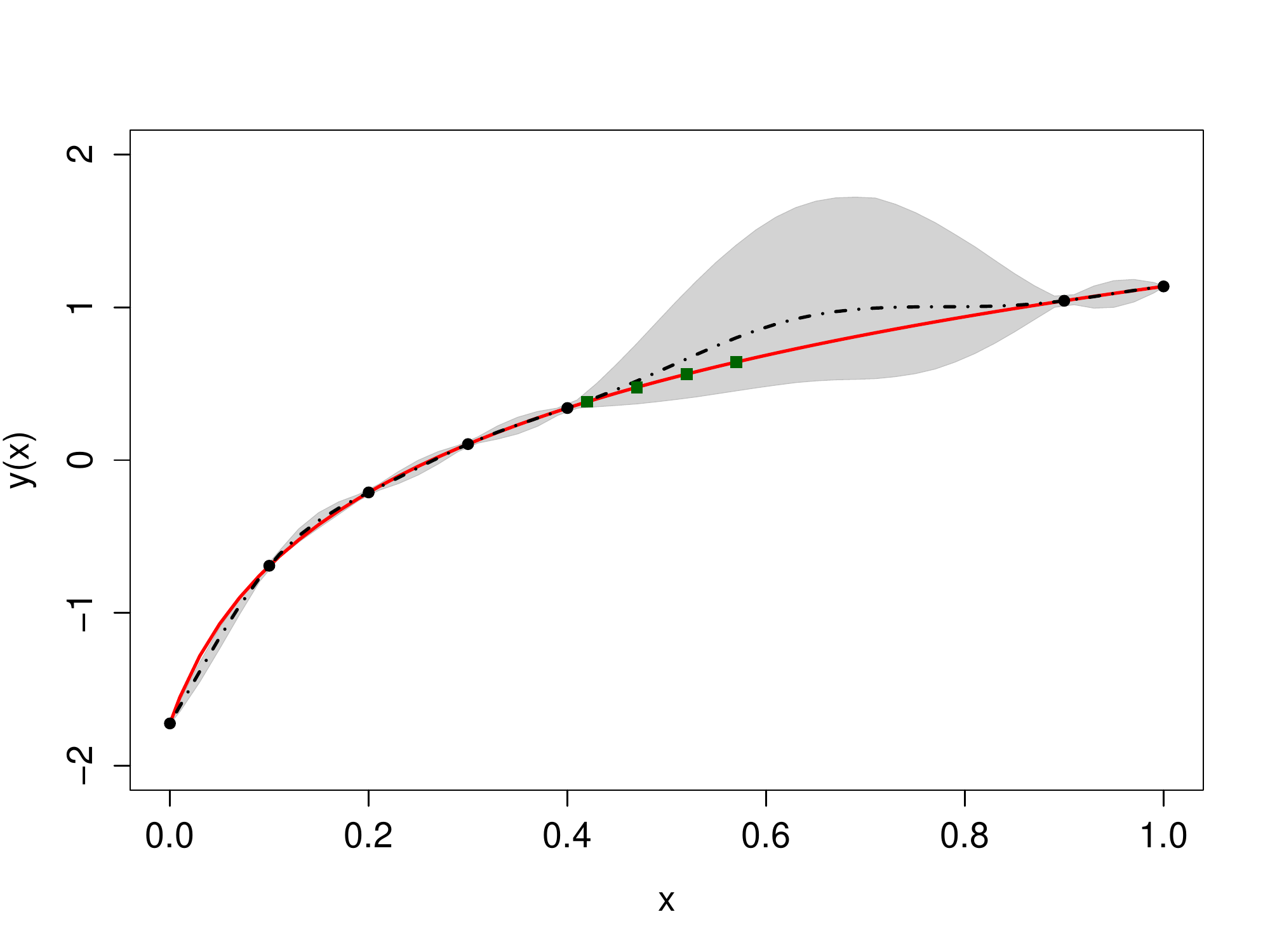}
                \caption{}
                \label{tau3}
        \end{subfigure}
 \begin{subfigure}[b]{0.23\textwidth}
                \centering
                \includegraphics[width=\textwidth]{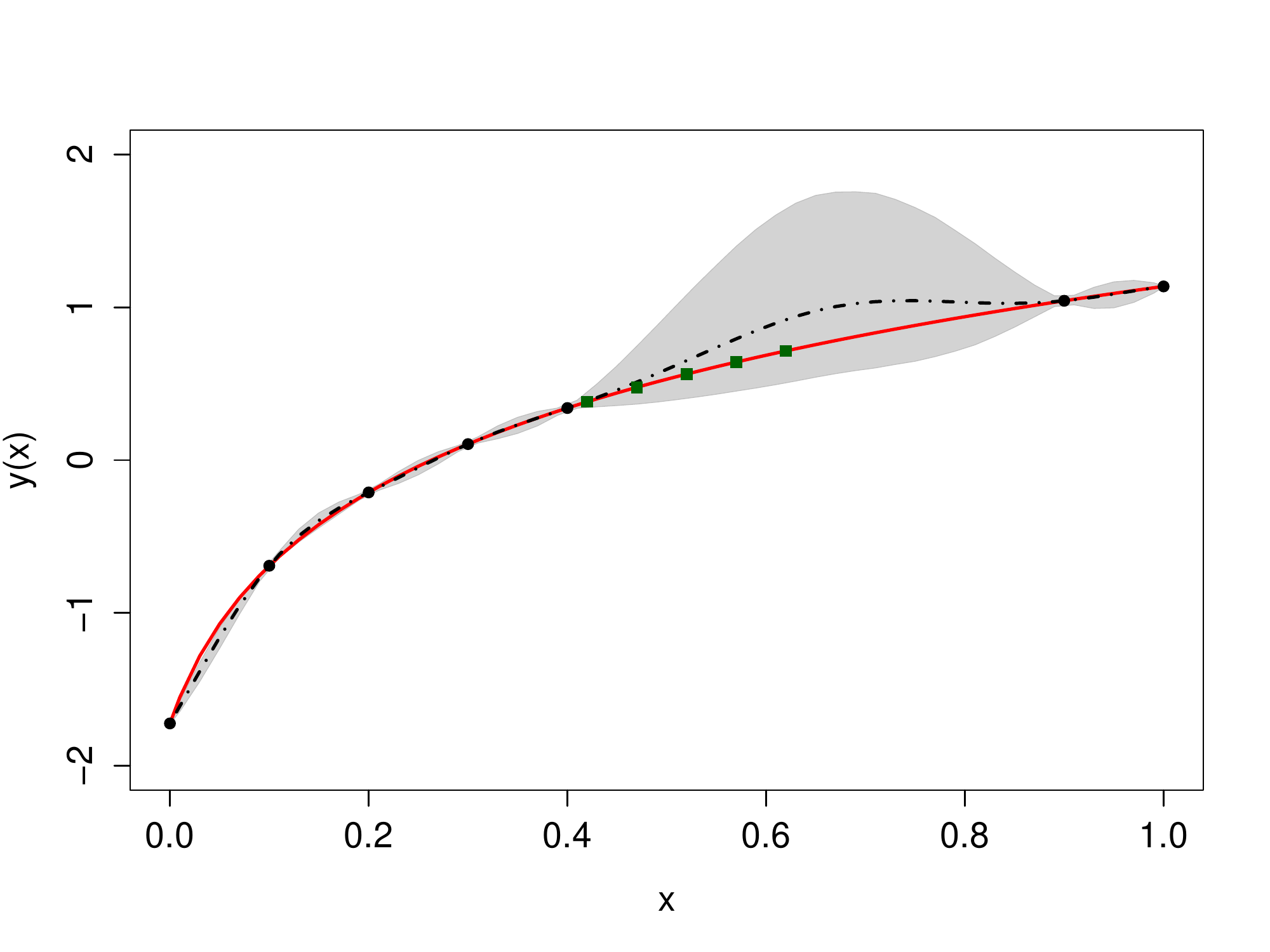}
                \caption{}
                \label{tau4}
        \end{subfigure}
        \begin{subfigure}[b]{0.23\textwidth}
                \centering
                \includegraphics[width=\textwidth]{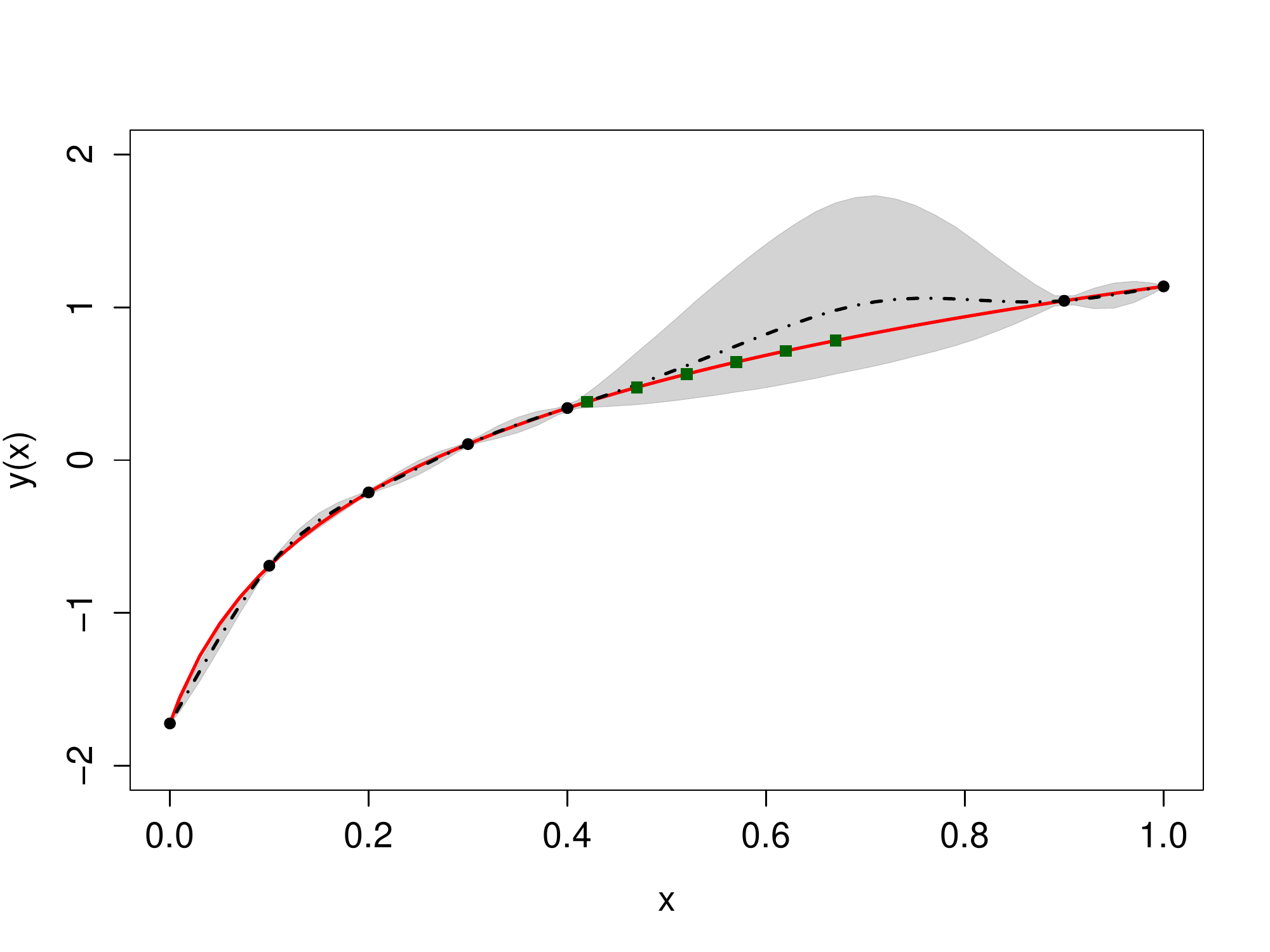}
                \caption{}
                \label{tau5}
        \end{subfigure}
 \begin{subfigure}[b]{0.23\textwidth}
                \centering
                \includegraphics[width=\textwidth]{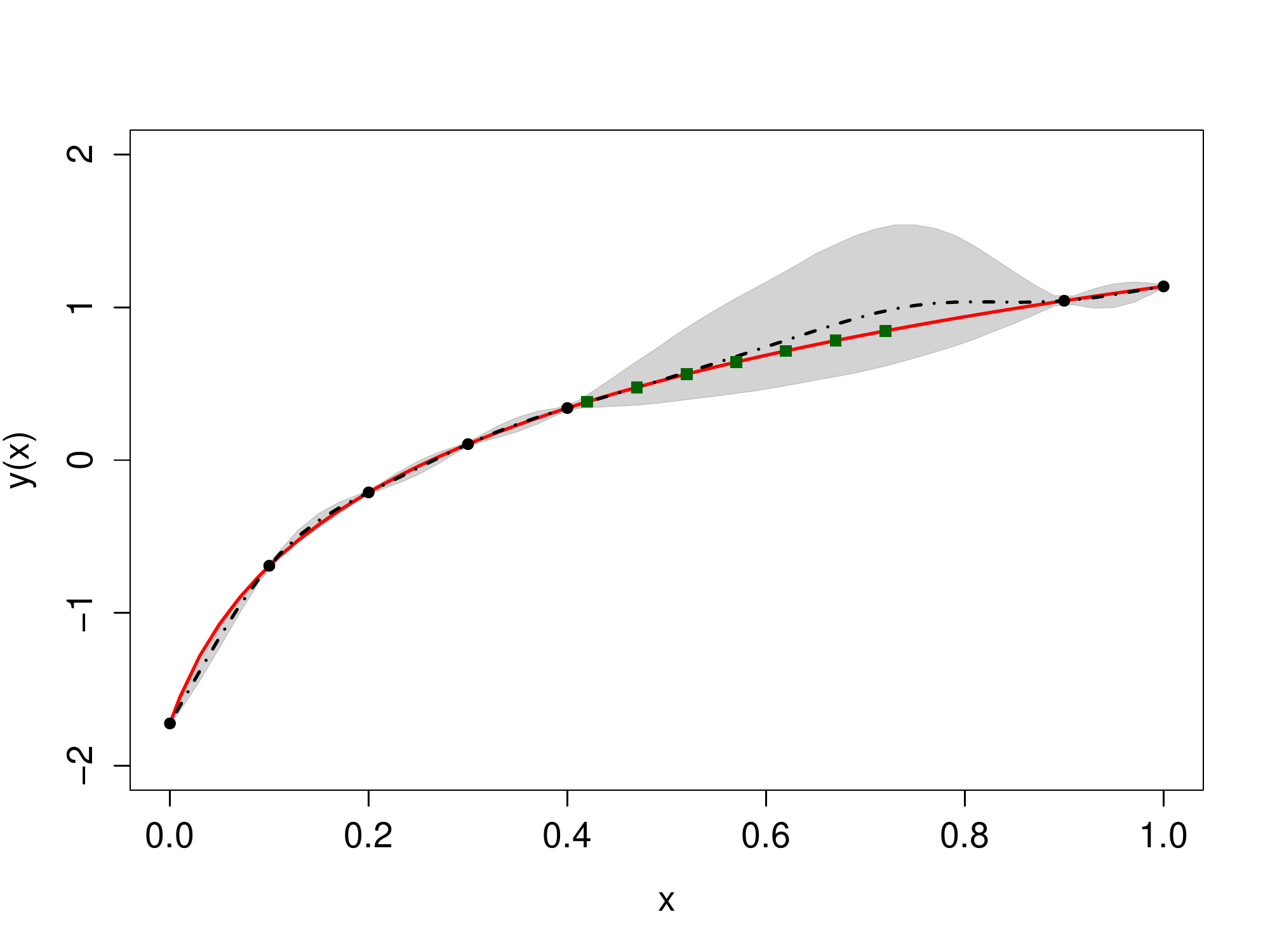}
                \caption{}
                \label{tau6}
        \end{subfigure}\\
        \begin{subfigure}[b]{0.23\textwidth}
                \centering
                \includegraphics[width=\textwidth]{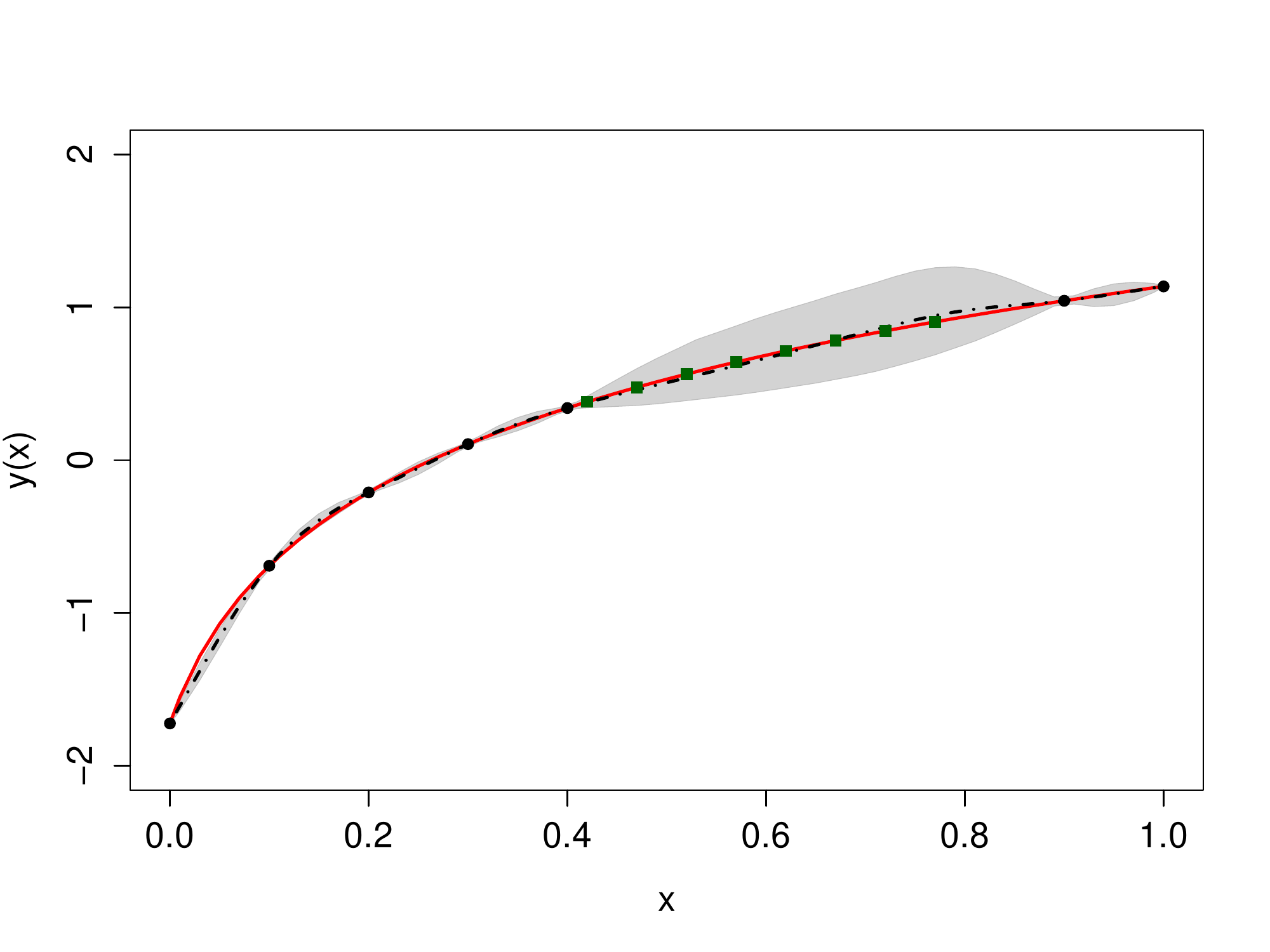}
                \caption{}
                \label{tau7}
        \end{subfigure}
 \begin{subfigure}[b]{0.23\textwidth}
                \centering
                \includegraphics[width=\textwidth]{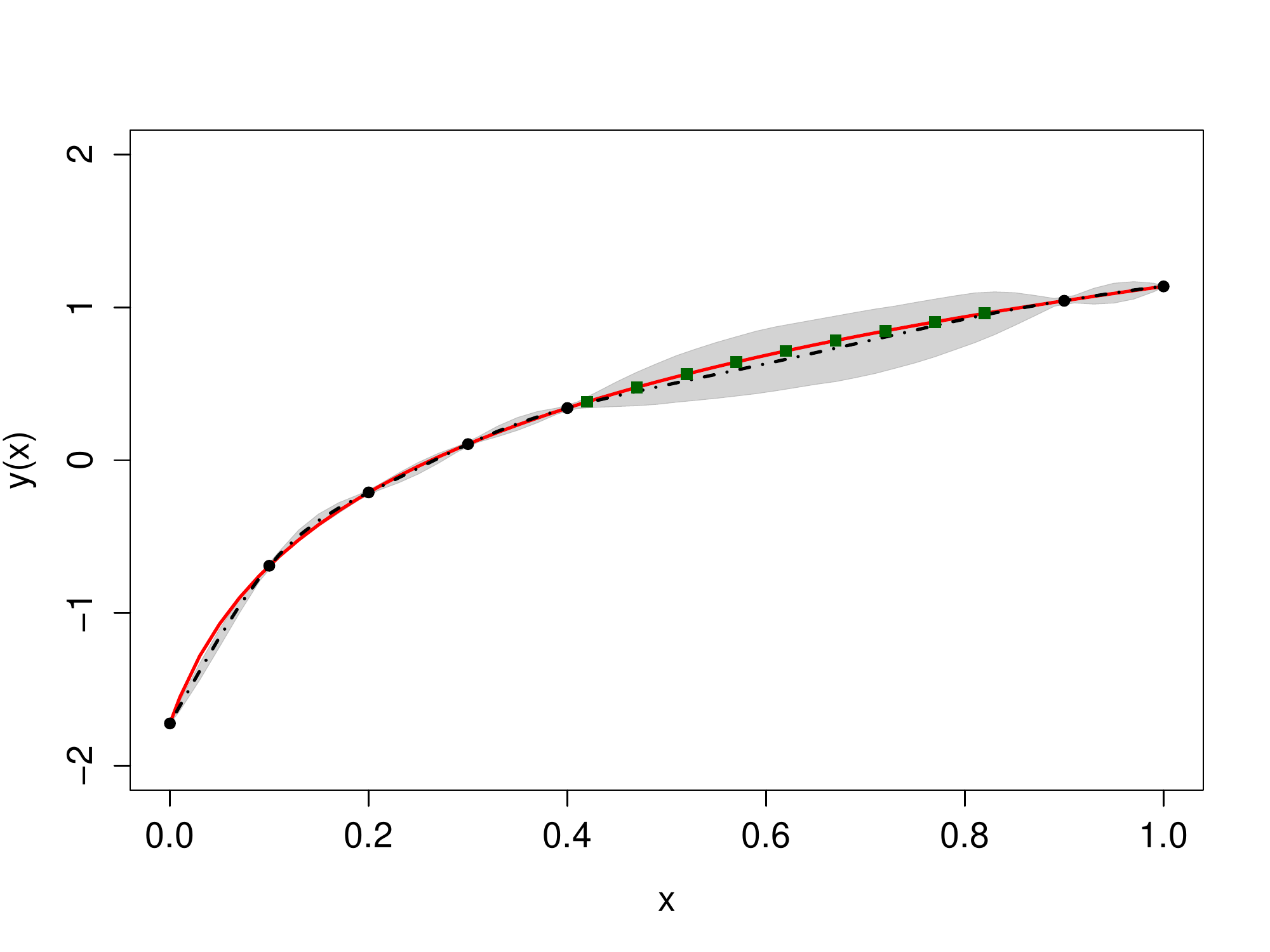}
                \caption{}
                \label{tau8}
        \end{subfigure}
        \begin{subfigure}[b]{0.23\textwidth}
                \centering
                \includegraphics[width=\textwidth]{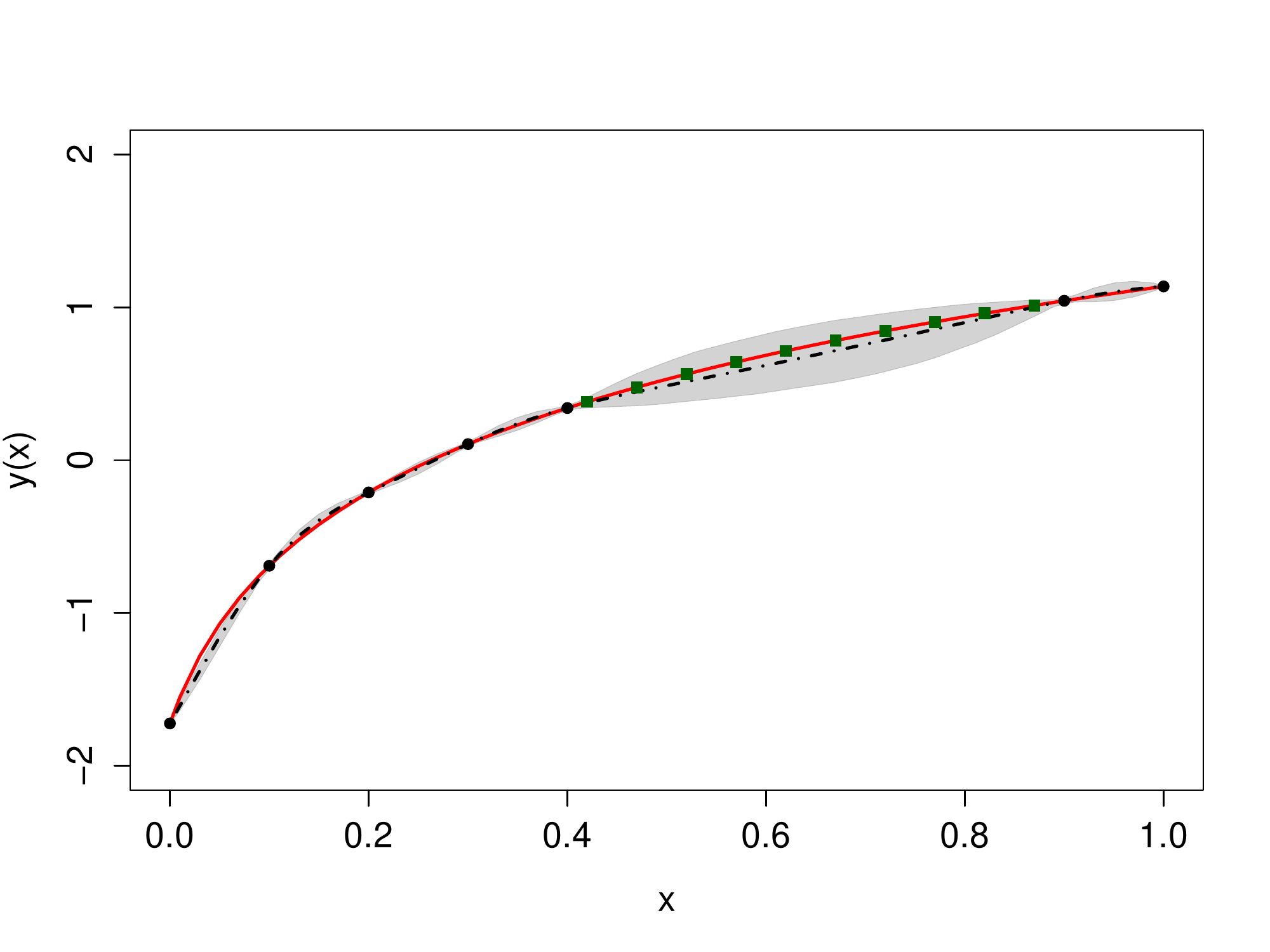}
                \caption{}
                \label{t10}
        \end{subfigure}
     \begin{subfigure}[b]{0.23\textwidth}
                \centering
                \includegraphics[width=\textwidth]{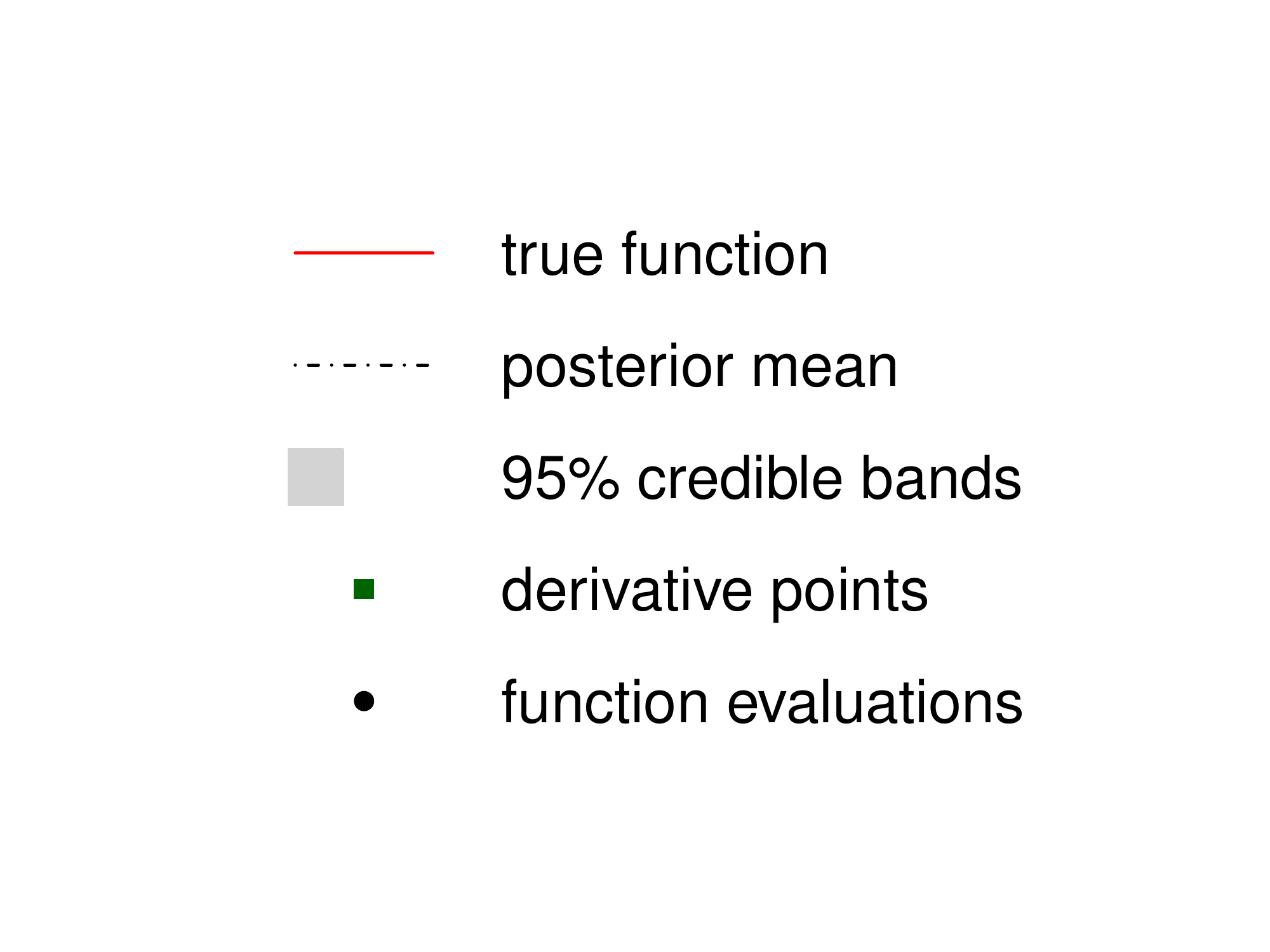}
                \label{}
        \end{subfigure}
     
        \caption{Example 1. The effect of sequential addition of derivative points on 95\% credible intervals; the posterior mean and credible bands obtained by (a) unconstrained GP model and (b-k) constrained models, together with the true function}\label{seq-der}
\end{figure}

Figure~\ref{seq-der} displays a sequence of point-wise 95\% credible intervals for Example 1, as derivative locations are added. Starting from an unconstrained GP (Figure~\ref{t0}) derivatives are constrained at locations added in the gap one at a time from left to right. Non-eligible sample paths are filtered from the posterior by addition of each derivative point to eventually obtain a trimmed collection of posterior sample paths that satisfy positivity of derivatives at ten locations (Figure~\ref{t10}).

The choice of the derivative input set can be extended to the case where monotonicity is required in two or more inputs. We argue that the problem is no more complex than the one-dimensional case since it can be tackled dimension-wise. As mentioned earlier in Section~\ref{two-dim-der} the derivative locations do not have to be the same when taking the partial derivatives with respect to different dimensions. Therefore, we will use a different derivative input set, $\mathbf{X}'_k$, $k=1,\ldots ,d_m$ for each of the $d_m$ input dimensions in which the underlying function is assumed to be monotone. 

As a straightforward extension to the one-dimensional case we place the partial derivatives in the neighborhood of the prediction point, on the corresponding slice, parallel to the corresponding axes (see, for e.g., Figure~\ref{design01}). Placement of the derivatives in this manner encourages {\it local,  dimension-wise} monotonicity.

\section{Examples}
\label{examples}
In this section, two examples are used to illustrate the performance of the proposed method. The first is the example illustrated in Figure 1, and the second is used to demonstrate the methodology in the two-dimensional case.  Comparisons are made  with the  Bayesian GP model that ignores monotonicity information. In our examples the design points are purposely chosen to create situations in which inference about the underlying function is challenging.

{\bf Example 1.} Consider the monotone increasing function $y(x)=\log(20x+1)$ shown in Figure 1. Let $\mathbf{X}=(0,0.1,0.2,0.3,0.4,0.9,1)$ be the locations at which the function is evaluated. The large gap in the design between 0.4 and 0.9 will be a challenge for the unconstrained GP which will tend to revert to the overall mean response with no data. 

As mentioned in Section 3.5, ideally, the derivative input set is chosen to uniformly inform the model about monotonicity over the input space. However, where the function evaluations are densely located enforcing the constraints is a waste of computation since negative derivatives are unlikely to occur in these regions (see Figure~\ref{logder}). Consequently, we choose a derivative set containing ten equally spaced points in the gap: $\mathbf{X}'=(0.42, 0.47, 0.52, 0.57, 0.62, 0.67, 0.72, 0.77, 0.82, 0.87)$. To evaluate the global predictive capability of the methods, the prediction set, $\mathbf{X}^*$, is a fine grid of size 50 on $[0,1]$.

To specify the monotone model, prior distributions and the target value of the monotonicity parameter, $\tau_T$, that governs the strictness of the monotonicity restriction must be determined.  The monotonicity parameter is chosen to be $\tau_T=10^{-6}$.  This allows a fairly strict monotonicity restriction.  The prior distributions on components of $\mathbf{\boldsymbol{\it l}}$ are such that $\frac{\sqrt{2\lambda}}{l_k}$ have chi-squared distributions with one degree of freedom, and the prior on $\sigma^2$ is a chi-squared distribution with five degrees of freedom. In each case, the specification allows for a weakly informative prior. 

To sample from the target posterior using Algorithm~\ref{alg1}, proposal distributions and the increasing sequence $\{\tau_{t}\}_{t=1}^{T-1}$ need to be specified. Components of $\mathbf{\boldsymbol{\it l}}$ are proposed independently using a random walk scaled to provide satisfactory acceptance rates.  For the variance, $\sigma^2$, a chi-squared distribution whose degrees of freedom is the current value of this parameter (i.e., $\chi^2(\sigma^{2(j-1)})$) is used.   We use the same proposal distributions through the filtering sequence (i.e., $Q_{1t}=Q_1$ and $Q_{2t}=Q_2$ for all $t$), but adjust the step size for the random walk and let $q_{t}$ vary to obtain a reasonable acceptance rate in the sampling step of the algorithm. Posterior sampling is performed through $T=20$ steps of the SCMC algorithm. The number of particles used is $N=40,000$. The same choices are made for our examples and simulation studies for the rest of the paper; only the proposal step sizes are adjusted accordingly in each case.

Figures~\ref{polyb} and~\ref{polyc} show the mean and 95\% credible intervals for the set, $\mathbf{X}^*$, obtained using the unconstrained GP model and the proposed model, respectively. Notice that the posterior sample obtained by the usual GP model includes non-monotone predictions (Figure~\ref{polya}) and the uncertainty is quite large. However, looking at the derivative-constrained prediction intervals in Figure~\ref{polyc}, we see a substantial decrease in uncertainty. While the improvement in accuracy is most evident in the region with no training data, we also see improvement in the area with more closely spaced points.

\begin{figure}[h!]
  \centering
    \includegraphics[width=.7\textwidth]{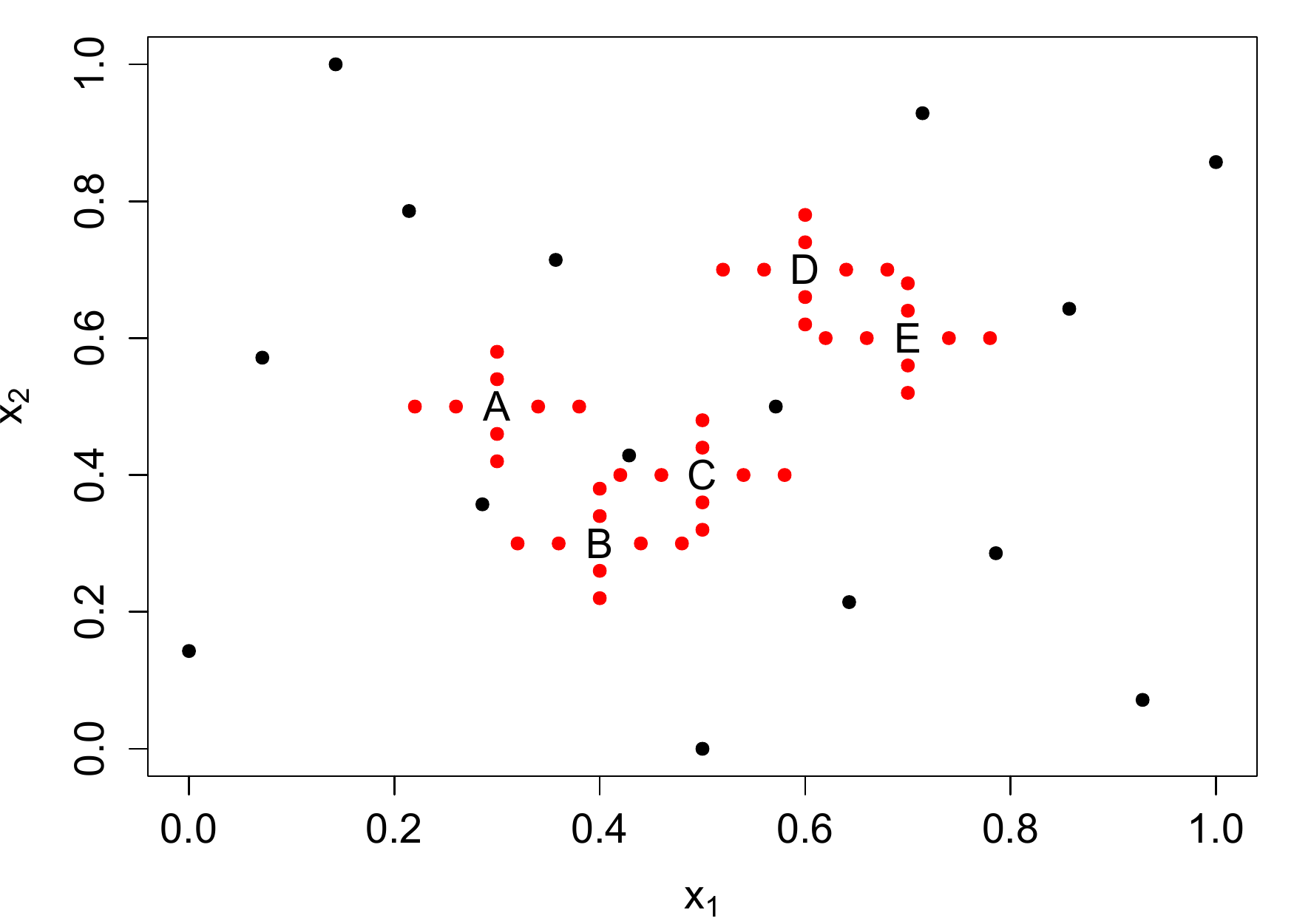}
  \caption{Example 2. Input sets; training set (black), prediction set (letters), derivative set (red)}
\label{design01}
\end{figure}

{\bf Example 2.} Consider the function, $y(x_1,x_2)=11x_1^{10}+10x_2^9+9x_1^8+8x_2^7+7x_1^6$, evaluated at 15 locations specified by a Latin hypercube design in the unit square, displayed in Figure~\ref{design01}.  A GP model is fit to the given evaluations to estimate five points ($A-E$) in the interior of the input space.   Following the intuitive justifications for placement of derivatives in more than one dimension in Section~\ref{der-design}, the partial derivatives are enforced to be positive for the constrained model at 40 locations: 20 locations along each dimension.  The derivative points are arranged in a ``$+$" shape around each prediction location. Along the horizontal part of the ``$+$", derivative information with respect to $x_1$ is provided. Along the vertical part of the ``$+$" derivative information with respect to $x_2$ is provided (Figure~\ref{design01}).

\begin{figure}[htbp]
        \centering

        \begin{subfigure}[b]{0.3\textwidth}
                \centering
                \includegraphics[width=\textwidth]{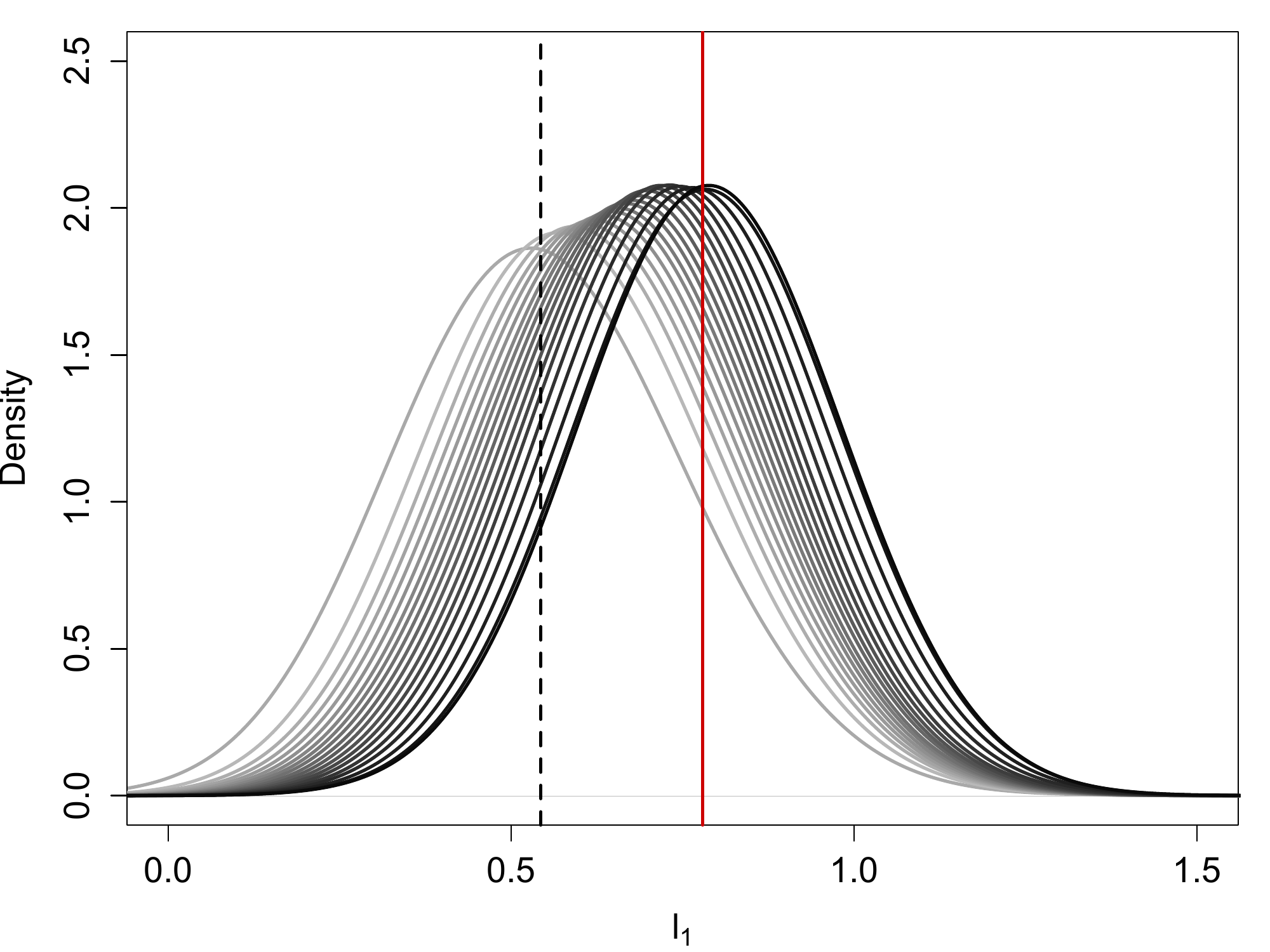}
                \caption{$\pi_t(l_1)$}
                \label{l1}
        \end{subfigure}
        \begin{subfigure}[b]{0.3\textwidth}
                \centering
                \includegraphics[width=\textwidth]{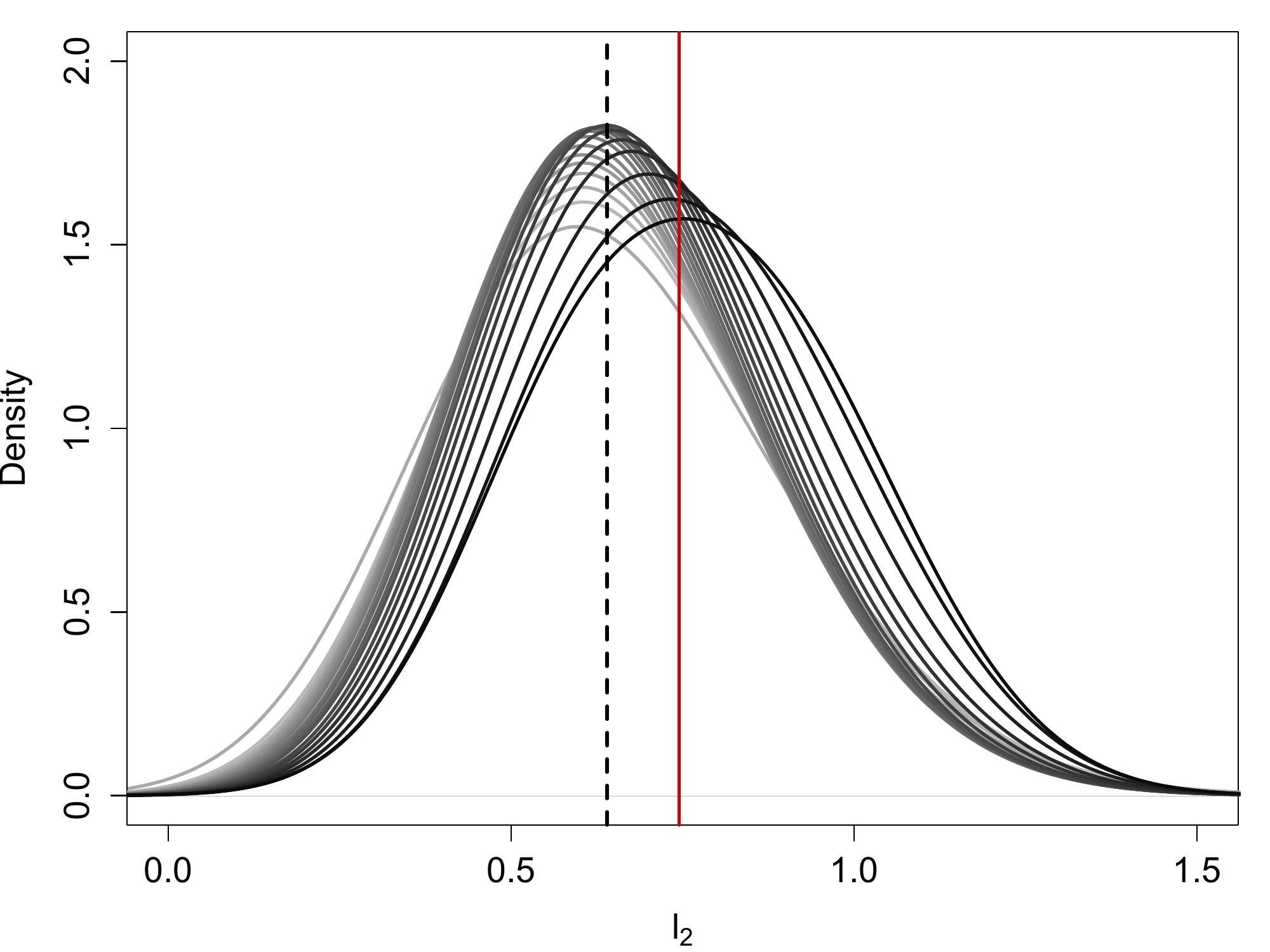}
               \caption{$\pi_t(l_2)$}
                \label{l2}
        \end{subfigure}
 \begin{subfigure}[b]{0.3\textwidth}
                \centering
                \includegraphics[width=\textwidth]{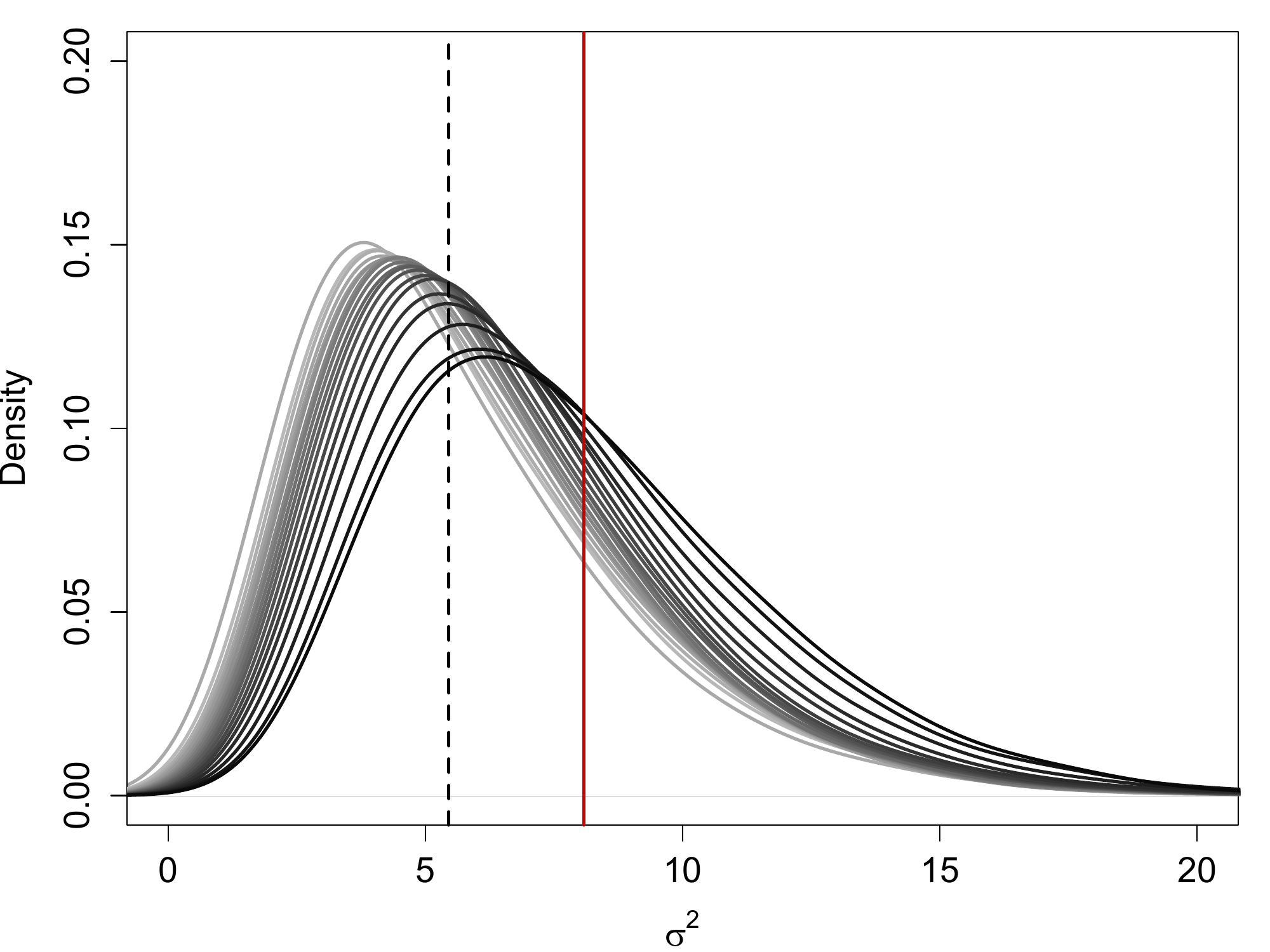}
               \caption{$\pi_t(\sigma^2)$}
                \label{sig2}
        \end{subfigure}

        \caption{Monotone emulation example; SCMC evolution of GP hyper-parameters; kernel density estimates of the posterior at times $t=0,1,\ldots,T$, the color of the curves grows darker with time; the posterior means for times $t=0$ (black) and $t=T$ (red) are plotted for each parameter; (a) length scale in the first dimension (b) length scale in the second dimension (c)  variance parameter.}\label{GP-par}
\end{figure}

\begin{figure}[htbp]
        \centering
        \begin{subfigure}[b]{0.3\textwidth}
                \centering
                \includegraphics[width=\textwidth]{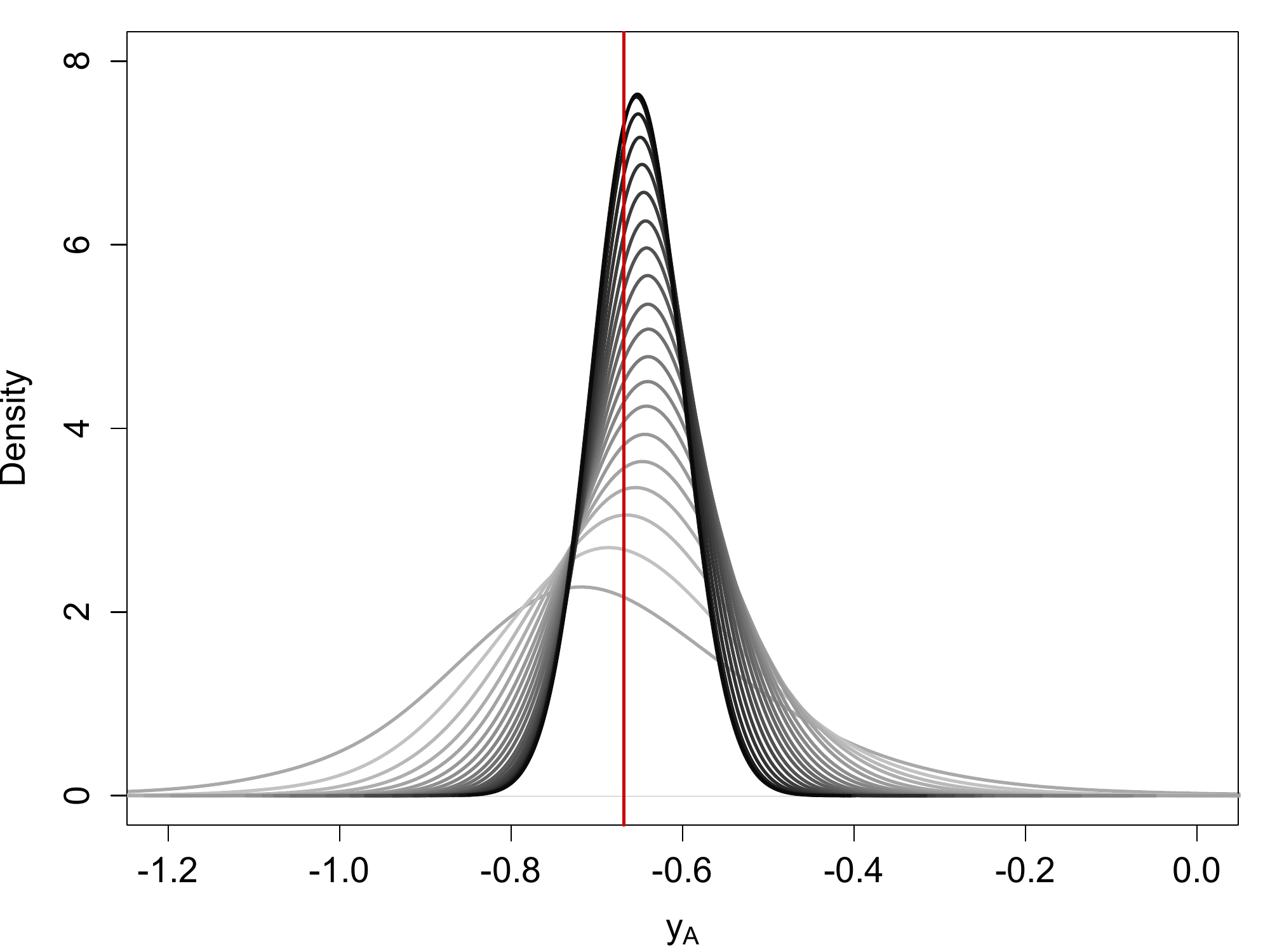}
                \caption{$\pi_t\left(y_A\right)$}
                \label{yA}
        \end{subfigure}
        \begin{subfigure}[b]{0.3\textwidth}
                \centering
                \includegraphics[width=\textwidth]{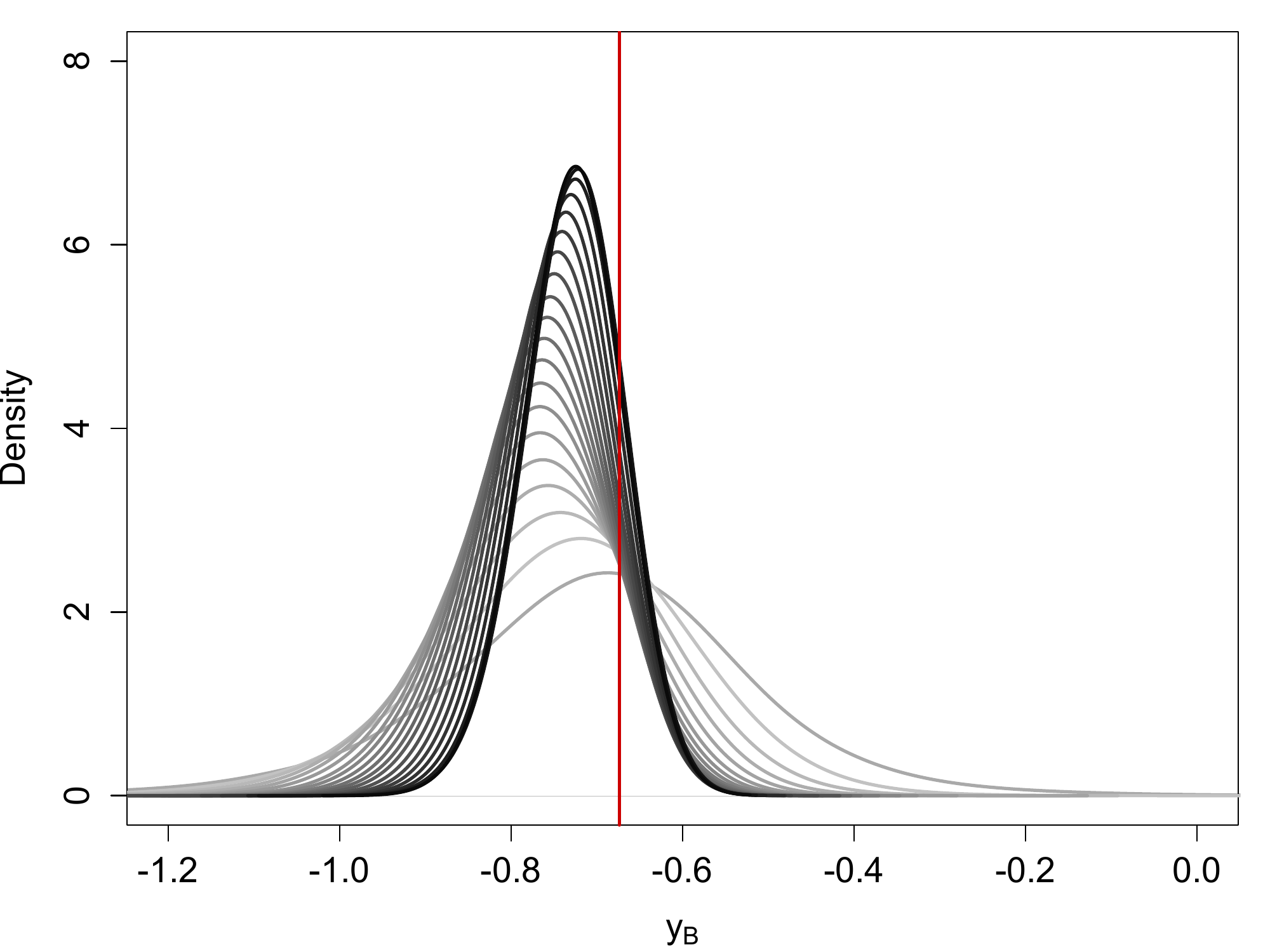}
                \caption{$\pi_t\left(y_B\right)$}
                \label{yB}
        \end{subfigure}
        \begin{subfigure}[b]{0.3\textwidth}
                \centering
                \includegraphics[width=\textwidth]{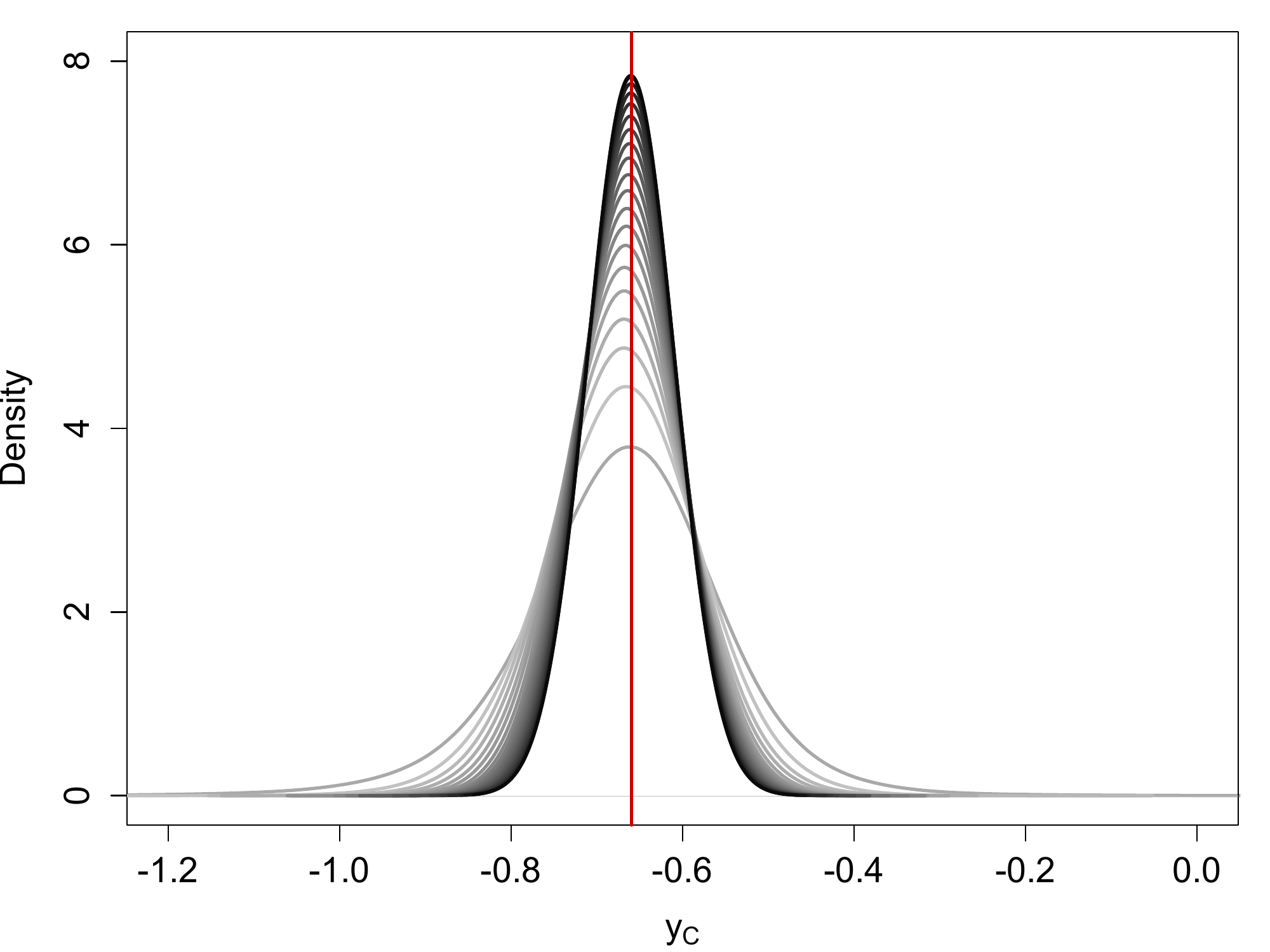}
               \caption{$\pi_t\left(y_C\right)$}
                \label{yC}
        \end{subfigure}
 \begin{subfigure}[b]{0.3\textwidth}
                \centering
                \includegraphics[width=\textwidth]{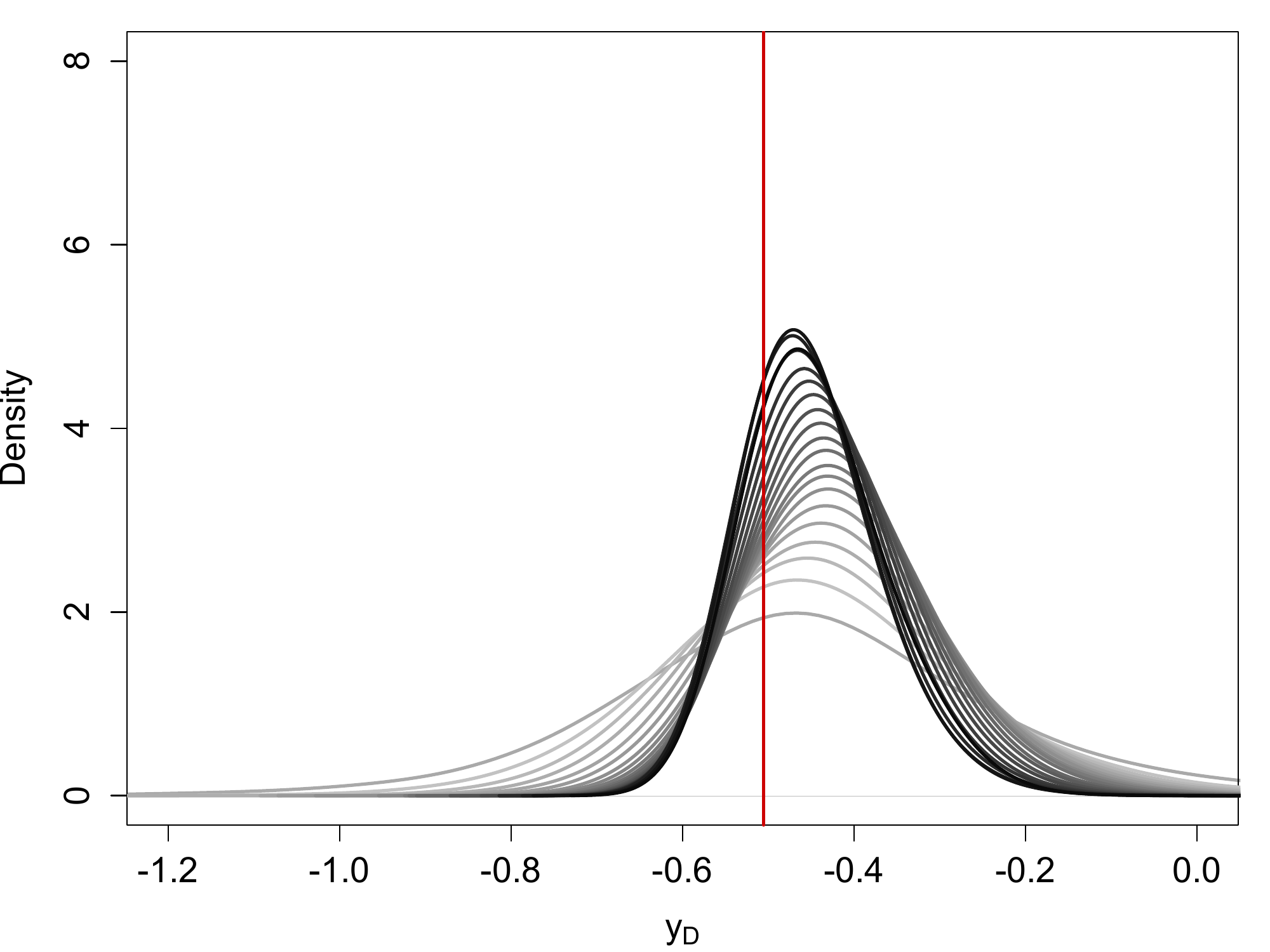}
               \caption{$\pi_t\left(y_D\right)$}
                \label{yD}
        \end{subfigure}
     \begin{subfigure}[b]{0.3\textwidth}
                \centering
                \includegraphics[width=\textwidth]{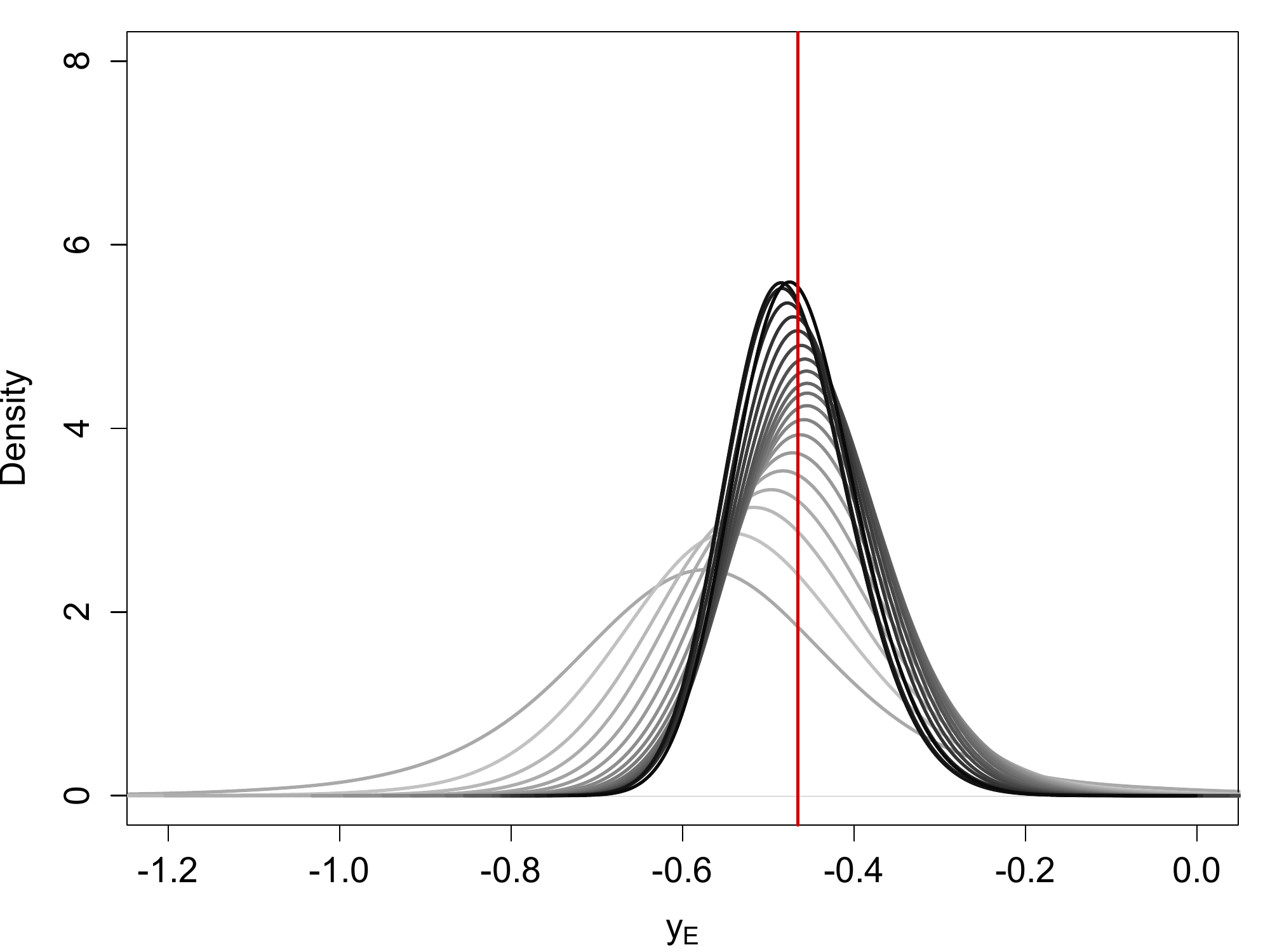}
               \caption{$\pi_t\left(y_E\right)$}
                \label{yE}
        \end{subfigure}
     
        \caption{Monotone emulation example; evolution of predictions at points A-E; kernel density estimates of the posterior predictive distribution at times $t=0,1,\ldots,T$, the color of the curves grows darker with time; the red vertical lines show the true function values.}\label{pred-ev}
\end{figure}

The results are presented as the kernel density estimates of the posterior of the GP hyper-parameters (Figure~\ref{GP-par}) and the predictions (Figure~\ref{pred-ev}) as the posterior samples evolve towards the target posterior in twenty steps of the SCMC algorithm. While the light grey curves corresponding to the earlier steps of the sampler are diffuse due to large prediction uncertainty, the posterior becomes more focused about the true values as the monotonicity constraint is imposed more strictly (darker curves).

\section{Simulation study}

\label{simulation}
In this section we describe a simulation study that demonstrates the performance of the methodology to predict monotone polynomials. The monotone and unconstrained GP models are compared in terms of the root mean squared error, the average width of the 95\% credible intervals and coverage probability for 100 simulated data sets.  

\begin{figure}[h!]
        \centering
        \begin{subfigure}[b]{0.4\textwidth}
                \centering
                \includegraphics[width=\textwidth]{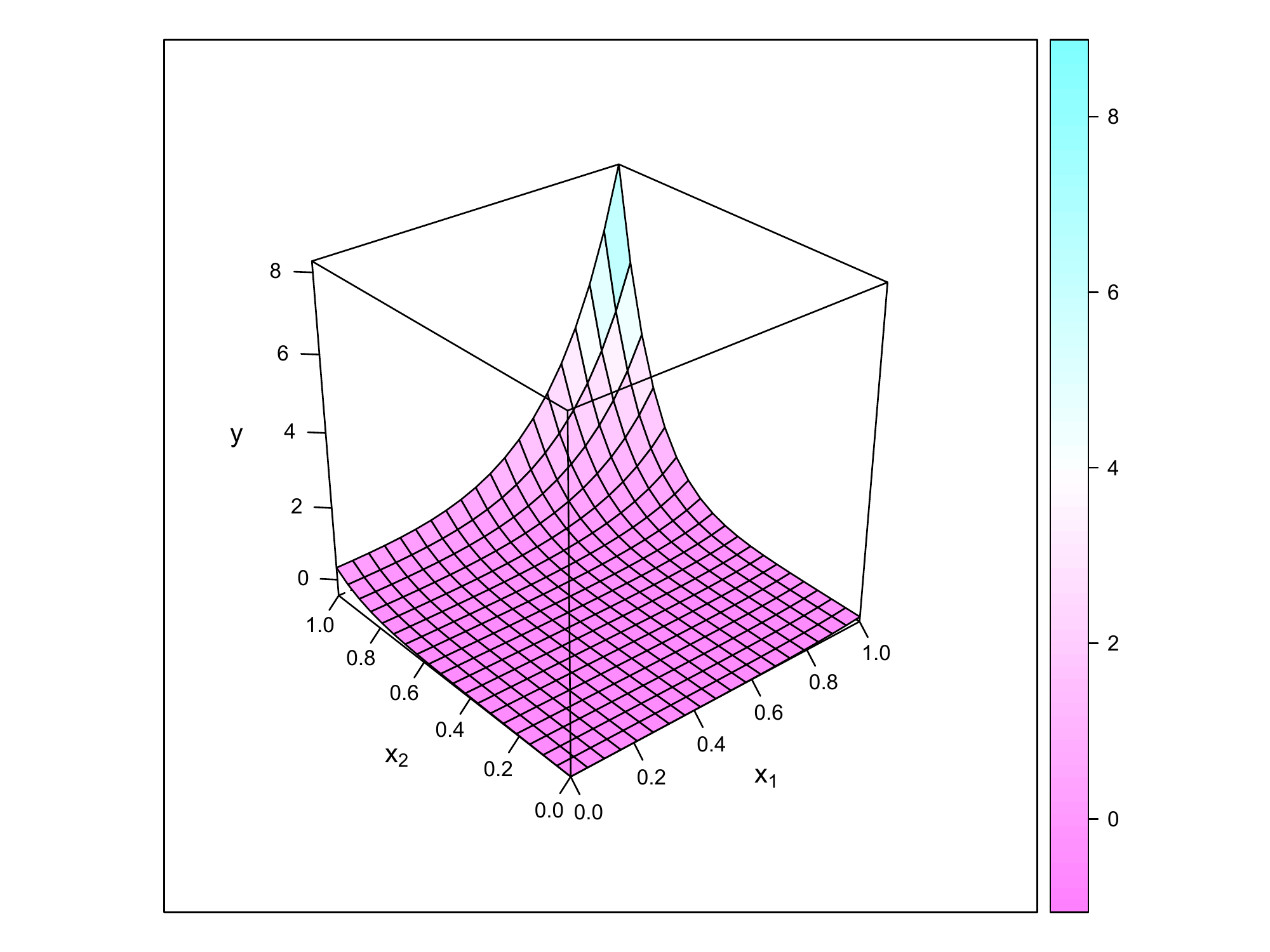}
                \caption{}
                \label{nu1}
        \end{subfigure}
        ~ 
        \begin{subfigure}[b]{0.4\textwidth}
                \centering
                \includegraphics[width=\textwidth]{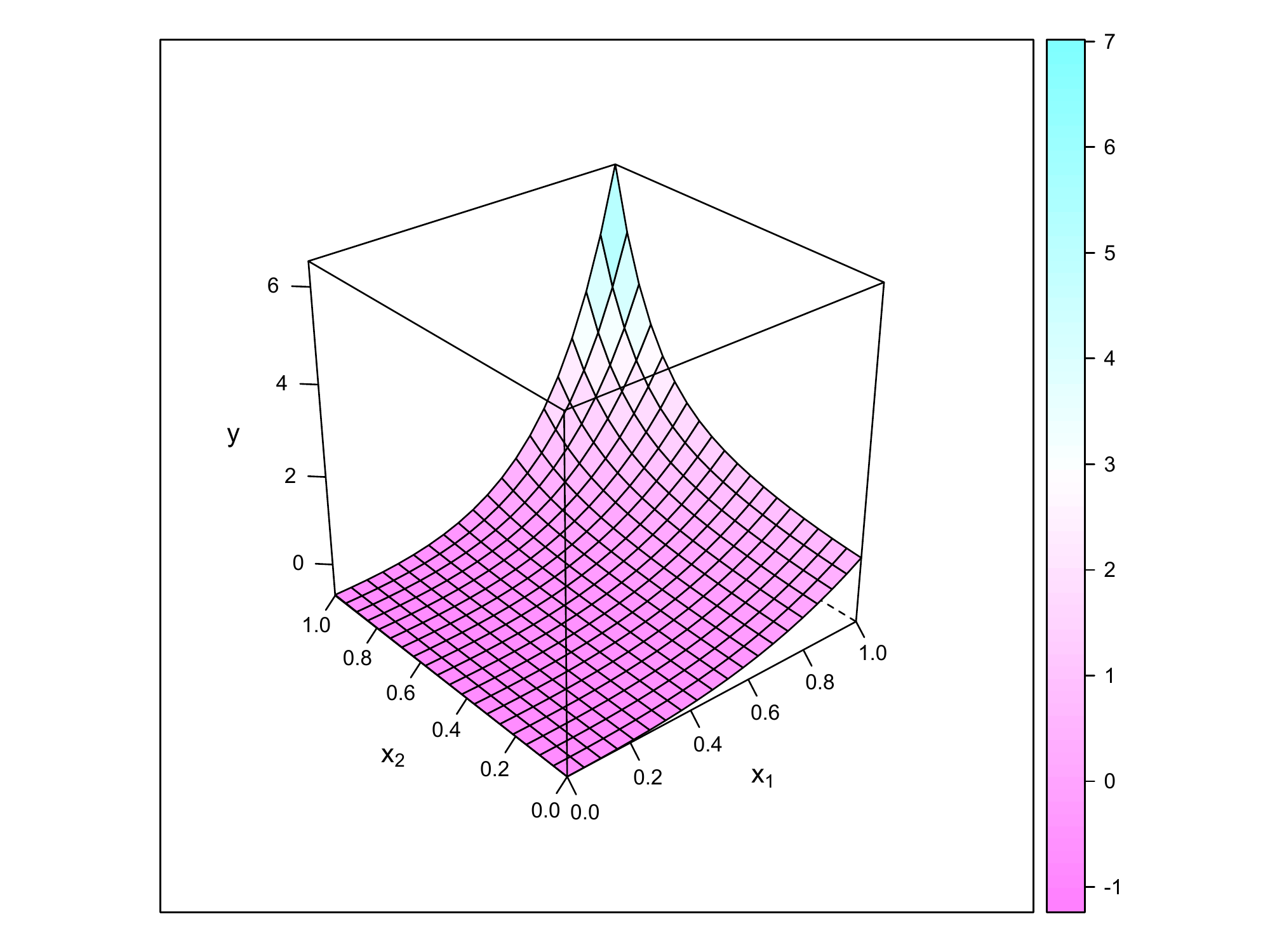}
                \caption{}
                \label{nu2}
        \end{subfigure}
         \begin{subfigure}[b]{0.4\textwidth}
                \centering
                \includegraphics[width=\textwidth]{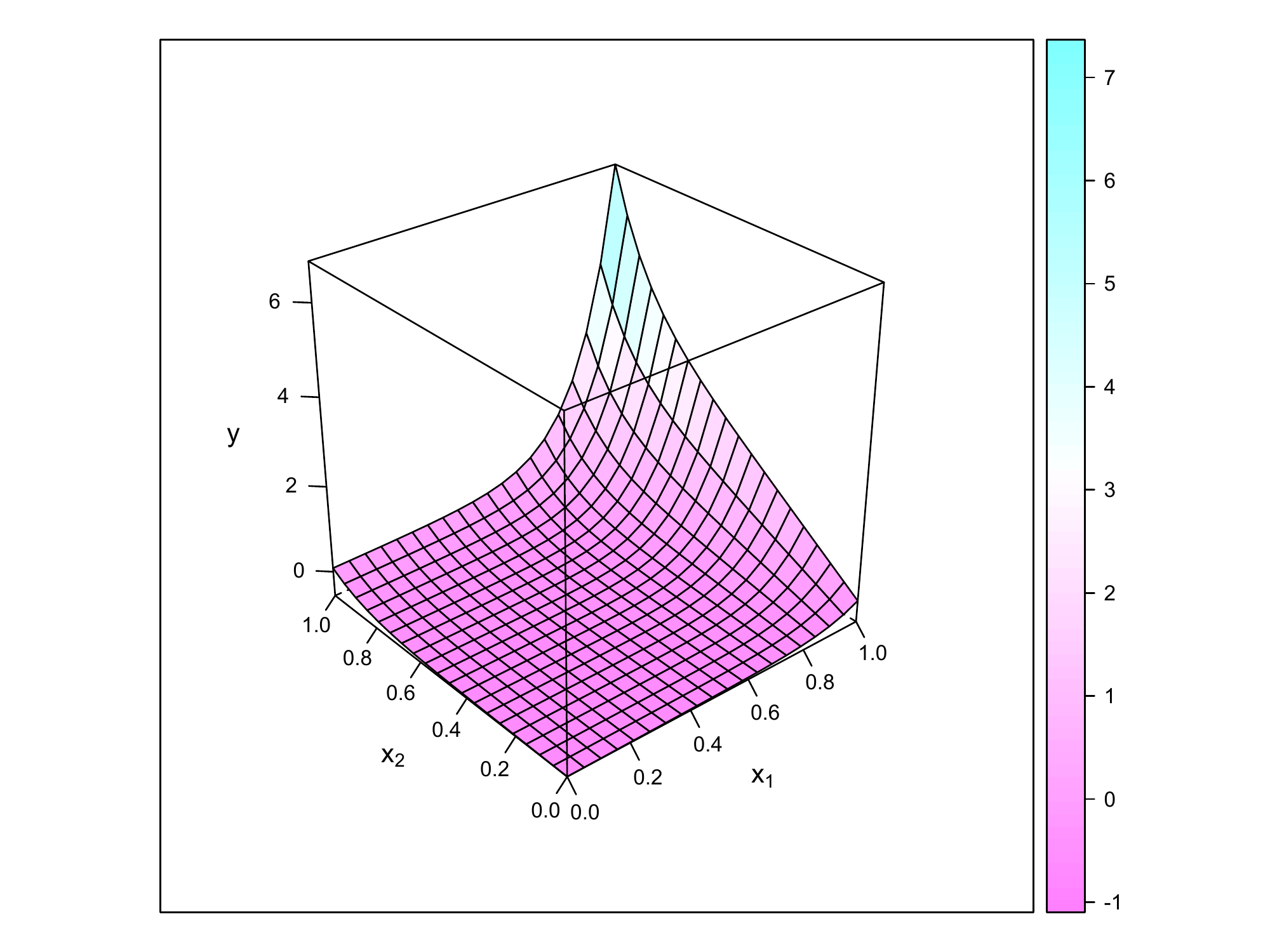}
                \caption{}
                \label{nu3}
        \end{subfigure}
         \begin{subfigure}[b]{0.4\textwidth}
                \centering
                \includegraphics[width=\textwidth]{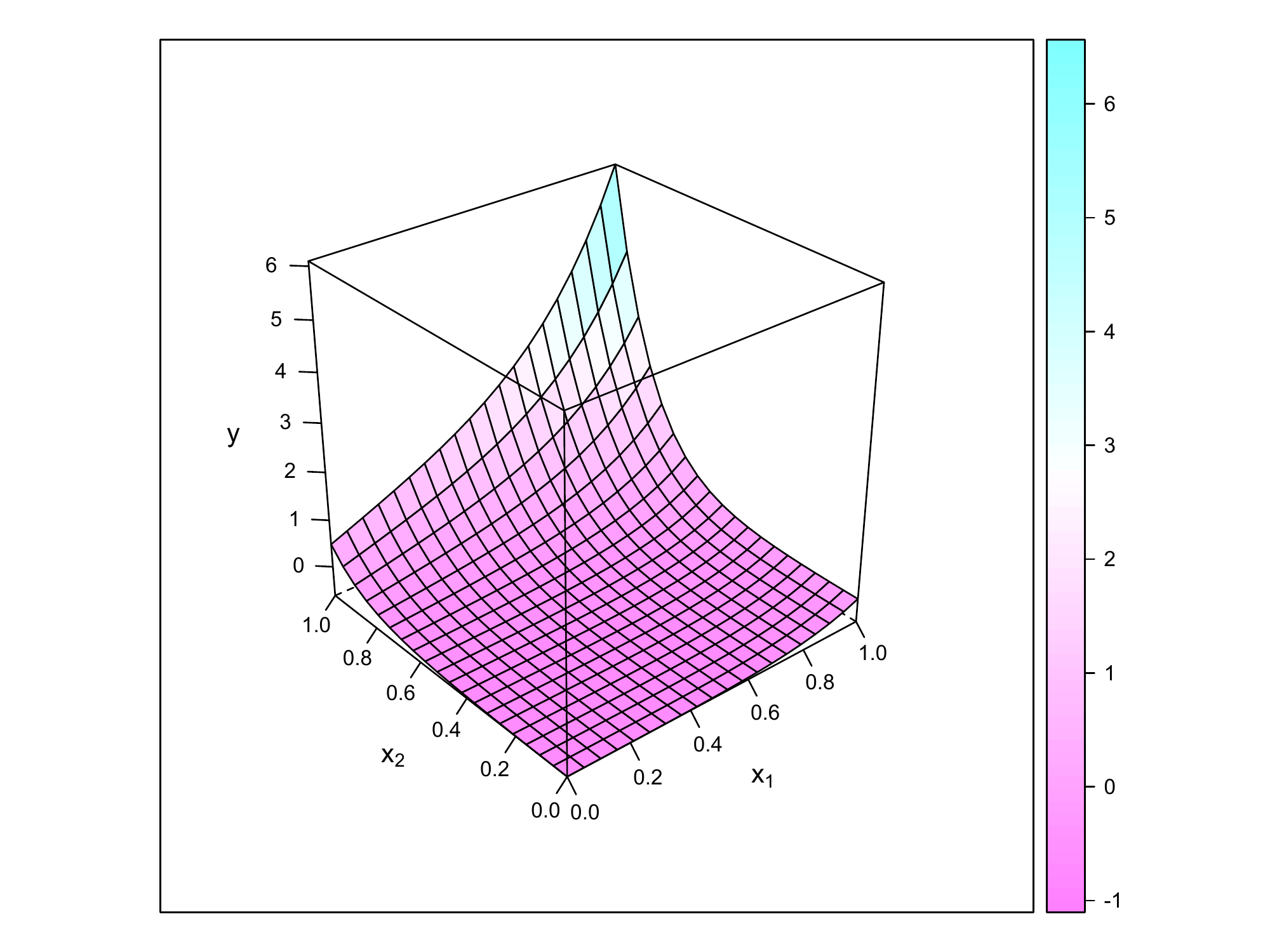}
                \caption{}
                \label{nu4}
        \end{subfigure}
        
        \caption{Simulation study: examples of polynomials with random coefficients generated from a gamma(.01,1)}\label{surfaces}
\end{figure}

The underlying model from which data are simulated is a 20-th order polynomial function of two inputs $x_1$ and $x_2$ where all the main effects and interactions are included with positive coefficients that are randomly generated, i.e., 

$$y(x_1,x_2)=\sum_{i=0}^{10}\sum_{j=0}^{10}\gamma_{ij}x_1^ix_2^j,$$
where $\gamma_{ij}>0$ are gamma random variables. To ensure equal contributions to the response surface from each term, the coefficients are scaled as follows; let $\gamma_{i0}$ and $\gamma_{j0}$ be the coefficients of $x_1^i$ and $x_1^j$, respectively. We wish to choose the coefficients such that 
$$\text{E}(\gamma_{i0} x_1^i)=\text{E}(\gamma_{t0}x_1^t).$$
Since $x_k\sim \text{Uniform}(0,1)$, $k=1,2$, i.e., $\text{E}(x_k^i)=\frac{1}{i+1}$, the coefficients must be chosen such that 
$$\frac{\text{E}(\gamma_{i0})}{\text{E}(\gamma_{t0})}=\frac{i+1}{j+1}.$$
Therefore, we let,
$$\gamma_{ij}=(i+1)(j+1)\beta,$$
where $\beta$ is generated from a Gamma distribution with shape parameter 0.01 and rate parameter 1. Figure~\ref{surfaces} shows four such generated surfaces.

\begin{figure}[h!]
        \centering
        \begin{subfigure}[b]{0.4\textwidth}
                \centering
                \includegraphics[width=\textwidth]{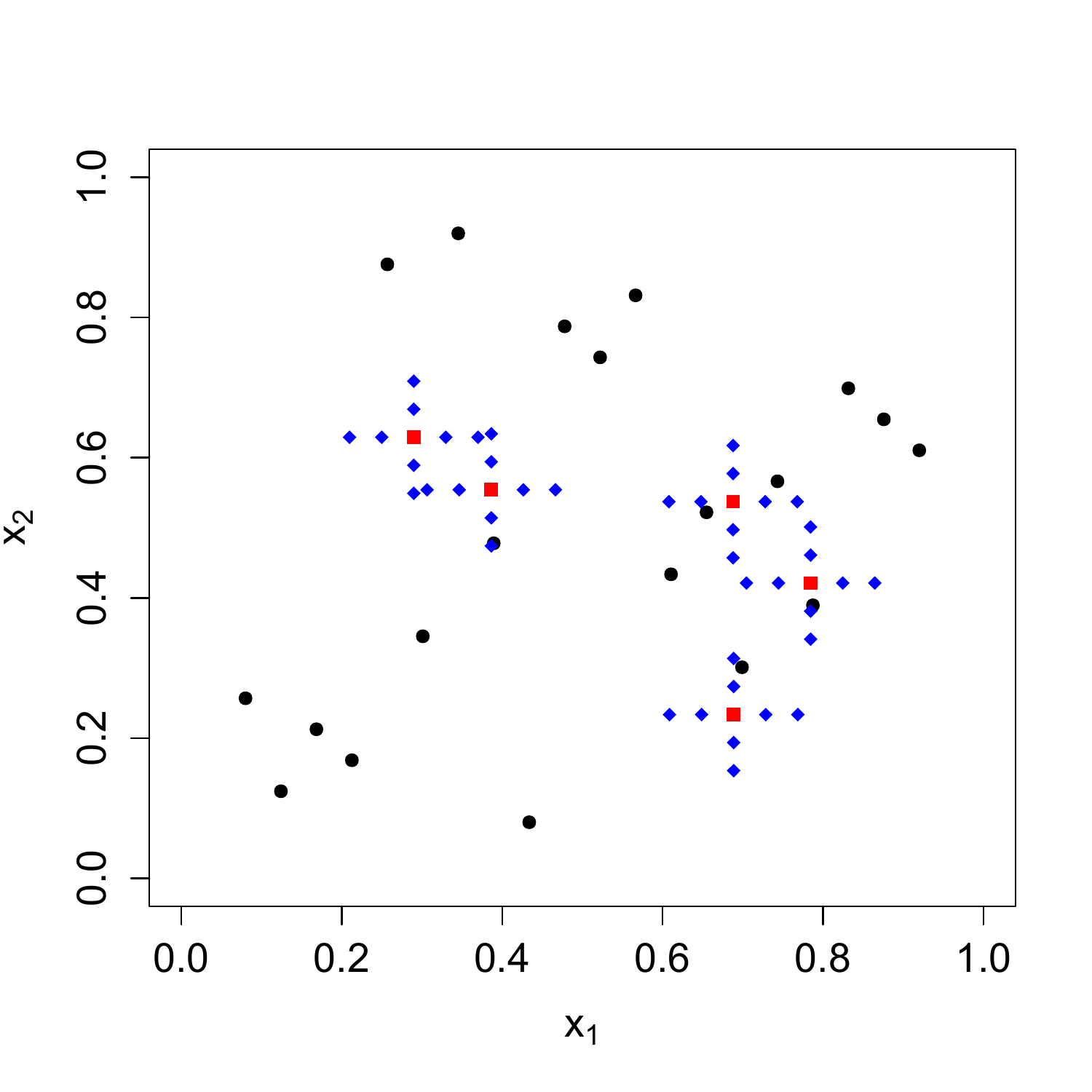}
                \caption{}
                \label{d1}
        \end{subfigure}
        ~ 
        \begin{subfigure}[b]{0.4\textwidth}
                \centering
                \includegraphics[width=\textwidth]{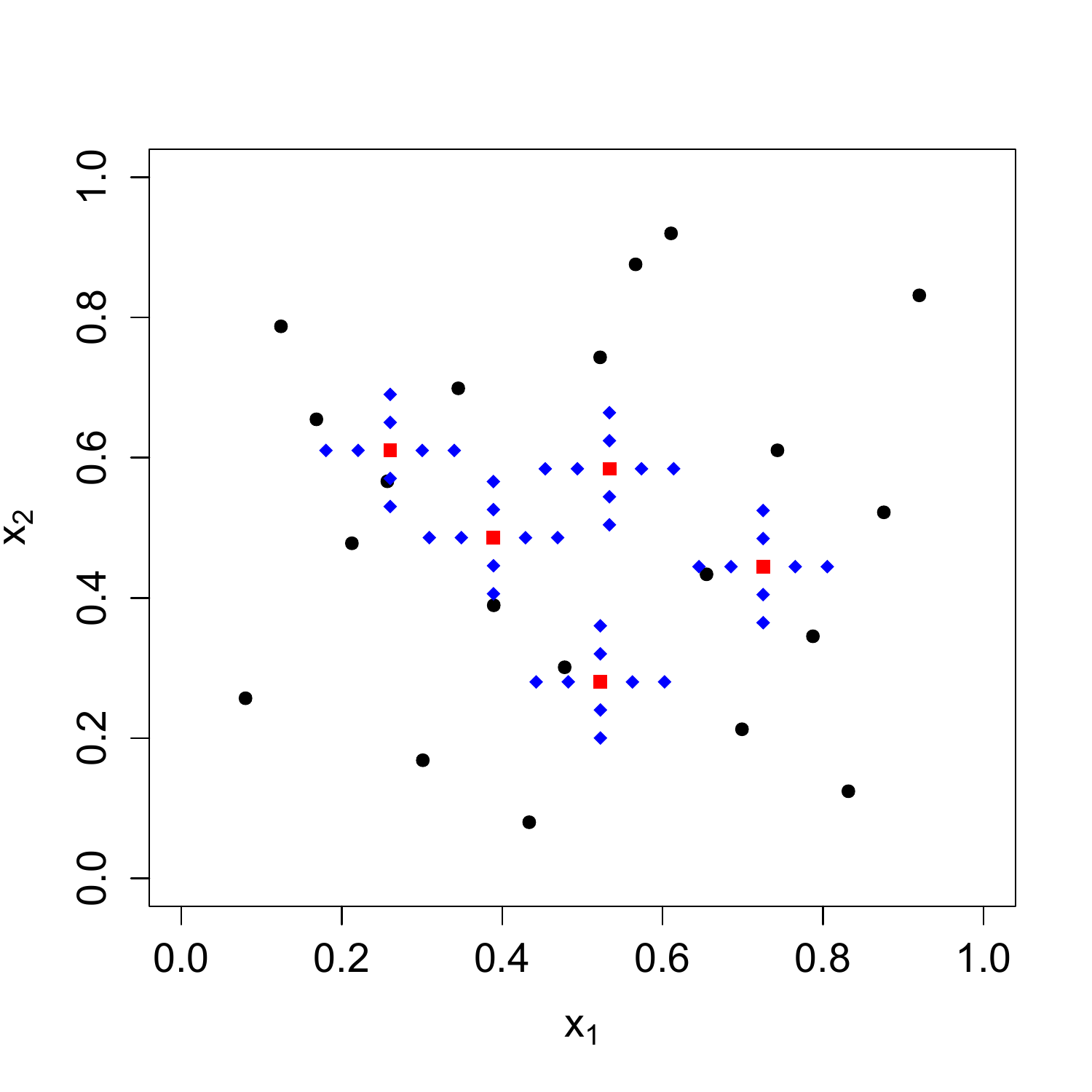}
                \caption{}
                \label{d2}
        \end{subfigure}
         \begin{subfigure}[b]{0.4\textwidth}
                \centering
                \includegraphics[width=\textwidth]{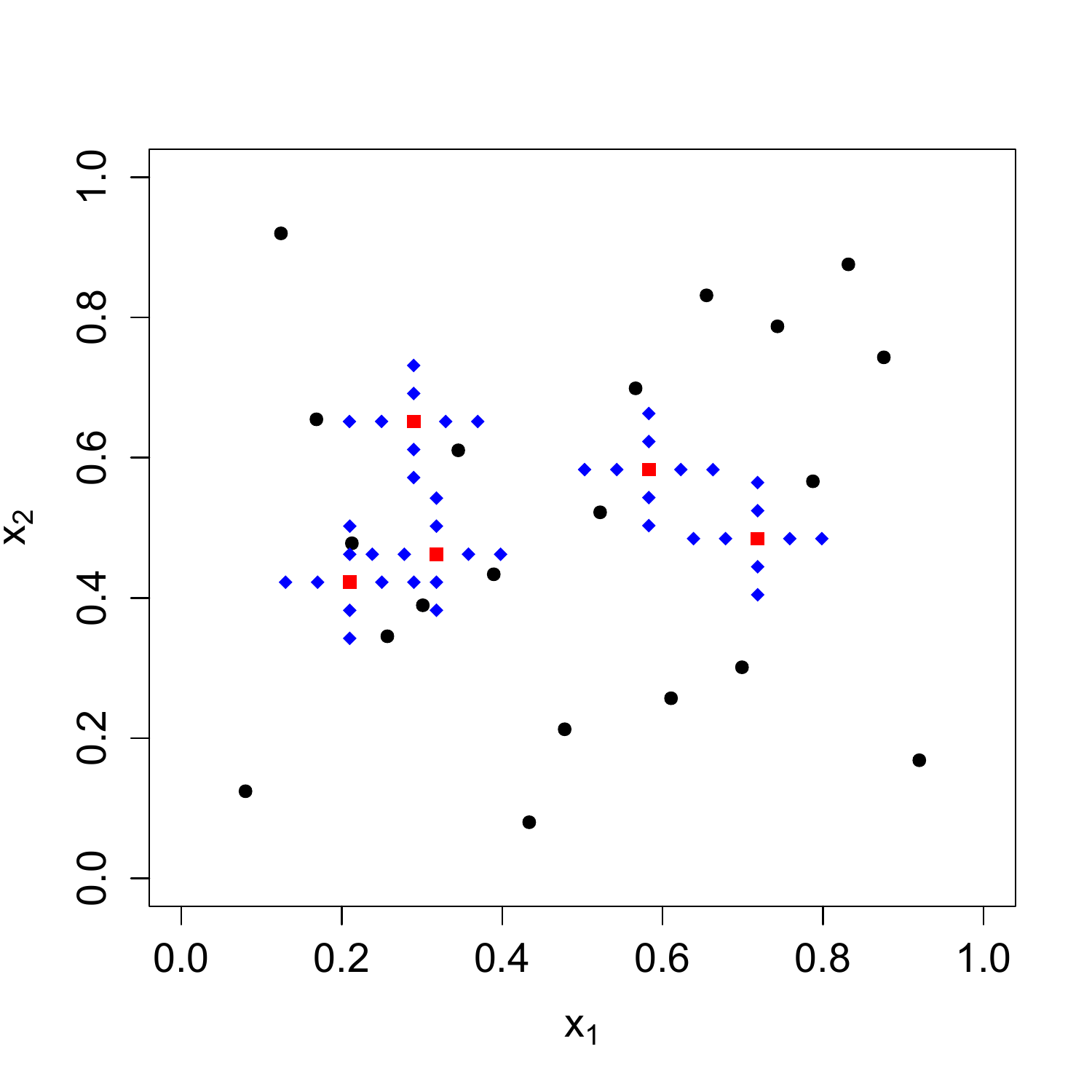}
                \caption{}
                \label{d3}
        \end{subfigure}
         \begin{subfigure}[b]{0.4\textwidth}
                \centering
                \includegraphics[width=\textwidth]{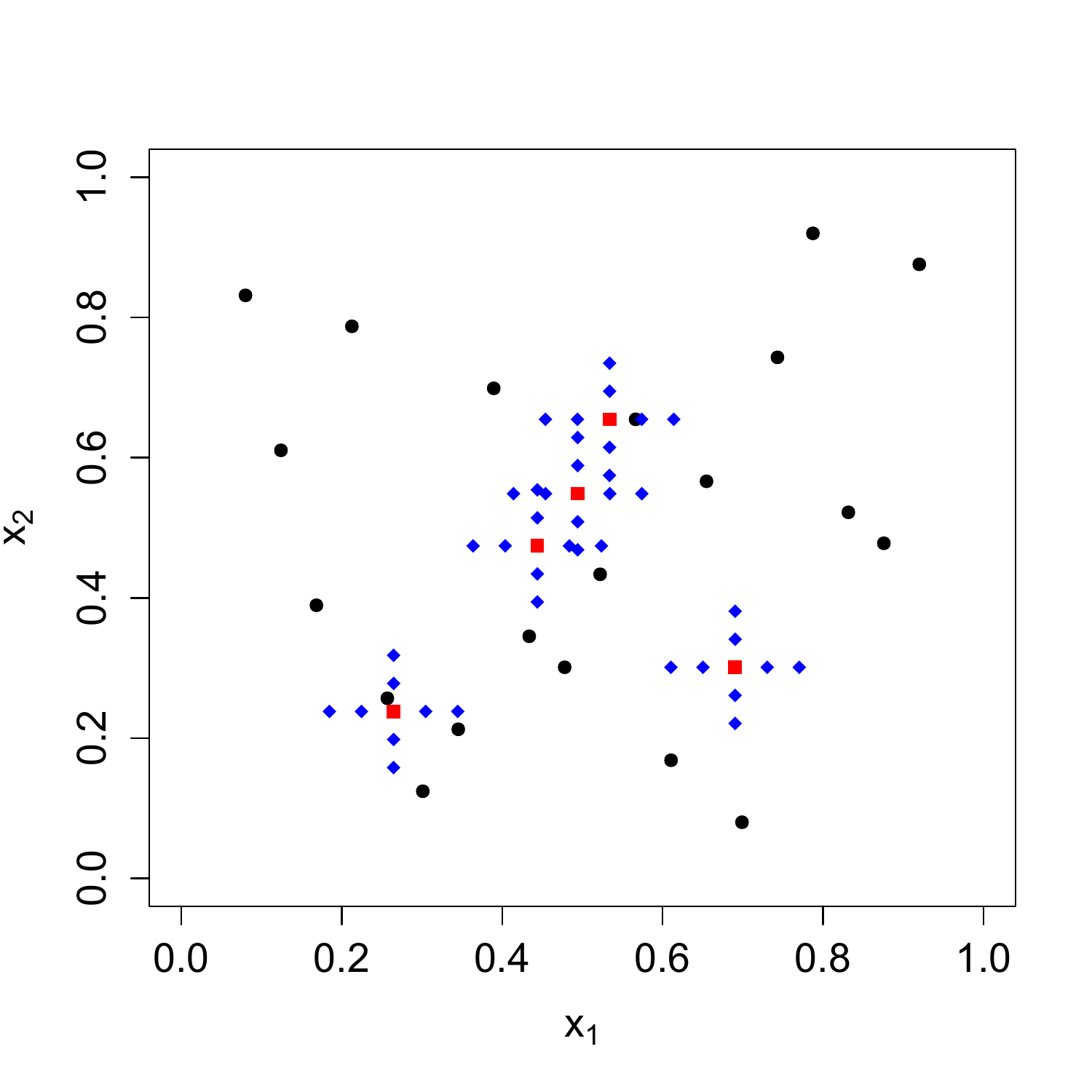}
                \caption{}
                \label{d4}
        \end{subfigure}
        
        \caption{Simulation input sets; training set (black dots), prediction set (red squares), derivative set (blue diamonds)}\label{simdesign}
\end{figure}

Each polynomial is evaluated at 25 points on a random Latin hypercube design. These 25 points are randomly divided into a training set of size $n=20$ and a test set of size five. As in Example 2, derivative points are placed on both sides of the prediction locations along each axis. Figure \ref{simdesign} shows four of these random designs, with a total of $p=40$ derivative locations. 

For each generated data set, an unconstrained GP model and the proposed monotone model are used to make predictions at five test points. The following criteria are calculated for the two models: the root mean squared error (RMSE) for each simulated data set,
$$\text{RMSE}=\sqrt{\frac{1}{5}\sum_{i=1}^{5}(\hat{y_i}-y_{\text{true},i})^2},$$
where $\hat{y_i}$ is the posterior mean for the $i$-th predicted value and $y_{\text{true}}$ is the true value of the function at the same point; the average width of the 95\% credible intervals (AWoCI) for each data set,
$$\text{AWoCI}=\frac{1}{5}\sum_{i=1}^{5}(Q^{(i)}_{0.975}-Q^{(i)}_{0.025}),$$
where $Q^{(i)}_p$ is the $p$-th posterior sample quantile; and the coverage probability over the 500 predicted points (5 test points and 100 data set realizations),
$$\text{cp}=\frac{1}{500}\sum_{i=1}^{500}\delta_{\{y_{\text{true},i}\in (Q^{(i)}_{0.025},Q^{(i)}_{0.975})\}}.$$

\begin{figure}[h!]
        \centering
        \begin{subfigure}[b]{0.4\textwidth}
                \centering
                \includegraphics[width=\textwidth]{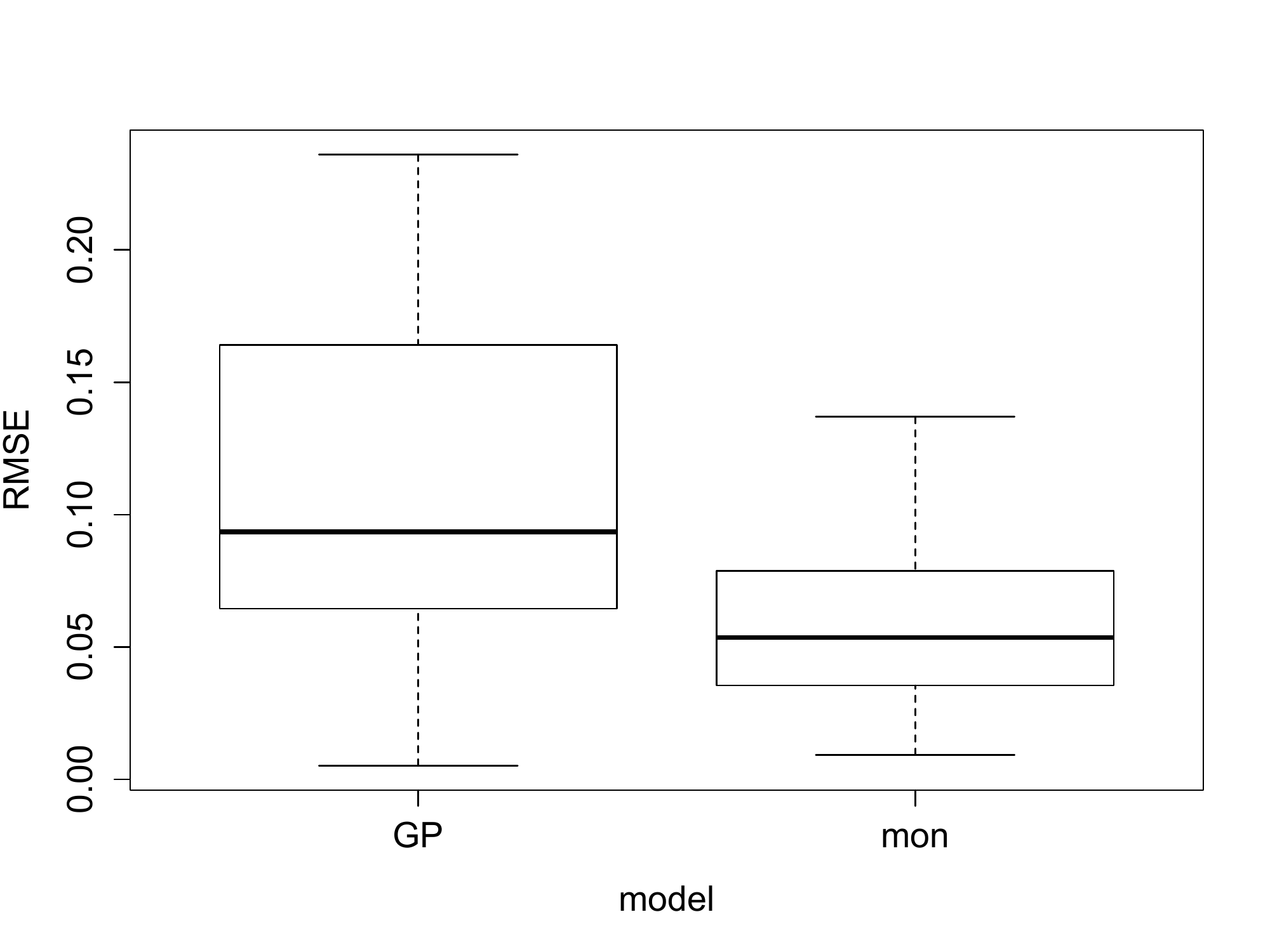}
                \caption{}
                \label{MSEbox}
        \end{subfigure}
        ~ 
        \begin{subfigure}[b]{0.4\textwidth}
                \centering
                \includegraphics[width=\textwidth]{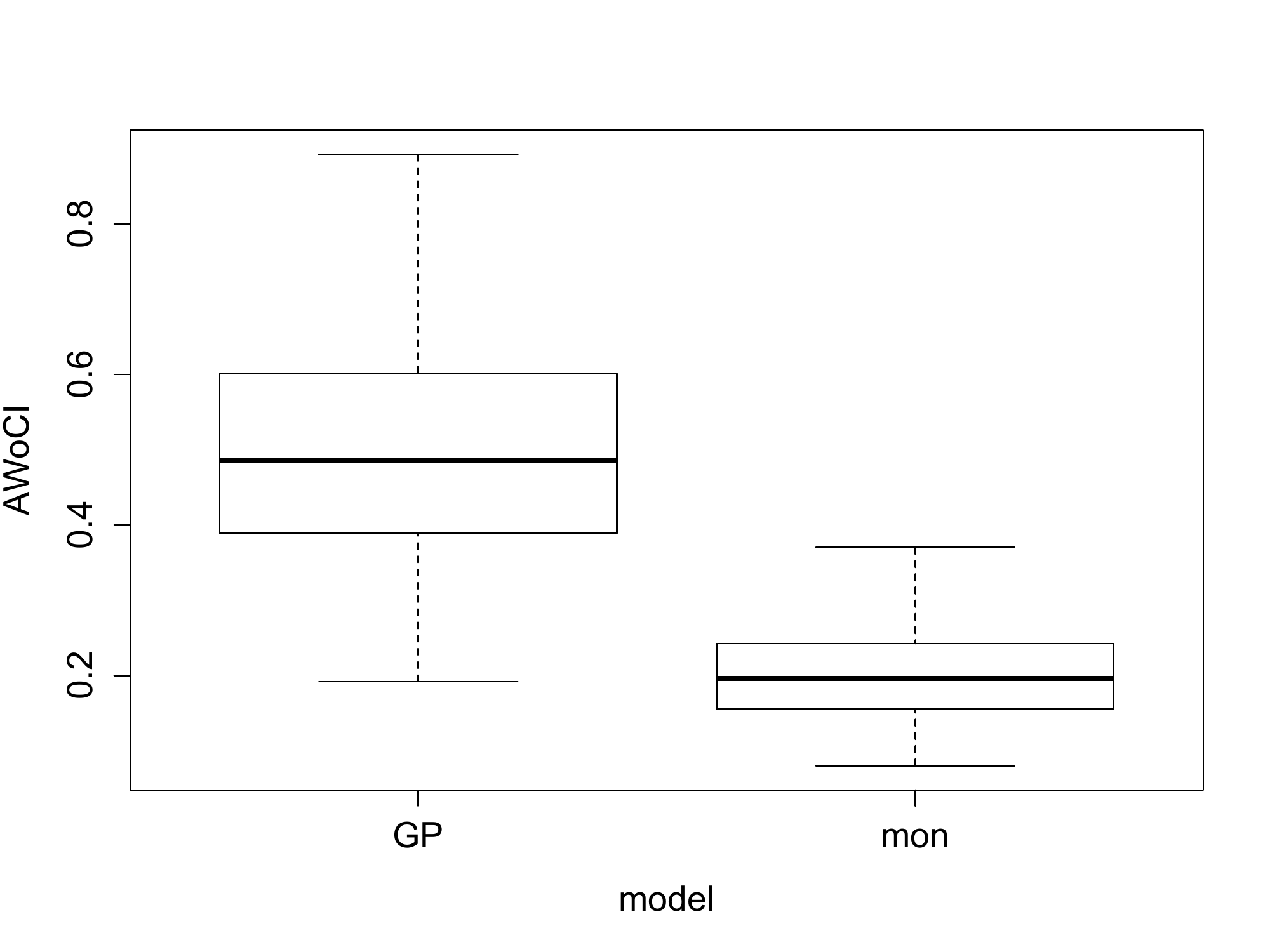}
                \caption{}
                \label{volbox}
        \end{subfigure}

        \caption{Simulation results: side by side boxplots of the (a) calculated RMSEs and (b) average width of the 95\% credible intervals for the unconstrained GP model and the monotone model}\label{simresults}
\end{figure}

The comparison results are illustrated in forms of side by side boxplots of the calculated RMSEs and the average width of the 95\% credible intervals for the two models in Figures~\ref{MSEbox} and \ref{volbox}, respectively.

The decrease in RMSE as a result of using the monotone model is evident from the boxplots.  Also, the 95\% credible intervals are considerably narrower for the monotone model. The reduction in prediction uncertainty for the monotone model is consistent with the examples in Section~\ref{examples}. The observed coverage probabilities of the 95\% credible intervals for the GP model and the constrained model are 0.960 and 0.908, respectively. 

\section{Queuing application}

\begin{figure}[h!]
  \centering
    \includegraphics[width=.6\textwidth]{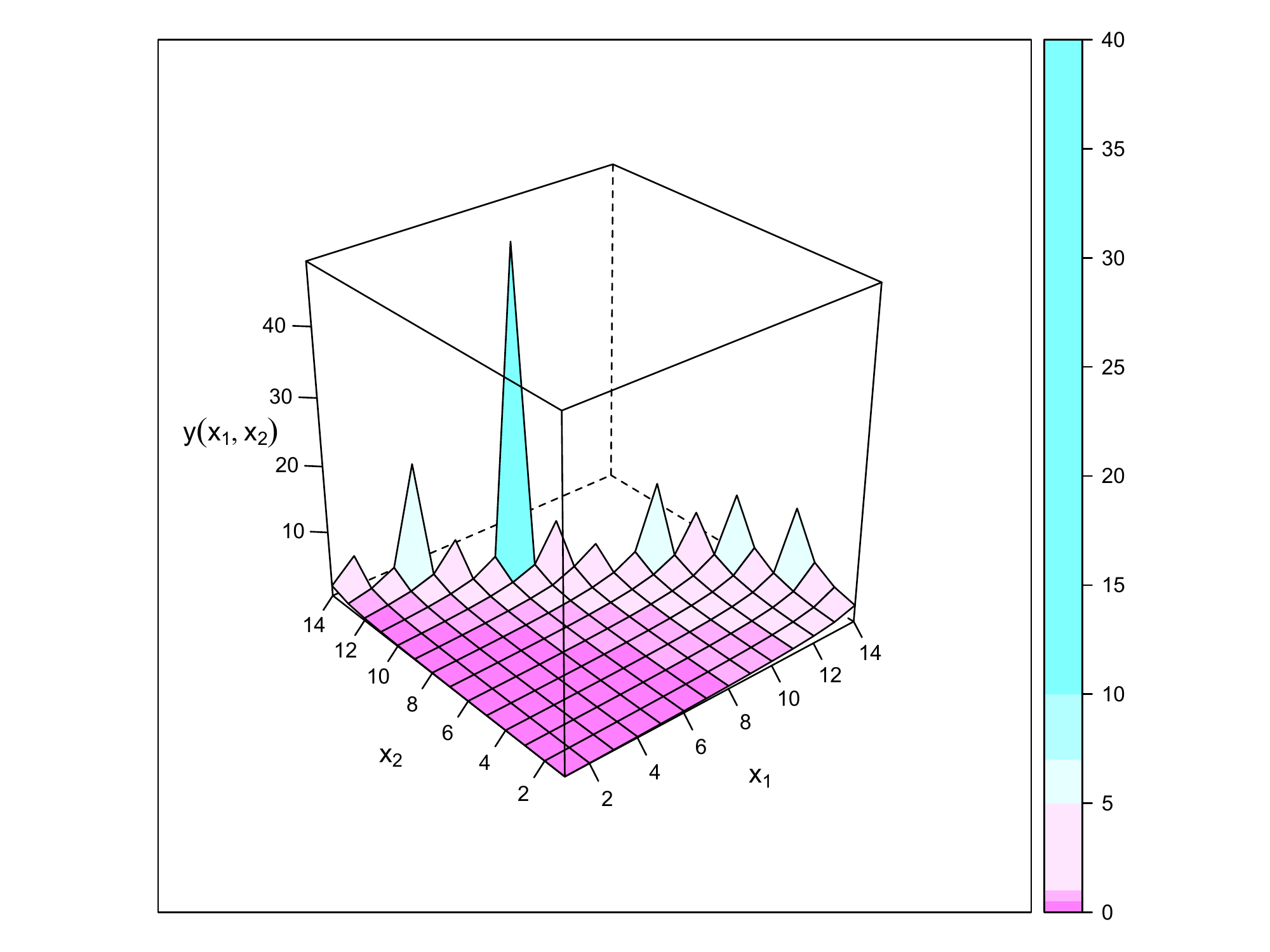}
  \caption{queuing application: the average delay as a function of the two job arrival rates}
\label{queue-surface}
\end{figure}

The queuing application from Section~\ref{queue} is now re-visited.  The two inputs to the computational model are the arrival rates, $x_1$ and $x_2$.  The response variable is the average delay - or average time spent waiting in the queue for others to be served.  The true response surface is plotted in Figure~\ref{queue-surface}.  The system response is monotone in each of the inputs: one would expect to wait longer if the arrival rate in the queue is larger.

\begin{figure}[h!]

  \centering
    \includegraphics[width=.5\textwidth]{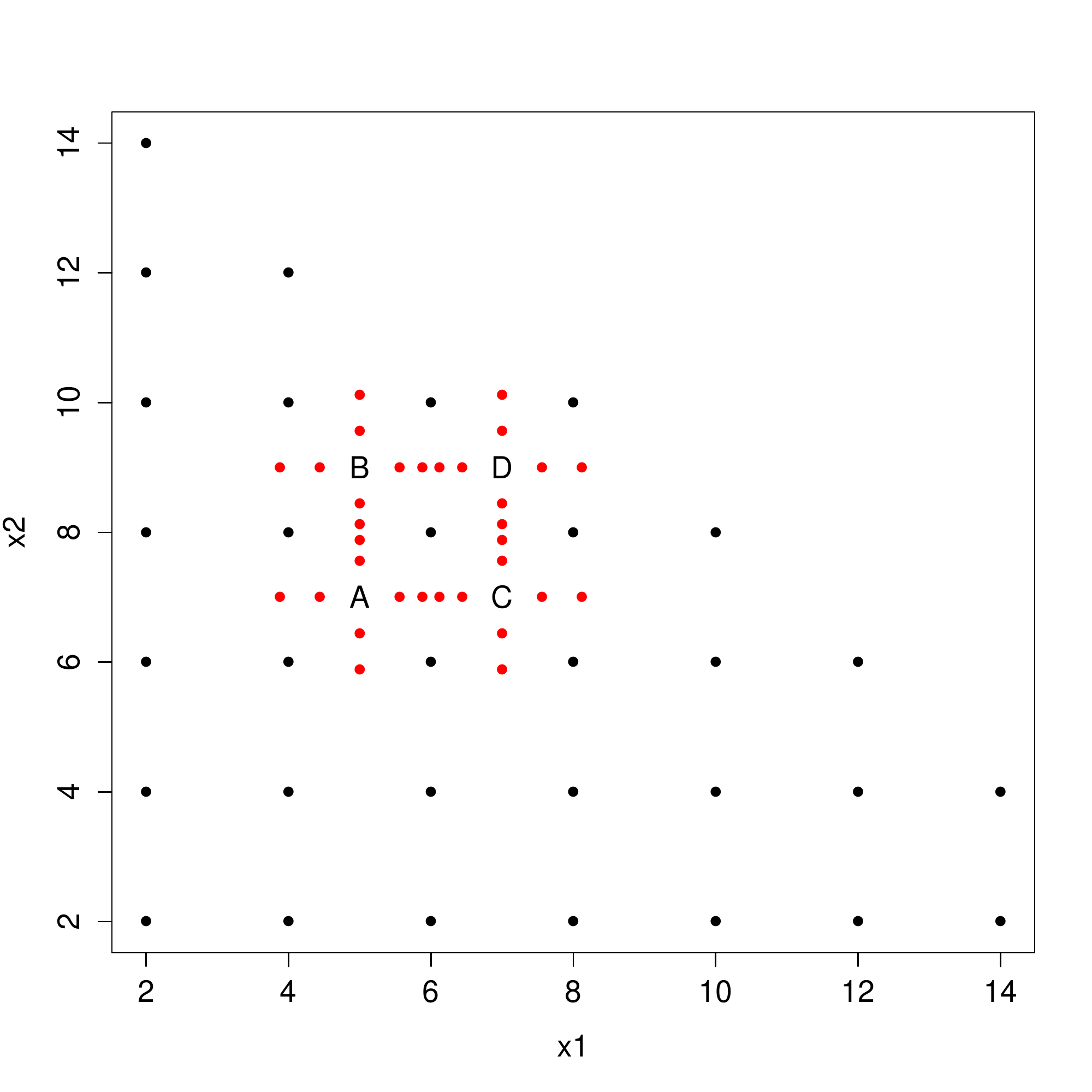}
  \caption{Input sets; training set (black), prediction set (letters), derivative set (red)}
\label{qdatanewdes}
\end{figure}

The input region to be investigated is not rectangular \citep{Ranjan08}.  Instead the region of interest, after scaling, is a subset of the unit square where the expected delay is finite. The proposed methodology is evaluated using this setting. The design is shown in Figure~\ref{qdatanewdes}. The evaluation set contains 32 points (the black dots in Figure~\ref{qdatanewdes})  on a grid in the two dimensional input space. Predictions at four input locations, $ A-D$, are made and compared to the true system response.  We chose these points because they are in a region of the  design space where the response surface changes quite rapidly.  This is a challenging emulation problem because such non-stationary behavior is difficult to model with a stationary GP.  The derivatives are constrained at the locations shown by the red dots in Figure~\ref{qdatanewdes} - that is four points along each axis in the neighborhood of the prediction locations.  

\begin{figure}[h!]
        \centering
        \begin{subfigure}[b]{0.4\textwidth}
                \centering
                \includegraphics[width=\textwidth]{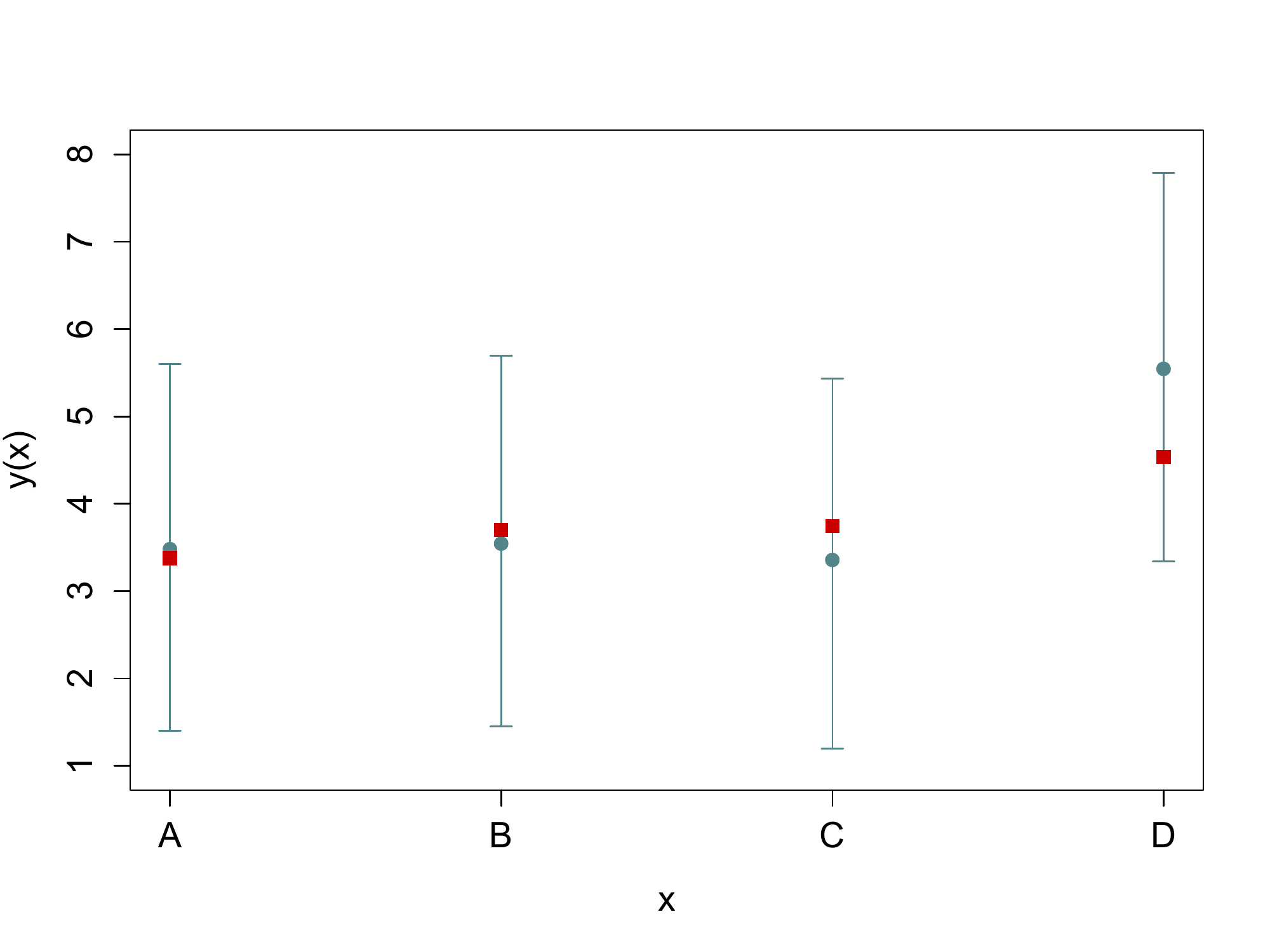}
                \caption{}
                \label{qdatanewa}
        \end{subfigure}
        ~ 
        \begin{subfigure}[b]{0.4\textwidth}
                \centering
                \includegraphics[width=\textwidth]{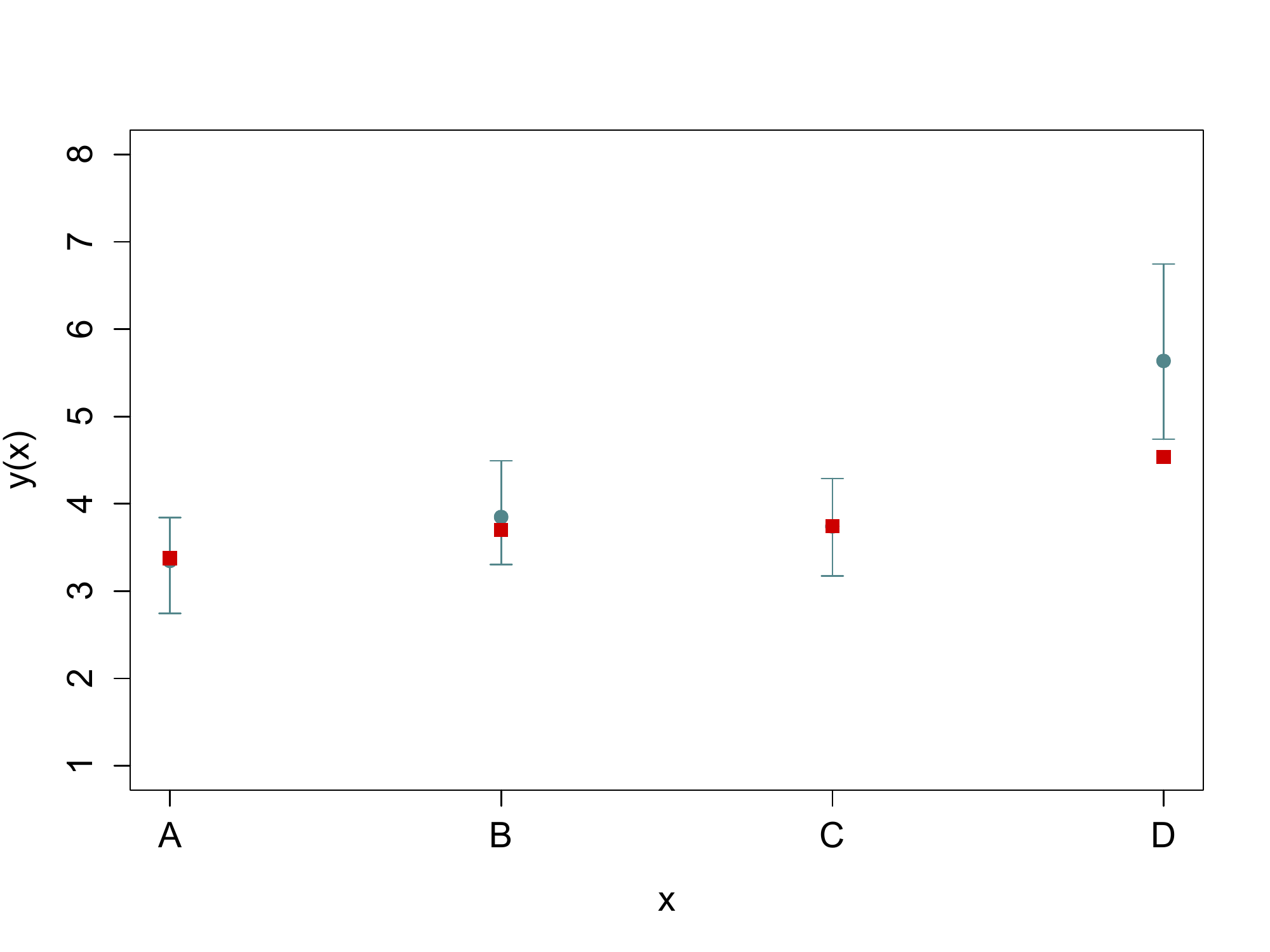}
                \caption{}
                \label{qdatanewb}
        \end{subfigure}
        
        \caption{Posterior mean and 95\% credible intervals obtained by (a) unconstrained Bayesian GP model (b) GP model with monotonicity constraints; the red squares show the true function values.}\label{qdatanew}
\end{figure}

The proposed monotone interpolator and the unconstrained GP are applied to this application. The posterior mean and 95\% credible intervals obtained from the unconstrained GP and the monotone model are given in Figure~\ref{qdatanewa}.  From the values of the posterior mean for each model, at each location, we see that both approaches are doing fairly well.  However, the posterior uncertainty is so large for the unconstrained GP as to be almost uninformative.  Notice that both approaches overestimate the response at prediction point $D$. This is due to the rapid change in the response.

Both models find this queuing application challenging.  The response surface (Figure \ref{queue-surface}) is very flat for much of the input space and increases rapidly near the boundary of the input region.  This is not the behavior of a simulator that is easily modeled by a stationary GP.  Overall, taking monotonicity into account helped reduce the predictive MSE and also the uncertainty in the predictions.

\section{Conclusion}

In this article we have developed a methodology for the incorporation of monotonicity information into the GP emulation of a deterministic computer model.  Through the mechanism of a link function, monotonicity information is encoded via virtual derivatives (binary indicators of derivative sign) at points in a derivative input set.   Some guidelines for construction of the derivative input set are proposed; an online specification of the derivative design that can be incorporated into the sampling algorithm is currently under investigation.  Full inference for unsampled values of the function and its derivatives are available via the constrained SMC algorithm.  Notably, by sampling GP parameters, $\boldsymbol{\it l}$ and $\sigma^2$, a more realistic representation of uncertainty is provided than would be obtained from plug-in estimates of $\boldsymbol{\it l}$ and $\sigma^2$.  Through examples, a set of simulations, and a queuing application, we demonstrate improvement in prediction uncertainty while respecting monotonicity information.

The proposed methodology will be most effective in situations where monotonicity of the function is not clear from the data.  As Example 1 indicates, when the training set has gaps or holes, the lack of nearby data points may lead conventional GP models to estimate a non-monotone relationship.  As the queuing application suggests, another challenging situation arises when the true function is monotone but nearly constant.  In both these situations, incorporating monotonicity information into the model is helpful, since the conventional GP model may infer a non-monotone relationship.

In some prediction scenarios (e.g. point D in Figure~\ref{qdatanew}), the monotonicity constraint may impact coverage of credible intervals.  Although credible intervals from the unconstrained model have higher coverage, they are of little value because the intervals are wide enough to be uninformative.  One may gain back some prediction uncertainty by relaxing the model assumptions in the proposed method. In our framework, this can be done by either relaxing the monotonicity assumption to some extent or relaxing the interpolation requirement. The former is done by choosing a smaller monotonicity parameter allowing for (small) positive probability for the occurrence of negative derivatives while the latter is performed by adding a nugget which allows for positive uncertainty at the evaluation points.

As input dimension increases, efficient computation will become more of a consideration.  If derivatives are constrained in $d_m$ input dimensions, then monotonicity information is required in each of these inputs. Increasing $d_m$ has the effect of  increasing the size of the covariance matrix in (\ref{covmat}) that has to be inverted with each evaluation of the likelihood. This can slow down the computation considerably. However, the SMC, in comparison to MCMC, has the advantage of being easily parallelizable and is shown to be stable as the dimensionality increases, thereby, being able to handle fairly high dimensional scenarios.
\newpage
\appendix

\section{Derivatives of the covariance function}
\label{derivatives}
The first and second derivatives of then Matern Covariance function with $\lambda=\frac{5}{2}$ are given by the following expressions, respectively,
$$g_{\lambda=\frac{5}{2}}(|x_{ik}-x_{jk}|,\theta)=(1+\theta |x_{ik}-x_{jk}|+\frac{1}{3}\theta^2 |x_{ik}-x_{jk}|^2)\exp(-\theta |x_{ik}-x_{jk}|).$$
\begin{align*}
\frac{\partial g_{\lambda=\frac{5}{2}}(|x_{ik}-x_{jk}|,\theta)}{\partial x_{ik}}=-\frac{1}{3}\theta^2|x_{ik}-x_{jk}|(1+\theta|x_{ik}-x_{jk}|)\exp(-\theta |x_{ik}-x_{jk}|)\text{sign}(x_{ik}-x_{jk})
\end{align*}

\begin{align*}
\frac{\partial^2 g_{\lambda=\frac{5}{2}}(|x_{ik}-x_{jk}|,\theta)}{\partial x_{ik} \partial x_{jk}}=\frac{1}{3}\theta^2(1+\theta|x_{ik}-x_{jk}|-\frac{\theta^2|x_{ik}-x_{jk}|^2}{2})\exp(-\theta |x_{ik}-x_{jk}|)\text{sign}(x_{ik}-x_{jk})
\end{align*}
where $$\theta= \frac{\sqrt{2\lambda}}{l}.$$

\newpage
\bibliographystyle{apalike}
\bibliography{refs}

\newpage
\listofcomments

\end{document}